\def\thecol{1}
\if\thecol1
\documentclass[12pt, draftclsnofoot, onecolumn]{IEEEtran}
\else
\documentclass[journal,10pt]{IEEEtran}
\fi

\usepackage{enumitem}
\usepackage{amssymb}
\usepackage{xcolor}

\usepackage{graphicx}
\usepackage{epsfig}
\usepackage{stfloats}
\ifCLASSOPTIONcompsoc
\usepackage[caption=false,font=normalsize,labelfon
t=sf,textfont=sf]{subfig}
\else
\usepackage[caption=false,font=footnotesize]{subfig}
\fi

\usepackage{amsfonts}
\usepackage{amsmath}
\usepackage{amsthm}
\usepackage{autobreak}
\usepackage{algorithm}
\usepackage{algorithmic}

\usepackage[style=ieee,url=false,doi=false,isbn=false,eprint=false]{biblatex}
\usepackage[colorlinks,linkcolor=blue,citecolor=blue]{hyperref}

\usepackage{booktabs}

\usepackage{amsthm}
\theoremstyle{definition}
\newtheorem{lem}{Lemma}
\newtheorem{prop}{Proposition}

\newtheorem{rem}{Remark}

\newtheorem{deff}{Definition}

\newcounter{eqlabel}[subsection]
\newcommand{\equlabel}{\refstepcounter{eqlabel}(\alph{eqlabel})}

\begin{document}

\title{Hierarchical-Absolute Reciprocity Calibration for Millimeter-wave Hybrid Beamforming Systems}

\author{Li Chen,~Rongjiang Nie,~Yunfei Chen,~and~Weidong Wang
\thanks{
		Li Chen, Rongjiang Nie, and Weidong Wang are with the CAS Key Laboratory of Wireless Optical Communication, University of Science and Technology of China, Hefei 230027, China (e-mail:chenli87@ustc.edu.cn; johnnrj@mail.ustc.edu.cn;  wdwang@ustc.edu.cn).}
\thanks{Yunfei Chen is with the School of Engineering, University of Warwick, Coventry CV4 7AL, U.K. (e-mail: Yunfei.Chen@warwick.ac.uk).}
}
\maketitle


\begin{abstract}
	In time-division duplexing (TDD) millimeter-wave (mmWave) massive multiple-input multiple-output (MIMO) systems, the reciprocity mismatch severely degrades the  performance of the hybrid beamforming (HBF). In this work, to mitigate the detrimental effect of the reciprocity mismatch, we investigate reciprocity calibration for the mmWave-HBF system with a fully-connected phase shifter network. To reduce the overhead and computational complexity of reciprocity calibration, we first decouple digital radio frequency (RF) chains and analog RF chains with beamforming design. Then, the entire calibration problem of the HBF system is equivalently decomposed into two subproblems corresponding to the digital-chain calibration and analog-chain calibration. To solve the calibration problems efficiently, a closed-form solution to the digital-chain calibration problem is derived, while an iterative-alternating optimization algorithm for the analog-chain calibration problem is proposed. To measure the performance of the proposed algorithm, we derive the Cram\'er-Rao lower bound on the errors in estimating mismatch coefficients. The results reveal that the estimation errors of mismatch coefficients of digital and analog chains are uncorrelated, and that the mismatch coefficients of receive digital chains can be estimated perfectly. Simulation results are presented to validate the analytical results and to show the performance of the proposed calibration approach.
\end{abstract}

\begin{refsection}
\section{Introduction}
In time-division duplexing (TDD) massive multiple-input multiple-output (MIMO) systems, the base station (BS) estimates the downlink channel state information (CSI) by exploiting the reciprocity of the wireless channel, to relieve the overhead of acquiring CSI\cite{Larsson2014Massive,Flordelis2018Massive}.

In practice, the estimated CSI is composed of not only the wireless propagation channel response but also the radio frequency (RF) response of RF chains \cite{Shan2018General}. The transmit and receive RF chains consist of different RF components. The transmit chain is composed of a digital-to-analog converter, a power amplifier,  etc., while the receive RF chain consists of an analog-to-digital converter, a low noise amplifier, etc. Due to the different compositions, the RF responses of transmit and receive chains are generally asymmetric, which results in the reciprocity mismatch of the uplink and downlink channels \cite{Nie2021Impact}.

The study of reciprocity mismatch has attracted extensive attention in the past decade and can be mainly classified into impact analysis and calibration design. To examine the impact of the reciprocity mismatch, some theoretical analyses have been provided for massive MIMO systems with linear precoding techniques, e.g., the zero-forcing (ZF) and matched filter (MF).  W. Zhang \emph{et al.} in \cite{Zhang2015LargeScale} studied the performance of the multi-user massive MIMO system with regularized ZF and MF precoding. They found that the reciprocity mismatch hardly caused any performance loss in the low signal-to-noise ratio (SNR) regime but severe performance loss in the high SNR regime. The theoretical results in \cite{Wei2016Impact} revealed that the reciprocity mismatch at the BS side was the key contributing factor to the multi-user interference and led to severe system performance degradation for the ZF precoding, while the mismatch at the user equipment (UE) side only led to very slight performance loss. Further, the theoretical comparison of MF and ZF precoding in \cite{Raeesi2018Performance,Luo2016Multiuser} indicated that the ZF-precoded system was more sensitive to the reciprocity mismatch than the MF-precoded system. The experimental results in \cite{Jiang2015MIMOTDD} verified the theoretical conclusions of the system performance in the presence of the reciprocity mismatch.

Since the reciprocity mismatch causes severe system performance degradation, the reciprocity calibration plays an essential role in the deployment of the massive MIMO system. \textcolor{black}{Unlike the CSI estimation error \cite{Mi2017Massive}, which changes with each channel realization, the reciprocity mismatch coefficients remain constant over hours or even days, and reciprocity calibration can be performed infrequently, e.g., once an hour.} Reciprocity calibration techniques can be mainly classified into two categories, which are hardware-based calibration and over-the-air (OTA) calibration. The hardware-based calibration utilizes the auxiliary circuits and components to connect the transmit RF chains and the receive RF chains. A real-time hardware-based calibration was first proposed in \cite{Nishimori2001Automatic} for narrowband conventional MIMO systems, where the transmitted data signals were used to calibrate the antenna array. Then, A. Bourdoux \emph{et al.} in \cite{Bourdoux2003Nonreciprocal} proposed a calibration approach which calibrated the different subcarriers respectively for wideband systems. To reduce transceiver interconnection effort, a daisy chain interconnection structure of the hardware circuits was proposed in \cite{Benzin2017Internal}, which also reduces the hardware cost of realizing reciprocal calibration to a certain extent. To study the trade-off between the connection structure and performance of the hardware-based calibration, X. Luo \emph{et al.} in \cite{Luo2019Massive} proposed an optimal interconnection by minimizing the Cram\'er-Rao lower bound (CRLB) of mismatch coefficients, which revealed that the star structure of hardware circuits was optimal. The hardware cost and circuit complexity of these hardware-based calibration methods increase with the number of antennas and may be unaffordable in the massive MIMO systems.

Different from the hardware-based calibration, the OTA calibration is based on the software and protocol design, which only utilizes air-interface signals between uncalibrated antennas to compute the calibration coefficients \cite{Kaltenberger2010Relative}. OTA calibration approaches can be divided into the full-end OTA calibration which was mainly used for conventional MIMO systems, and the partial-end OTA calibration which was designed for massive MIMO systems. In conventional systems, the OTA calibration requires both the BS and UE to get involved in the operation, and is therefore known as the full-end calibration. The full-end reciprocity calibration was first proposed in \cite{Guillaud2005practical}, and the total least squares (LS) algorithm was applied to solve calibration coefficients. Then, in \cite{Petermann2010Lowcomplexity}, the full-end calibration was extended to OFDM systems with each subcarrier calibrated independently in the frequency domain. In this case, the overhead and complexity of the reciprocity calibration increased with the number of subcarriers. To reduce the overhead and complexity, B. Kouassi \emph{et al.} in \cite{Kouassi2012Estimation} proposed a time-domain calibration for OFDM systems because the number of coefficients in the time domain was much less than those in the frequency domain. Since the overhead of channel feedback increases with the antenna number, the full-end calibration would produce heavy overhead pressure in massive MIMO systems. 

Thanks to the theoretical and experimental results that the reciprocity mismatch at the single-antenna UE only causes minor performance loss, the OTA calibration only needs to be performed at the BS side, which is known as the partial-end calibration or one-side calibration. C. Shepard \emph{et al.} in \cite{Shepard2012Argos} proposed a simple one-side calibration for the massive MIMO Argos prototype, which was sensitive to the fading channel and the location of the reference antenna. To avoid the issue of the Argos calibration, a partial-end calibration based on the strong mutual coupling between the adjacent antennas was presented in \cite{Wei2016Mutual}. By summarizing existing partial-end calibration approaches, X. Jiang \emph{et al.} proposed an OTA calibration framework in \cite{Jiang2018Framework}. Compared with co-located system, the calibration in distributed systems needs to gather the CSI from access points (APs).  To reduce the overhead of gathering the CSI, R. Roganlin \emph{et al.} in \cite{Rogalin2013Hardwareimpairment} proposed a hierarchical calibration which consisted of the intra-calibration and inter-calibration of AP. In \cite{Su2014Retrieving}, an OTA calibration with supporters was proposed for coordinated multi-point transmission systems to improve the SNR of calibration signals. To combat the path loss between the APs, our work in \cite{Nie2020Relaying} proposed a beamforming-based OTA calibration for distribution MIMO relaying systems.

\textcolor{black}{
Although the reciprocity calibration designs for full-digital beamforming (DBF) MIMO systems have been extensively investigated in recently, they can not be applied to the hybrid analog-digital beamforming (HBF) systems. Due to the more complex structure than DBF systems, the reciprocity calibration in HBF systems is more challenging. On one hand, a typical HBF transceiver possesses a hierarchical structure consisting of the digital precoder, digital RF chains, the analog precoder, and analog RF chains\cite{Molisch2017Hybrid}, which results in more complex modeling of the uplink-downlink channel reciprocity mismatch. On the other hand, the digital RF chains and the analog RF chains are coupled with the analog precoder, e.g., a phase-shifter network\cite{Wei2020Calibration}, so that the digital chains and analog chains can not transmit signals independently. A reciprocity calibration for the sub-connected phase-shifter network HBF system was proposed in \cite{Jiang2018Channel}, which transformed the sub-connected HBF transceiver to a DBF transceiver by virtually changing the position of the RF components to the front end near the antennas. When it is applied to the fully-connected HBF system, the dimension of the equivalent channel matrix after the transformation becomes much larger than the realistic channel, which results in a large overhead of the calibration. Additionally, since this reciprocity calibration can only acquire the ratio of coefficients of transmit and receive RF chains, mmWave channel estimation approaches, e.g., the approaches in \cite{Guo2017MillimeterWave,Venugopal2017Channel,Zhang2022MMVBased} which requires mismatch coefficients rather their ratios, remain unusable.  To reduce the overhead and recover the mmWave channel estimation, a relative reciprocity calibration approach was proposed for fully-connected mmWave HBF system in \cite{Wei2021TxRx}. Although this approach can reduce the calibration overhead to a certain extent, it requires UE to feed back received downlink calibration signals to BS, which can still causes large overhead. Further, since the relative calibration can not construct the equivalent channel, some existing hybrid beamforming designs, e.g. the designs proposed in \cite{Sohrabi2016Hybrid,Yu2016Alternating,Lin2019Hybrida}, cannot be applied in the calibrated systems.
}

Motivated by the above observations, we investigate the reciprocity calibration for TDD mmWave-HBF systems with the fully-connected phase shifter network. To reduce the overhead and complexity of the reciprocity mismatch in the fully-connected HBF system, hierarchical ideology is employed to calibrate digital and analog RF chains. Since digital and analog RF chains are physically coupled via a phase shifter network, we propose a beamforming design to virtually decouple the reciprocity calibration of digital and analog RF chains. Based on the decoupling operation, the entire reciprocity calibration problem is decoupled into two subproblems corresponding to the calibrations of digital RF chains and analog RF chains. To guarantee the application of mmWave channel estimation approaches, we propose an absolute reciprocity calibration approach to estimate the mismatch coefficients of transmit and receive RF chains. The mismatch coefficients of digital RF chains are solved from the closed-form expression of the solution to the digital-chain problem, while the mismatch coefficients of analog chains are jointly estimated with mmWave channel coefficients. Finally, the CRLB of the mismatch coefficients is derived to measure the performance of the proposed calibration. The main contributions of this work can be summarized as follows. 
\begin{itemize}
	\item \textbf{Reciprocity mismatch decoupling.} Since digital and analog RF chains are physically coupled via a phase shifter network, we propose a beamforming design to virtually decouple the digital and analog RF chains. Then, the entire reciprocity mismatch calibration problem of the HBF system is decomposed into two separate problems of digital-chain calibration and analog-chain calibration.
	\item \textbf{Absolute reciprocity calibration.} To guarantee the efficacy of mmWave-channel estimation approaches, we propose novel estimating methods to acquire the mismatch coefficients of RF chains. Specifically, the closed-form expression of digital-chain mismatch coefficients is derived, and an iterative-alternating estimation algorithm is proposed for analog-chain mismatch coefficients.
	\item \textbf{CRLB for estimating mismatch coefficients.} To measure the performance of the proposed algorithms, we derive the CRLB for the mismatch coefficient estimation. The CRLB reveals that the errors in estimating mismatch coefficients of digital chains and analog chains are independent of each other, and the mismatch coefficients of receive digital chains can be estimated perfectly. 
\end{itemize}

The rest of the paper is organized as follows. Section II describes the system model. The hierarchical-absolute reciprocity calibration for the mmWave-HBF system is proposed in Section III. In Section IV, the performance including the overhead, computational complexity, and CRLB of the proposed calibration is derived. Simulation results are given in Section V, and the conclusion is given in Section VI. For readability, some proofs are deferred to the supplementary material.

Throughout the paper, vectors and matrices are denoted in bold lowercase and uppercase respectively, e.g., $ \mathbf{a} $ and $ \mathbf{A} $. Let $ \mathbf{A}^{T} $, $ \mathbf{A}^{H} $, and $ \mathbf{A}^{-1} $ denote the transpose, conjugate transpose, and inverse of a matrix $ \mathbf{A}, $ respectively. $ \mathrm{tr}(\cdot) $, $ \mathbb{E}(\cdot) $, and $ \mathrm{vec}(\cdot) $ stand for the trace operator, the expectation operation, and column vectorization. Let $ |a| $ and $ \angle a $ denote the amplitude and phase of the complex number $ a $, and $ \|\cdot\|_{\mathrm{F}} $ denotes the Frobenius norm. $ \mathrm{diag}(a_1,\cdots,a_N) $ denotes an $ N $ by $ N $ diagonal matrix with diagonal entries given by $ a_1,\cdots,a_N $, and $ \mathrm{blkdiag}(\mathbf{a}_1,\cdots,\mathbf{a}_N) $ represents a block diagonal matrix. $ \otimes $, $ \odot $, and $ \circ $ represent the Kronecker product, Khatri–Rao product, and Hadamard product, respectively.  $\mathbb{C}$ and $\mathbb{R}$ stand for the complex numbers and real numbers, respectively. Let $[1:N]$ denote the set $\left\lbrace1,2,\cdots,N\right\rbrace$, and $ a\%b $ denote the remainder of $ a $ divided by $ b $. 





\section{System Model}

\if\thecol1
\begin{figure}
	\centering
	\includegraphics[width=0.85\linewidth]{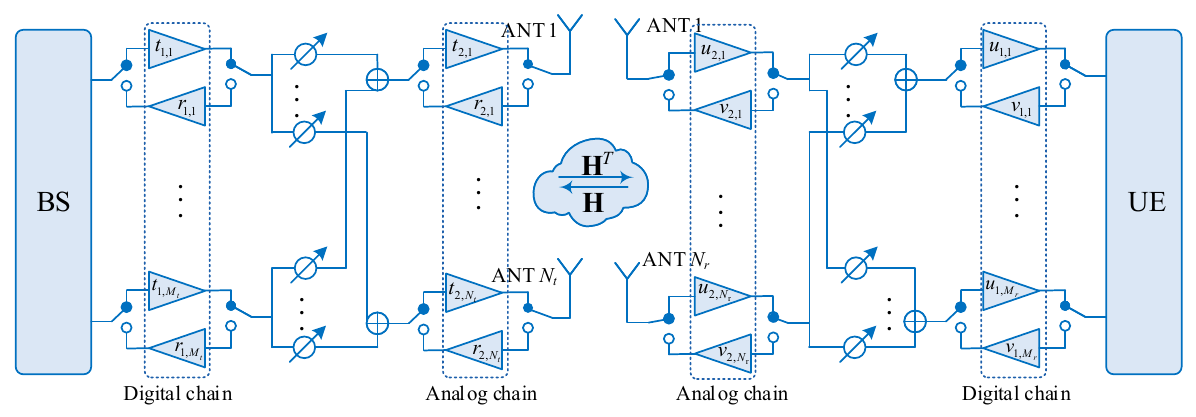}
	\caption{Hybrid beamforming massive MIMO with reciprocity mismatch.}
	\label{fig:RCMisHBF}
\end{figure}
\fi

\if\thecol2
\begin{figure*}
	\centering
	\includegraphics[width=0.75\linewidth]{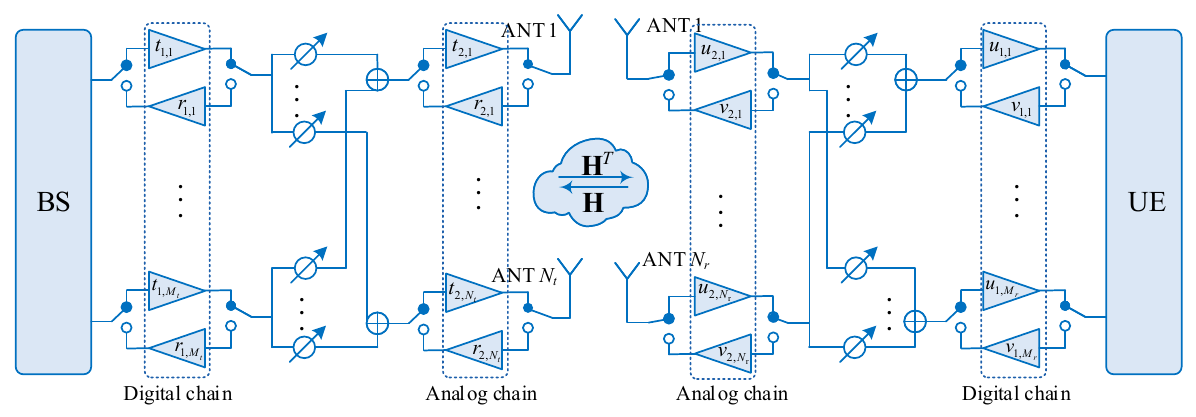}
	\caption{Hybrid beamforming massive MIMO with reciprocity mismatch.}
	\label{fig:RCMisHBF}
\end{figure*}
\fi

We consider an mmWave massive MIMO system as illustrated in Fig. \ref{fig:RCMisHBF}, where the BS is assumed to communicate with a single UE. The BS is quipped with $M_{\mathrm{t}}$ digital RF chains and $N_{\mathrm{t}}$ analog RF chains, and the UE is equipped with $M_{\mathrm{r}}$ digital RF chains and $N_{\mathrm{r}}$ analog RF chains. In both BS and UE, each analog chain is connected to an antenna in the uniform linear array (ULA), and the digital chains are connected to the analog chains via a fully-connected phase shift network.


In mmWave systems, the wireless channel is generally considered to possess limited scattering. Thus, we adopt a geometric channel model with $ K\ (K\ll N_{\mathrm{t}}, N_{\mathrm{r}}) $ scatters, and each scatter contributes to a single propagation path between the BS and UE. Based on these assumptions, the wireless channel between the BS and the UE can be modeled as
\begin{equation}
	\mathbf{H}=\sqrt{\frac{N_{\mathrm{t}}N_{\mathrm{r}}}{K}}\sum_{k=1}^{K}\alpha_{k}\mathbf{a}_{\mathrm{t}}(\theta_{k})\mathbf{a}_{\mathrm{r}}^T(\phi_k),
	\label{eq:wirelesschannel}
\end{equation}
where $\alpha_{k}\sim \mathcal{CN}(0,\sigma_{\alpha}^2)$ is the complex gain of the $k$-th path, $\mathbf{a}_{\mathrm{t}}(\theta_{k})\in \mathbb{C}^{N_{\mathrm{t}}}$ and $\mathbf{a}_{\mathrm{r}}(\phi_{k})\in \mathbb{C}^{N_{\mathrm{r}}}$ denote the array steering vectors of the BS and UE which are given by
\begin{equation}
	\begin{split}
		\mathbf{a}_{\mathrm{t}}(\theta_{k})&=\left[1,e^{-j\frac{2\pi d}{\lambda}\sin\theta_{k}},\cdots,e^{-j\frac{2\pi d}{\lambda}(N_\mathrm{t}-1)\sin\theta_{k}}\right]^T,\\
		\mathbf{a}_{\mathrm{r}}(\phi_{k})&=\left[1,e^{-j\frac{2\pi d}{\lambda}\sin\phi_{k}},\cdots,e^{-j\frac{2\pi d}{\lambda}(N_\mathrm{r}-1)\sin\phi_{k}}\right]^T,
	\end{split}
\end{equation}
$\lambda$ is the wavelength of the carrier, and $d$ is the distance of the adjacent antenna set to $ \lambda/2 $, $ \theta_{k}\in[-\pi/2,\pi/2) $ and $ \phi_{k}\in[-\pi/2,\pi/2) $ are the azimuth angles of arrival or departure (AoAs/AoDs) of the BS and MS.

\textcolor{black}{
In a practical system, the receive and transmit RF chains are generally asymmetric. Let $ \mathbf{T}_1/\mathbf{R}_1 $ and $ \mathbf{T}_2/\mathbf{R}_2 $ represent the mismatch matrices of the transmit/receive digital and analog RF chains of the BS, and denote $ \mathbf{V}_1/\mathbf{U}_1 $ and $ \mathbf{V}_2/\mathbf{U}_2 $ as the mismatch matrices of the transmit/receive digital and analog RF chains of the UE. All of these matrices are diagonal and defined as
\begin{equation}
	\begin{split}
		&\mathbf{T}_{1}=\mathrm{diag}(t_{1,1},\cdots,t_{1,M_{\mathrm{t}}}),\ \mathbf{T}_{2}=\mathrm{diag}(t_{2,1},\cdots,t_{2,N_{\mathrm{t}}}),\\
		&\mathbf{R}_{1}=\mathrm{diag}(r_{1,1},\cdots,r_{1,M_{\mathrm{t}}}),\ \mathbf{R}_{2}=\mathrm{diag}(r_{2,1},\cdots,r_{2,N_{\mathrm{t}}}),\\
		&\mathbf{V}_{1}=\mathrm{diag}(v_{1,1},\cdots,v_{1,M_{\mathrm{r}}}),\ \mathbf{V}_{2}=\mathrm{diag}(v_{2,1},\cdots,v_{2,N_{\mathrm{r}}}),\\
		&\mathbf{U}_1=\mathrm{diag}(u_{1,1},\cdots,u_{1,M_{\mathrm{r}}}),\ \mathbf{U}_{2}=\mathrm{diag}(u_{2,1},\cdots,u_{2,N_{\mathrm{r}}})
	\end{split}
\end{equation}
where $ t_{1,m}/r_{1,m} $ denotes the mismatch coefficient of the $ m $-th ($ m\in[1:M_{\mathrm{t}}] $) transmit/receive digital RF chain of the BS, and $ t_{2,i}/r_{2,i} $ denotes the mismatch coefficient of the $ i $-th ($ i\in[1:N_{\mathrm{t}}] $) transmit/receive analog RF chain of the BS. At the UE side, $ v_{1,\bar{m}}/u_{1,\bar{m}} $ denotes the mismatch coefficient of the $ \bar{m} $-th ($ \bar{m}\in[1:M_{\mathrm{r}}] $) transmit/receive digital RF chain, and $ v_{2,\bar{i}}/u_{2,\bar{i}} $ denotes the mismatch coefficient of the $ \bar{i} $-th ($ \bar{i}\in[1:N_{\mathrm{r}}] $) transmit/receive analog RF chain.
}

\textcolor{black}{
As depicted in Fig. \ref{fig:RCMisHBF}, the overall channel observed by the baseband processor is the combination of the wireless channel, the digital RF chain, the phase shifter network, and the analog RF chain, which can be expressed by
\begin{equation}
	\tilde{\mathbf{H}}_{\mathrm{UL}}=\tilde{\mathbf{R}}\mathbf{H}\tilde{\mathbf{V}},\quad
	\tilde{\mathbf{H}}_{\mathrm{DL}}=\tilde{\mathbf{U}}\mathbf{H}^T\tilde{\mathbf{T}},
	\label{eq:realchannel}
\end{equation}
where $\tilde{\mathbf{R}}=\mathbf{R}_2\mathbf{F}_{\mathrm{r}}^T\mathbf{R}_1$, $\tilde{\mathbf{T}}=\mathbf{T}_1\mathbf{F}_{\mathrm{t}}\mathbf{T}_2$, $\mathbf{F}_{\mathrm{r}}\in\mathbb{C}^{N_{\mathrm{t}}\times M_{\mathrm{t}}}$ and $\mathbf{F}_{\mathrm{t}}\in\mathbb{C}^{N_{\mathrm{t}}\times M_{\mathrm{t}}}$ are the analog receive and transmit beamforming matrices of the BS, $\tilde{\mathbf{V}}=\mathbf{V}_1\mathbf{B}_{\mathrm{t}}\mathbf{V}_2$, $\tilde{\mathbf{U}}=\mathbf{U}_2\mathbf{B}_{r}^T\mathbf{U}_1$, $\mathbf{B}_{\mathrm{t}}\in\mathbb{C}^{N_{\mathrm{r}}\times M_{\mathrm{r}}}$ and $\mathbf{B}_{\mathrm{r}}\in\mathbb{C}^{N_{\mathrm{r}}\times M_{\mathrm{r}}}$ are the analog beamforming matrices of the UE.
}

Based on the channel modeling and system setting, the downlink transmission signal received by the UE can be denoted as
\begin{equation}
	\mathbf{y}=\mathbf{D}_{\mathrm{r}}^T\mathbf{U}_1\mathbf{B}_{\mathrm{r}}^T\mathbf{U}_2\mathbf{H}^T\mathbf{T}_2\mathbf{F}_{\mathrm{t}}\mathbf{T}_1\mathbf{W}_{\mathrm{t}}\mathbf{s}+\mathbf{D}_{\mathrm{r}}^T\mathbf{U}_{\mathrm{1}}\mathbf{B}_{\mathrm{r}}^T\mathbf{n},
	\label{eq:downlinktrans}
\end{equation}
where $\mathbf{D}_{\mathrm{r}}\in\mathbb{C}^{M_{\mathrm{r}}\times M_{\mathrm{r}}}$ is the digital combining matrix of the UE, $\mathbf{W}_{\mathrm{t}}\in\mathbb{C}^{M_{\mathrm{t}}\times M_{\mathrm{t}}}$ denotes the digital precoding matrix of the BS, $ \mathbf{s}\in\mathbb{C}^{N_{\mathrm{s}}} $ denotes the data vector satisfying $ \mathbb{E}\left\lbrace \mathbf{s}\mathbf{s}^H \right\rbrace=\rho_{\mathrm{d}}\mathbf{I}_{N_{\mathrm{s}}}$, $ \rho_{\mathrm{d}} $ denotes the average transmit power, $ N_{\mathrm{s}} $ is the number of data streams, and $ \mathbf{n}\in\mathbb{C}^{N_{\mathrm{r}}} $ represents the additive white Gaussian noise (AWGN) vector with distribution $ \mathbf{n}\sim \mathcal{CN}(\mathbf{0},\sigma_{\mathrm{n}}^2\mathbf{I}_{N_{\mathrm{r}}}) $. 

In TDD mode, the digital and analog beamforming matrices are computed by the BS based on the knowledge of uplink CSI. According to \eqref{eq:realchannel}, the estimated uplink CSI is unequal to the downlink channel response at all, which is known as the reciprocity mismatch of the uplink and downlink channel. With the reciprocity mismatch, the existing beamforming approaches for HBF systems, e.g., \cite{Ayach2014Spatially}, fail to achieve satisfactory performance. Further, due to the uncertainty of the reciprocity mismatch coefficients, mmWave channel estimation approaches like \cite{Guo2017MillimeterWave} are invalid. Accordingly, the reciprocity calibration is essential for mmWave-HBF systems.


\section{Reciprocity Calibration for mmWave-HBF System}
In this section, the reciprocity calibration approach is proposed. We first introduce an existing reciprocity calibration approach for HBF systems and discuss its limitation in applying to the fully-connected structure. Then, an absolute reciprocity calibration for the mmWave-HBF system is proposed, which takes advantage of the particularity of the fully-connected structure to decouple the calibrations of digital RF chains and analog RF chains.

\subsection{Conventional Reciprocity Calibration Approach of HBF System}

The conventional reciprocity calibration (CRC) of HBF was proposed in \cite{Jiang2018Channel}, which is an extension of the relative calibration of the full-digital MIMO system. The CRC treats the HBF system as a virtual full-digital MIMO with $ N_{\mathrm{t}}M_{\mathrm{t}}$ virtual antennas and applies OTA signals to estimate the ratio of the transmit and receive mismatch coefficients, which are also called relative calibration coefficients.

\textcolor{black}{
In the CRC, the equivalent transmit and receive mismatch coefficients of the BS are defined as $\mathbf{T}_{\mathrm{eq}}=\mathbf{T}_1\otimes \mathbf{T}_2$ and $\mathbf{R}_{\mathrm{eq}}=\mathbf{R}_1\otimes \mathbf{R}_2$, and the equivalent mismatch coefficients of the UE are defined as $ \mathbf{V}_{\mathrm{eq}}=\mathbf{V}_1\otimes \mathbf{V}_{2}$ and $ \mathbf{U}_{\mathrm{eq}}=\mathbf{U}_1\otimes \mathbf{U}_{2} $. The equivalent uplink and downlink channels are defined as $\mathbf{H}_{\mathrm{UL,eq}}=(\mathbf{r}\otimes \mathbf{R}_2)\mathbf{H}(\mathbf{v}_1\otimes \mathbf{V}_2)$ and $\mathbf{H}_{\mathrm{DL,eq}}=(\mathbf{u}_1\otimes \mathbf{U}_2)\mathbf{H}^T(\mathbf{t}_1\otimes \mathbf{T}_2)$, where $ \mathbf{t}_1 $, $ \mathbf{r}_1 $, $ \mathbf{v}_1 $, and $ \mathbf{u}_1 $ consist of the diagonal entries of $ \mathbf{T}_1 $, $ \mathbf{R}_1 $, $ \mathbf{V}_1 $, and $ \mathbf{U}_1 $, respectively.
}
Based on these definitions, the equation of CRC can be denoted as
\begin{equation}
	\mathbf{H}_{\mathrm{DL,eq}}=\underbrace{\mathbf{U}_{\mathrm{eq}}\mathbf{V}_{\mathrm{eq}}^{-1}}_{\mathbf{C}_{\mathrm{UE}}^{-1}}\mathbf{H}_{\mathrm{UL,eq}}^T\underbrace{\mathbf{R}_{\mathrm{eq}}^{-1}\mathbf{T}_{\mathrm{eq}}}_{\mathbf{C}_{\mathrm{BS}}},
	\label{eq:crcequchanel}
\end{equation}
where $\mathbf{C}_{\mathrm{BS}}$ and $\mathbf{C}_{\mathrm{UE}}$ represent the relative calibration matrices of the BS and UE.


To obtain the relative calibration coefficients, it is necessary to acquire the equivalent uplink and downlink CSI. To estimate the equivalent downlink CSI, the BS transmits $L_{\mathrm{crc}}$-length pilots by using $Q_{\mathrm{crc}}$ transmit beamforming matrices, and the UE receives the pilots with $P_{\mathrm{crc}}$ receive beamforming matrices, where $P_{\mathrm{crc}}Q_{\mathrm{crc}}=L_{\mathrm{crc}}$. Assume that a $Q_{\mathrm{crc}}$-length pilot sequence is denoted as $\{\mathbf{x}_{1},\mathbf{x}_{2},\cdots,\mathbf{x}_{Q_{\mathrm{crc}}}\}$, where $\mathbf{x}_{q}$ denotes the pilot during the $[(p-1)Q_{\mathrm{crc}}+q]$-th transmission $(p\in[1:P_{\mathrm{crc}}])$ satisfying $ \mathbb{E}\left\lbrace \mathbf{x}_{q}\mathbf{x}_{q}^H\right\rbrace=\rho_{\mathrm{c}}\mathbf{I}_{M_{\mathrm{r}}} $. During the training, the digital precoding and combining matrices are set as identity matrices, i.e.,  $\mathbf{D}_{\mathrm{r}}=\mathbf{I}_{M_{\mathrm{r}}}$ and $\mathbf{W}_{\mathrm{t}}=\mathbf{I}_{M_\mathrm{t}}/\sqrt{M_{\mathrm{t}}}$, and the signal received by the UE can be denoted as
\begin{equation}
	\begin{split}
		\mathbf{y}_{\mathrm{UE},p,q}&=\mathbf{U}_1\mathbf{B}_{\mathrm{r},p}\mathbf{H}_{\mathrm{DL}}\mathbf{F}_{\mathrm{t},q}\mathbf{T}_1\mathbf{x}_{q}+\mathbf{U}_1\mathbf{B}_{\mathrm{r},p}\mathbf{n}_{\mathrm{UE},p,q}\\
		&=\mathbf{B}_{\mathrm{eq},p}\mathbf{H}_{\mathrm{DL,eq}}\mathbf{F}_{\mathrm{eq},q}\mathbf{x}_q+\mathbf{n}_{\mathrm{eq},p,q},
	\end{split}
\end{equation}
where $ \mathbf{H}_{\mathrm{DL}}=\mathbf{U}_2\mathbf{H}^T\mathbf{T}_2 $, $\mathbf{F}_{\mathrm{t},q}$ denotes the analog beamforming matrix at the BS, $\mathbf{B}_{\mathrm{r},q}$ is the analog combining matrix at the UE, $\mathbf{F}_{\mathrm{eq},q}=\mathrm{blkdiag}(\mathbf{f}_{q,1},\mathbf{f}_{q,2},\cdots,\mathbf{f}_{q,M_{\mathrm{t}}})$, $\mathbf{f}_{q,m}$ denotes the $m$-th column of $\mathbf{F}_{\mathrm{t},q}$, $\mathbf{B}_{\mathrm{eq},p}=\mathrm{blkdiag}(\mathbf{b}_{p,1},\mathbf{b}_{p,2},\cdots,\mathbf{b}_{p,M_{\mathrm{r}}})$, and $\mathbf{b}_{p,m}$ is the $m$-th row of $\mathbf{B}_{\mathrm{r},p}$.

By stacking all $L_{\mathrm{crc}}$-length signals in matrix form denoted as $ \mathbf{Y}_{\mathrm{UE}}=[\bar{\mathbf{Y}}_{\mathrm{UE},1}^T,\cdots, \bar{\mathbf{Y}}_{\mathrm{UE},P_{\mathrm{crc}}}^T]^T\in\mathbb{C}^{M_{\mathrm{r}}P_{\mathrm{crc}}\times Q_{\mathrm{crc}}}$ and $ \bar{\mathbf{Y}}_{\mathrm{UE},p}=[\mathbf{y}_{\mathrm{UE},p,1},\cdots,\mathbf{y}_{\mathrm{UE},p,Q_{\mathrm{crc}}}]$, the received signal model can be given by
\begin{equation}
	\mathbf{Y}_{\mathrm{UE}}=\tilde{\mathbf{B}}_{\mathrm{eq}}\mathbf{H}_{\mathrm{DL,eq}}\tilde{\mathbf{F}}_{\mathrm{eq}}+\mathbf{N}_{\mathrm{eq}},
\end{equation}
where $ \tilde{\mathbf{B}}_{\mathrm{eq}}= [\mathbf{B}_{\mathrm{eq},1}^T,\mathbf{B}_{\mathrm{eq},2}^T,\cdots,\mathbf{B}_{\mathrm{eq},P_{\mathrm{crc}}}^T]^T \in\mathbb{C}^{M_{\mathrm{r}}P_{\mathrm{crc}}\times N_{\mathrm{r}}}$, $ \tilde{\mathbf{F}}_{\mathrm{eq}}=[\mathbf{F}_{\mathrm{eq},1}\mathbf{x}_1,\mathbf{F}_{\mathrm{eq},2}\mathbf{x}_2,\cdots,\mathbf{F}_{\mathrm{eq},Q_{\mathrm{crc}}}\mathbf{x}_{Q_{\mathrm{crc}}}] \in\mathbb{C}^{N_{\mathrm{t}}\times Q_{\mathrm{crc}}}$, $\mathbf{N}_{\mathrm{eq}}=[\bar{\mathbf{N}}_{\mathrm{eq},1}^T,\cdots,\bar{\mathbf{N}}_{\mathrm{eq},P_{\mathrm{crc}}}^T]^T$, and $ \bar{\mathbf{N}}_{\mathrm{eq},p}=[\mathbf{n}_{\mathrm{eq},p,1},\cdots,\mathbf{n}_{\mathrm{eq},p,Q_{\mathrm{crc}}}]$. By vectoring the matrix $\mathbf{Y}_{\mathrm{UE}}$, the received signal can be further denoted as
\begin{equation}
	\mathrm{vec}(\mathbf{Y}_{\mathrm{UE}})=\mathbf{B}_{\mathrm{crc}}\mathrm{vec}(\mathbf{H}_{\mathrm{DL,eq}})+\mathrm{vec}(\mathbf{N}_{\mathrm{eq}}),
\end{equation}
where $ \mathbf{B}_{\mathrm{crc}}=(\tilde{\mathbf{F}}_{\mathrm{eq}}^T\otimes \tilde{\mathbf{B}}_{\mathrm{eq}}) $. Using the LS approach\cite{Zhang2017Matrix}, the equivalent downlink channel is estimated as
\begin{equation}
	\mathrm{vec}(\mathbf{H}_{\mathrm{DL,eq}})=(\mathbf{B}_{\mathrm{crc}}^H\mathbf{B}_{\mathrm{crc}})^{-1}\mathbf{B}_{\mathrm{crc}}^H\mathrm{vec}(\mathbf{Y}_{\mathrm{UE}}).
	\label{eq:crcestichannel}
\end{equation}

Similarly, to estimate the equivalent uplink channel, the UE transmit the uplink training pilots to the BS. Further, to estimate the calibration coefficients, the UE feeds back the estimated downlink channel to the BS.

After the BS estimates the uplink channel and receives the downlink channel fed back from the UE, the calibration coefficients can be computed by the following proposition. 
\begin{prop}[CRC coefficients]
	With the knowledge of the equivalent uplink and downlink channels, the CRC coefficients can be computed by
	\begin{equation}
		\mathbf{c}=[1,-\mathbf{h}_{\mathrm{CRC},1}^T\mathbf{H}_{\mathrm{CRC},2}^*(\mathbf{H}_{\mathrm{CRC},2}^T\mathbf{H}_{\mathrm{CRC},2}^*)^{-1}]^T,
	\end{equation}
	where $\mathbf{h}_{\mathrm{CRC},1}$ is the first column of matrix $\mathbf{H}_{\mathrm{CRC}}$, $\mathbf{H}_{\mathrm{CRC},2}$ consists of the second to last columns of matrix $\mathbf{H}_{\mathrm{CRC}}$, $\mathbf{H}_{\mathrm{CRC}} $ is an $(N_{\mathrm{t}}M_{\mathrm{t}}+N_{\mathrm{r}}M_{\mathrm{r}})$-order square matrix defined as $ \mathbf{H}_{\mathrm{CRC}}=[\mathbf{I}_{M_{\mathrm{t}}N_{\mathrm{t}}}\odot \mathbf{H}_{\mathrm{UL,eq}}^T,-\mathbf{H}_{\mathrm{DL,eq}}^T\odot \mathbf{I}_{M_{\mathrm{r}}N_{\mathrm{r}}}] $.
\end{prop}

\begin{IEEEproof}
	The results can be derived based on \cite{Jiang2018Channel} by assuming that the antennas of the BS are divided into the group $ \mathcal{A} $ and the antennas of the UE are divided into the group $ \mathcal{B} $.
\end{IEEEproof}

To measure the complexity of the CRC, we further derive the overhead and computational complexity. The overhead of reciprocity calibration can be expressed by the count of channel use for transmitting calibration signals and feeding back the estimated CSI. The computational complexity can be measured by the times of multiplication for estimating the channel state information and computing the mismatch coefficients.

\begin{rem}[Overhead and complexity of CRC]
	According to \eqref{eq:crcestichannel}, $\tilde{\mathbf{F}}_{\mathrm{eq}}^T\otimes \tilde{\mathbf{B}}_{\mathrm{eq}}$ must be a full column rank matrix to estimate the downlink channel. Based on the property of the Kronecker product, the pilots must satisfy the condition that $Q_{\mathrm{crc}}\ge N_{\mathrm{t}}M_{\mathrm{t}}$ and $P_{\mathrm{crc}}\ge N_{\mathrm{r}}$, and $ L_{\mathrm{crc}}\geq N_{\mathrm{t}}M_{\mathrm{t}}N_{\mathrm{r}} $. This result means that the least overhead of downlink channel estimation is $ N_{\mathrm{t}}M_{\mathrm{t}}N_{\mathrm{r}}  $. Since the uplink channel estimation is similar to the downlink channel estimation, the entire overhead of the CRC can be denoted as $N_{\mathrm{t}}N_{\mathrm{r}}(M_{\mathrm{t}}+M_{\mathrm{r}}+1)$. Further, since the computation complexity is mainly determined by computing the inverse matrices of $ \mathbf{B}_{\mathrm{crc}}^H\mathbf{B}_{\mathrm{crc}}\in\mathbb{C}^{N_{\mathrm{t}}M_{\mathrm{t}}N_{\mathrm{r}}M_{\mathrm{r}}\times N_{\mathrm{t}}M_{\mathrm{t}}N_{\mathrm{r}}M_{\mathrm{r}}} $ and $ \mathbf{H}_{\mathrm{CRC},2}^T\mathbf{H}_{\mathrm{CRC},2}^*\in\mathbb{C}^{(N_{\mathrm{t}}M_{\mathrm{t}}+N_{\mathrm{r}}M_{\mathrm{r}})\times(N_{\mathrm{t}}M_{\mathrm{t}}+N_{\mathrm{r}}M_{\mathrm{r}})} $, the computation complexity of the CRC  is $\mathcal{O}(N_{\mathrm{t}}^3M_{\mathrm{t}}^3N_{\mathrm{r}}^3M_{\mathrm{r}}^3)$.	
\end{rem}


In the CRC, the dimensions of the equivalent channels $ \mathbf{H}_{\mathrm{UL,eq}} $ and $ \mathbf{H}_{\mathrm{DL,eq}} $ (see \eqref{eq:crcequchanel}) are much larger than that of the actual wireless channel matrix $ \mathbf{H} $ (see \eqref{eq:wirelesschannel}), which generates the heavy overhead of the channel estimation and high computational complexity. Further, since the CRC only estimates the ratio of the mismatch coefficients of transmit chains and receive chains, mmWave channel estimation of mmWave systems, which requires the knowledge of the individual mismatch coefficients, becomes invalid.  Thus, due to limitations of the CRC in fully connected mmWave-HBF systems, we propose a hierarchical-absolute calibration (HAC) approach, which decouples the reciprocity calibration of digital RF chains and analog RF chains and estimates the individual mismatch coefficients of transmit chains and receive chains, respectively.


\subsection{Decouple Principle of HAC}\label{sec:pilotdesign}


To reduce the overhead and complexity of the reciprocity calibration of fully connected HBF systems, digital RF chains and analog RF chains must be calibrated individually, which means hierarchical calibration. To adopt the mmWave channel estimation approaches of mmWave systems, the individual mismatch coefficients are required rather than the ratio of the mismatch coefficients, which can be addressed by applying the absolute reciprocity calibration.

\textcolor{black}{
However, due to the fully-connected structure of the HBF system, HAC encounters two challenges. On the one hand, the digital RF chains and analog RF chains are physically coupled via the fully-connected phase shift network, which results in the decoupling challenge. On the other hand, the fully-connected phase shift network causes that the RF chains can not transmit and receive signals independently, which results in the calibration challenge. To a certain degree, these problems can be addressed by using extra auxiliary circuits to assistant the reciprocity calibration, e.g. the calibration approaches presented in \cite{Mi2018SelfCalibration,Mi2020Massive}. But the auxiliary circuits may bring extra non-reciprocity, and the calibration accuracy of hardware-circuit calibration highly depends on the auxiliary circuits \cite{Wei2016Mutual}. Thus, we propose an OTA-based HAC for mmWave-HBF systems. Specifically, the calibrations of digital and analog RF chains are decoupled by a targeted beamforming scheme, and the mismatch coefficients are estimated by the OTA training signals between the BS and UE.
}

\textcolor{black}{
In the rest of this section, we will introduce the decoupling principle of the proposed HAC. Since the transmitter and receiver employ similar HBF structures, the decouple operation is first explained  in the multi-input single-output (MISO) system for clarity, and then we will propose the \emph{concrete design} for the general HBF-MIMO system.
}

\if\thecol1
\begin{figure}
	\centering\vspace{-2em}
	\subfloat[The HBF structure of transmited signals.]{
		\includegraphics[width=0.45\linewidth]{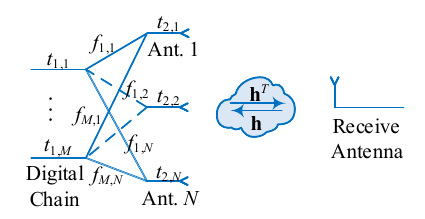}
		\label{fig:primaltr}
	}
	\subfloat[The virtual array of transmited signals.]{
		\includegraphics[width=0.45\linewidth]{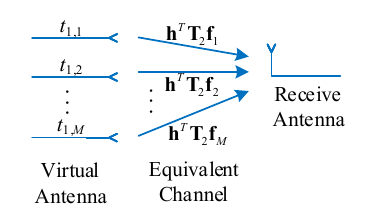}
		\label{fig:eqchtrRF}
	}
	\caption{Decouple the digital RF chains from the analog RF chains.}
	\label{fig:trdecuple}
	\vspace{-1em}
\end{figure}
\else
\begin{figure}
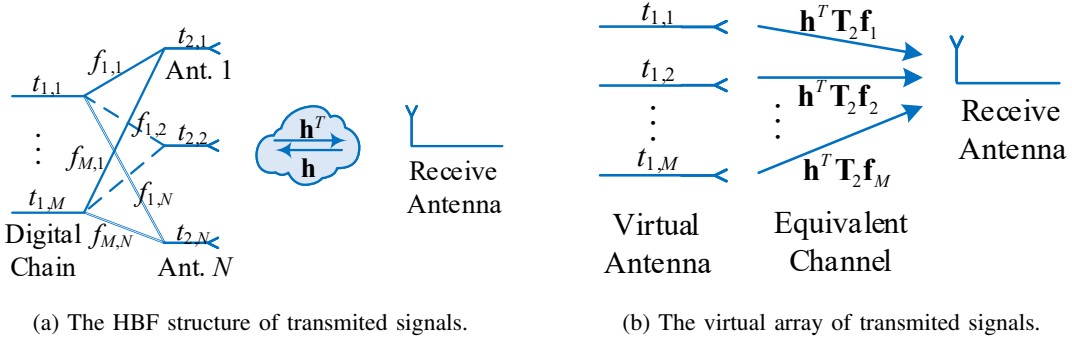

	\centering
	\subfloat[The HBF structure of transmited signals.]{
		\includegraphics[width=0.8\linewidth]{TransmitRFChain1.pdf}
		\label{fig:primaltr}
	}
\vfill
	\subfloat[The virtual array of transmited signals.]{
		\includegraphics[width=0.8\linewidth]{TransmitRFChain2.pdf}
		\label{fig:eqchtrRF}
	}
	\caption{Decouple the digital RF chains from the analog RF chains.}
	\label{fig:trdecuple}
\end{figure}
\fi


\textcolor{black}{
\textbf{MISO system for decoupling digital chains from the analog chains:}} We consider a MISO system where the transmitter is equipped with the fully-connected HBF structure and the receiver is equipped with a single antenna as illustrated in Fig. \ref{fig:primaltr}. Thanks to the fully-connected structure, the signal transmitted from each digital RF chain passes through all antennas in the transmitter and all wireless channels. For example, if the $m$-th transmit digital chain transmit a signal $x_m$ to the receiver, the received signal can be denoted as
\if\thecol1
\begin{equation}
		y_m=t_{1,m}\underbrace{\mathbf{h}^T\mathbf{T}_2\mathbf{f}_m}_{h_{\mathrm{eq},m}}x_m+n_m
		=t_{1,m}h_{\mathrm{eq},m}x_m+n_m,
\end{equation}
\else
\begin{equation}
	\begin{split}
		y_m&=t_{1,m}\underbrace{\mathbf{h}^T\mathbf{T}_2\mathbf{f}_m}_{h_{\mathrm{eq},m}}x_m+n_m\\
		&=t_{1,m}h_{\mathrm{eq},m}x_m+n_m,
	\end{split}
\end{equation}
\fi
where $h_{\mathrm{eq},m}$ denotes the virtual equivalent channel between the $m$-th transmit digital chain and the receive antenna. Based on this, the HBF MISO system can be virtually constructed as a DBF MISO system as illustrated in Fig. \ref{fig:eqchtrRF}, where the virtual antennas are the transmit digital chains, and the virtual-equivalent channels consist of the phase shift network, the analog RF chains, and the wireless channel. Further, it can be found that the virtual-equivalent channels equal to each other when the beamforming vectors are identical, i.e., $\mathbf{f}_1=\cdots=\mathbf{f}_M$. Thus, by applying this analog beamforming design, the digital chains can be decoupled from the analog chains, and the absolute reciprocity calibration can be considered as the calibration with known channel gains. 

\if\thecol2
\begin{figure}
	\centering
	\subfloat[The ABF structure of transmited signals.]{
		\includegraphics[width=0.8\linewidth]{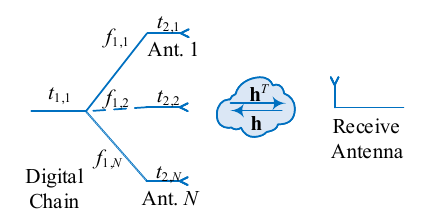}
		\label{fig:abfstruc}
	}
	\vfill
	\subfloat[The virtual array of transmited signals.]{
		\includegraphics[width=0.8\linewidth]{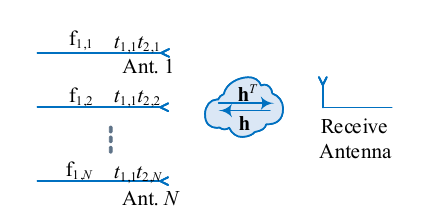}
		\label{fig:abs2digit}
	}
	\caption{Decouple the analog RF chains from digital RF chains.}
\end{figure}
\fi

\textcolor{black}{
\textbf{MISO system for decoupling the analog chains from the digital chains:}} Since each digital RF chain is connected to all analog RF chains via the fully-connected phase shifter network, the antenna array can be considered as an ABF system as shown in Fig. \ref{fig:abfstruc}. By using only one digital RF chain to transmit and receive calibration signals, the analog RF chains can be decoupled from the digital RF chains as illustrated in Fig. \ref{fig:abs2digit}. Based on this design, the analog RF chains can be calibrated with signal processing approaches.

\if\thecol1
\begin{figure}
	\centering
	\vspace{-2em}
	\subfloat[The ABF structure of transmited signals.]{
		\includegraphics[width=0.45\linewidth]{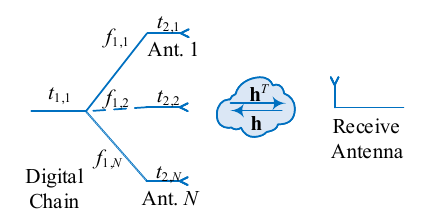}
		\label{fig:abfstruc}
	}
	\subfloat[The virtual array of transmited signals.]{
		\includegraphics[width=0.45\linewidth]{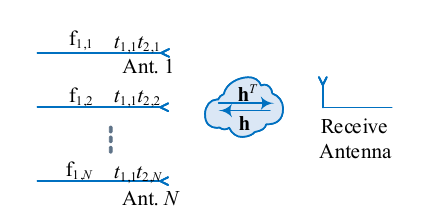}
		\label{fig:abs2digit}
	}
	\caption{Decouple the analog RF chains from digital RF chains.}
\end{figure}
\begin{figure}
	\centering
	\vspace{-1em}
	\includegraphics[width=0.6\linewidth]{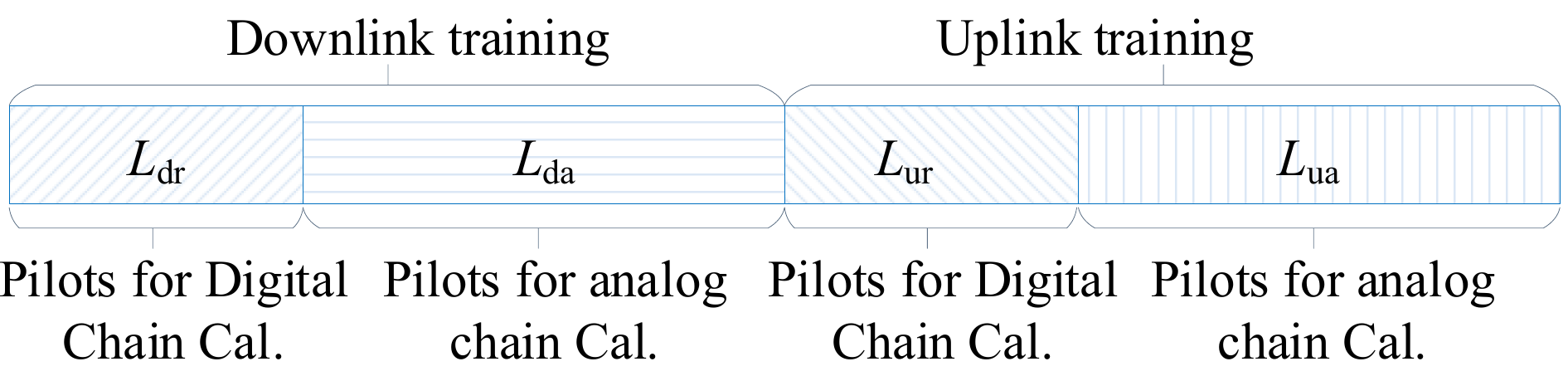}
	\vspace{-1em}
	\caption{The overall calibration process.}
	\label{fig:calprocess}
	\vspace{-1em}
\end{figure}
\else
\begin{figure}
	\centering
	\includegraphics[width=0.9\linewidth]{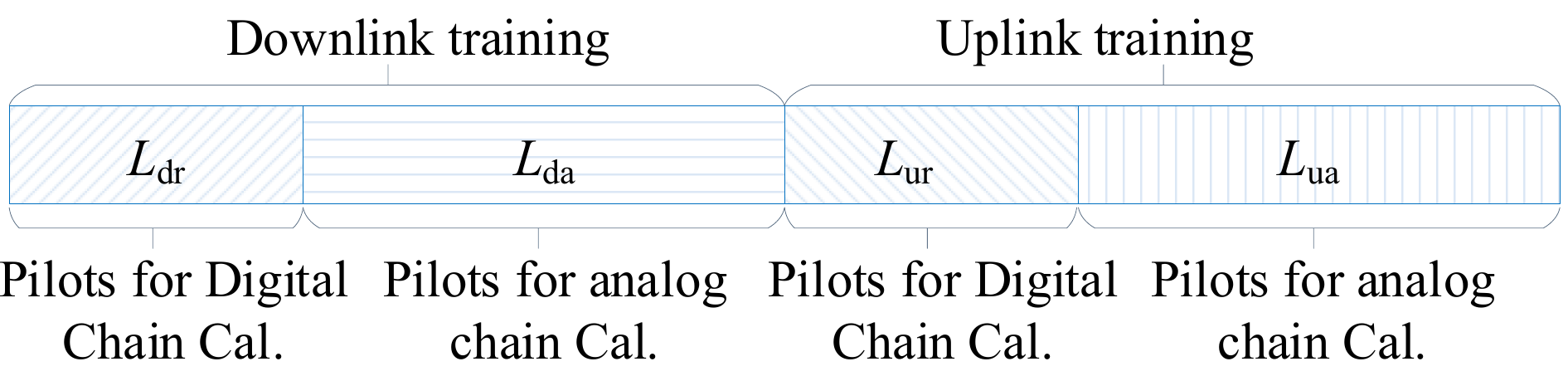}
	\caption{The overall calibration process.}
	\label{fig:calprocess}
\end{figure}
\fi

\textbf{Concrete design:} Based on the above MISO systems, the decoupling principle can be extended to a general case where both the BS and UE are equipped with multiple antennas and HBF structures as illustrated in Fig. \ref{fig:RCMisHBF}. For general point-to-point HBF MIMO systems, the overall HAC training phases can be divided into two phases, which are downlink training and uplink training as illustrated in Fig. \ref{fig:calprocess}. During the training processes, the calibration training signals and beamforming matrices of the BS and UE should be designed to decouple digital and analog RF chains. By taking the downlink training phase as an example, the concrete designs are given as follows.
\begin{itemize}
	\item \textbf{Downlink training pilots:} The entire $ L_{\mathrm{d}} $-length downlink training pilots consist of $ L_{\mathrm{dr}} $-length pilots for calibrating digital RF chains and $ L_{\mathrm{da}} $-length pilots for calibrating the analog RF chains, where $ L_{\mathrm{dr}}+L_{\mathrm{da}}=L_{\mathrm{d}} $. To increase the degree of freedom of received signals, the $ L_{\mathrm{dr}} $-length pilots are transmitted by using $ Q_{\mathrm{dr}} $ transmit beamforming matrices and received by using $ P_{\mathrm{dr}} $ receive beamforming matrices, where $ L_{\mathrm{dr}}=Q_{\mathrm{dr}}P_{\mathrm{dr}} $. The $ L_{\mathrm{da}} $-length pilots possess the homologous structure, i.e., $  L_{\mathrm{da}}=Q_{\mathrm{da}}P_{\mathrm{da}} $. By using $ \{\mathbf{x}_1,\cdots,\mathbf{x}_{Q_{\mathrm{max}}}\}$ to represent the pilots set, the transmitted pilot $ \mathbf{x}_{\mathrm{d},l} $ during the $ l $-th transmission can be denoted as $ \mathbf{x}_{\mathrm{d},l}=\mathbf{x}_{q} $, where $ \mathbb{E}\left\lbrace \mathbf{x}_{q}\mathbf{x}_{q}^H\right\rbrace=\rho_{\mathrm{c}}\mathbf{I}_{\mathrm{M}_{t}} $, $ q=l \% Q_{\mathrm{dr}} $ when $ l\le L_{\mathrm{dr}} $, and $ q=(l-L_{\mathrm{dr}}) \% Q_{\mathrm{da}} $ when $  L_{\mathrm{da}}<l\le L_{\mathrm{d}} $, $ Q_{\mathrm{max}}=\max\{Q_{\mathrm{dr}},Q_{\mathrm{da}}\} $.
	\item \textbf{Beamforming design for calibrating digital chains:} Let $ \mathbf{F}_{\mathrm{dr},q}\in \mathbb{C}^{N_\mathrm{t}\times M_{\mathrm{t}}} $ denote the  analog transmit beamforming matrix  and $ \mathbf{B}_{\mathrm{dr},p}\in \mathbb{C}^{N_\mathrm{r}\times M_{\mathrm{r}}} $ represent the receive beamforming matrix during the $ l $-th transmission, where $ l\le L_{\mathrm{dr}} $, $ q=l\% Q_{\mathrm{dr}} $, and $ p=l\% P_{\mathbf{dr}} $. To decouple the mismatch of the digital chains from the analog chains, analog beamforming matrices are designed as $ \mathbf{F}_{\mathrm{dr},1}=\cdots=\mathbf{F}_{\mathrm{dr},Q_{\mathrm{dr}}}=\mathbf{f}_{\mathrm{dr}}\mathbf{1}_{M_{\mathrm{t}}}^T $ and $ \mathbf{B}_{\mathrm{dr},1}=\cdots=\mathbf{B}_{\mathrm{dr},P_{\mathrm{dr}}}=\mathbf{b}_{\mathrm{dr}}\mathbf{1}_{M_{\mathrm{r}}}^T $, where each element of $ \mathbf{f}_{\mathrm{dr}}\in \mathbb{C}^{N_\mathrm{t}} $ and $ \mathbf{b}_{\mathrm{dr}}\in \mathbb{C}^{N_\mathrm{r}}$ possesses random phase. Further, the digital precoding matrix can be designed as $ \mathbf{W}_{\mathrm{dr},q}=\mathbf{I}_{M_\mathrm{t}}/\sqrt{M_{\mathrm{t}}} $ and the digital receive combining matrix is given by $ \mathbf{D}_{\mathrm{dr},p}=\mathbf{I}_{M_{\mathrm{r}}} $ during the downlink training phase.
	\item \textbf{Beamforming design for calibrating analog chains:} During the $ l $-th transmission ($ l> L_{\mathrm{dr}} $), the analog transmit beamforming matrix $ \mathbf{F}_{\mathrm{da},q} $ and receive beamforming matrix $ \mathbf{B}_{\mathrm{da},p} $ can be designed as random phase matrices, i.e., the elements of $ \mathbf{F}_{\mathrm{da},q} $ and $ \mathbf{B}_{\mathrm{da},p} $ possess random phases. Let $ \mathbf{W}_{\mathrm{da},q} $ denote the digital precoding matrix and $ \mathbf{D}_{\mathrm{da},p} $ denote the digital receive combining matrix, where $ q=(l-L_{\mathrm{dr}})\% Q_{\mathrm{da}} $, and $ p=(l-L_{\mathrm{dr}})\% P_{\mathrm{da}} $. To decouple the mismatch of analog RF chains from the mismatch of digital RF chains, the digital precoding matrix can be designed as $ \mathbf{W}_{\mathrm{da},q}=\mathrm{blkdiag}(1,\mathbf{0}_{M_{\mathrm{t}}-1,M_{\mathrm{t}}-1}) $, and the digital receive combining matrix can be given by $ \mathbf{D}_{\mathrm{da},p}=\mathrm{blkdiag}(1,\mathbf{0}_{M_{\mathrm{r}}-1,M_{\mathrm{r}}-1}) $.
\end{itemize}


\subsection{Problem Formulation and Decomposition of HAC}\label{sec:proforanddecom}
Since the mismatch coefficients of transmit chains have nothing to do with that of  receive chains, HAC can be divided into downlink HAC and uplink HAC. The downlink HAC is applied to calibrate the transmit chains of the BS and the receive chains of the UE, while the uplink HAC can calibrate the receive chains of the BS as well as the transmit chains of the UE. The uplink HAC is similar to the downlink HAC. Thus, we introduce the signal modeling, the problem formulation, and the problem decoupling by taking the downlink HAC as an example.

By considering the BS transmits $l$-th pilot to the UE, the signal received by the UE can be modeled as
\begin{equation}
	\mathbf{y}_{\mathrm{d},l}=\mathbf{D}_{\mathrm{r},l}^T\mathbf{U}_1\mathbf{B}_{\mathrm{r},l}^T\mathbf{U}_2\mathbf{H}^T\mathbf{T}_2\mathbf{F}_{\mathrm{t},l}\mathbf{T}_1\mathbf{W}_{\mathrm{t},l}\mathbf{x}_{\mathrm{d},l}+\tilde{\mathbf{n}}_{\mathrm{d},l},
	\label{eq:generalsgianl}
\end{equation}
where $ \tilde{\mathbf{n}}_{\mathrm{d},l}=\mathbf{D}_{\mathrm{r},l}^T\mathbf{U}_1\mathbf{B}_{\mathrm{r},l}^T\mathbf{n}_{\mathrm{d},l} $. When $ l\le L_{\mathrm{dr}} $, $ \mathbf{D}_{\mathrm{r},l}=\mathbf{D}_{\mathrm{dr},p} $, $ \mathbf{B}_{\mathrm{r},l}=\mathbf{B}_{\mathrm{dr},p} $, $ \mathbf{F}_{\mathrm{t},q} = \mathbf{F}_{\mathrm{dr},q} $, and $ \mathbf{W}_{\mathrm{t},l}=\mathbf{W}_{\mathrm{dr},q} $, where $ p=l\%P_{\mathrm{dr}} $, and $ q=l \% Q_{\mathrm{dr}} $. When $ l> L_{\mathrm{dr}} $, $ \mathbf{D}_{\mathrm{r},l}=\mathbf{D}_{\mathrm{da},p} $, $ \mathbf{B}_{\mathrm{r},l}=\mathbf{B}_{\mathrm{da},p} $, $ \mathbf{F}_{\mathrm{t},q} = \mathbf{F}_{\mathrm{da},q} $, and $ \mathbf{W}_{\mathrm{t},l}=\mathbf{W}_{\mathrm{da},q} $, where $ p=(l-L_{\mathrm{dr}})\%P_{\mathrm{da}} $, and $ q=(l-L_{\mathrm{dr}})\% Q_{\mathrm{da}} $. 

After the BS transmits $ L_\mathrm{d} $-length pilots to the UE, the optimization problem for jointly estimating $ \mathbf{U}_1,\ \mathbf{U}_2,\ \mathbf{T}_1,\ \mathbf{T}_2 $, and $ \mathbf{H} $ can be formulated as
\begin{equation}
	\min_{\mathbf{U}_{\mathrm{1}},\mathbf{T}_{\mathrm{1}},\mathbf{U}_2,\mathbf{T}_2,\mathbf{H}}\quad \sum_{l=1}^{L_{\mathrm{d}}}\left\|\mathbf{y}_{\mathrm{d},l}-\mathbf{D}_{\mathrm{r},l}^T\mathbf{U}_1\mathbf{B}_{\mathrm{r},l}^T\mathbf{H}_{\mathrm{DL}}\mathbf{F}_{\mathrm{t},l}\mathbf{T}_1\tilde{\mathbf{x}}_{\mathrm{d},l}\right\|_{\mathrm{F}}^2,
	\label{eq:priprobelm}
\end{equation}
where $ \mathbf{H}_{\mathrm{DL}}=\mathbf{U}_2\mathbf{H}^T\mathbf{T}_2 $, $ \tilde{\mathbf{x}}_{\mathrm{d},l}=\mathbf{W}_{\mathrm{t},l}\mathbf{x}_{\mathrm{d},l} $.
Thanks to the proposed pilots and training scheme design, the above joint optimization problem can be equivalently  decoupled into two subproblems demonstrated in the following proposition.

\begin{prop}[HAC problem decoupling]\label{prop:HACproblemdcp}
	Based on the specific pilots and training scheme design in Section \ref{sec:pilotdesign}, the problem of HAC in \eqref{eq:priprobelm} can be equivalently decoupled into two independent problems as
	\begin{align}
		&\mathcal{P}_{1}:\min_{\ \ \mathbf{u}_{\mathrm{1}},\mathbf{t}_{\mathrm{1}}\ }\quad \left\|\mathbf{Y}_{\mathrm{dr}}-(\mathbf{1}_{P_{\mathrm{dr}}}\otimes\mathbf{u}_1)\mathbf{t}_1^T\mathbf{X}_{\mathrm{dr}}\right\|_{\mathrm{F}}^2,\label{eq:hacrfcal}\\
		&\mathcal{P}_2:\min_{\mathbf{U}_{\mathrm{2}},\mathbf{T}_{\mathrm{2}},\mathbf{H}}\quad\left\|\mathbf{Y}_{\mathrm{da}}-\bar{\mathbf{B}}_{\mathrm{da}}^T\mathbf{U}_2\mathbf{H}^T\mathbf{T}_2\bar{\mathbf{F}}_{\mathrm{da}}\mathbf{X}_{\mathrm{da}}\right\|_{\mathrm{F}}^2,\label{eq:hacantcal}
	\end{align}
	where $ \mathbf{Y}_{\mathrm{dr}}=[\bar{\mathbf{Y}}_{\mathrm{dr},1}^T,\cdots,\bar{\mathbf{Y}}_{\mathrm{dr},P_{\mathrm{dr}}}^T]^T $, $ \bar{\mathbf{Y}}_{\mathrm{dr},p}=[\mathbf{y}_{\mathrm{d},(p-1)Q_{\mathrm{dr}}+1},\cdots,\mathbf{y}_{\mathrm{d},pQ_{\mathrm{dr}}}] $, $\mathbf{u}_1$ consists of the diagonal entries of $\mathbf{U}_1$, $\mathbf{t}_1$ is composed of the diagonal entries of $\mathbf{T}_1$,  $\mathbf{X}_{\mathrm{dr}}=[\mathbf{x}_{1},\cdots,\mathbf{x}_{Q_{\mathrm{dr}}}]$, $ \mathbf{Y}_{\mathrm{da}}=[\mathbf{y}_{\mathrm{da},1}^T,\cdots,\mathbf{y}_{\mathrm{da},P_{\mathrm{da}}}^T]^T $, $ \mathbf{y}_{\mathrm{da},p}=[y_{\mathrm{d},(p-1)Q_{\mathrm{da}}+1,1},\cdots,y_{\mathrm{d},pQ_{\mathrm{da}},1}] $, $ \bar{\mathbf{B}}_{\mathrm{da}}=[\mathbf{b}_{\mathrm{da},1,1},\cdots,\mathbf{b}_{\mathrm{da},P_{\mathrm{da}},1}] $, $ \bar{\mathbf{F}}_{\mathrm{da}}=[\mathbf{f}_{\mathrm{da},1,1},\cdots,\mathbf{f}_{\mathrm{da},Q_{\mathrm{da}},1}] $, $ \mathbf{X}_{\mathrm{da}}=\mathrm{diag}(x_{1,1},\cdots,x_{Q_{\mathrm{da}},1}) $, $ \mathbf{b}_{\mathrm{da},p,1} $ is the first column of $ \mathbf{B}_{\mathrm{da},p} $, $ \mathbf{f}_{\mathrm{da},q,1} $ denotes the first column of $ \mathbf{F}_{\mathrm{da},q} $, and $ x_{q,1} $ represents the first entry of $ \mathbf{x}_{q} $.
\end{prop}

\begin{IEEEproof}
	See Appendix \ref{proof:prodecople}.
\end{IEEEproof}

\begin{rem}[HAC decoupling] Since the problem $ \mathcal{P}_1 $ can solve the mismatch coefficients of the transmit digital RF chains of the BS and those of the receive digital RF chains of the UE,  it is known as the downlink calibration problem of digital RF chains. Similarly, the problem $ \mathcal{P}_2 $ is the downlink calibration problem of analog RF chains. Thus, Proposition \ref{prop:HACproblemdcp} indicates that HAC can be decoupled into the calibration of digital RF chains and the calibration of the analog RF chains, which is the purpose of the hierarchical calibration.
\end{rem}	

\subsection{Solution to Calibration Problem of HAC}\label{sec:solutionCal}
As $ \mathcal{P}_1 $ and $ \mathcal{P}_2 $ are independent of each other, we first find the solution to $ \mathcal{P}_1 $, then solve $ \mathcal{P}_2 $. As the objective of $\mathcal{P}_1$ is bilinear, it can be solved by iterative approaches but this is inefficient. To solve $ \mathcal{P}_1 $ efficiently, we propose a closed-form solution by regarding the first receive digital chain of the UE as the calibration reference. 

By using the auxiliary variables $ \bar{\mathbf{X}}_{\mathrm{dr},p}=\mathbf{u}_1 \mathbf{x}_{\mathrm{dt},p}^T $ and $ \mathbf{x}_{\mathrm{dt},p}=\mathbf{X}_{\mathrm{dr}}^T\mathbf{t}_1 $, the problem $\mathcal{P}_1$ can be further formulated by
\begin{equation}
	\mathcal{P}_{1.1}:\min_{\{\bar{\mathbf{X}}_{\mathrm{dr},p}\}_{p\in[1:P_{\mathrm{dr}}]}}\quad \sum_{p=1}^{P_{\mathrm{dr}}}\left\|\bar{\mathbf{Y}}_{\mathrm{dr},p}-\bar{\mathbf{X}}_{\mathrm{dr},p}\right\|_{\mathrm{F}}^2,
	\label{eq:problemxdp}
\end{equation}
By taking the derivative of the objective function of $ \mathcal{P}_{1,1} $, the solution can be given by\cite{Zhang2017Matrix}
\begin{equation}
	\bar{\mathbf{X}}_{\mathrm{dr},p} = \bar{\mathbf{Y}}_{\mathrm{dr},p},\quad \forall p\in[1:P_{\mathrm{dr}}].
	\label{eq:soltop12}
\end{equation}
Since the first receive digital RF chain of the UE is the reference, its mismatch coefficient can be treated as a known constant, e.g., $ u_{1,1}=c_{\mathrm{dr}}\neq 0 $. Based on this assumption and \eqref{eq:soltop12}, $ \mathbf{x}_{\mathrm{dt},p}^T $ equals to the first column of $ \bar{\mathbf{X}}_{\mathrm{dr},p} $, i.e., 
\begin{equation}
	\mathbf{x}_{\mathrm{dt},p}=\frac{1}{c_{\mathrm{dr}}}\mathbf{y}_{\mathrm{dr},(p-1)M_{\mathrm{r}}+1}^T,\quad \forall p\in[1:P_{\mathrm{dr}}],
	\label{eq:soltoxtn}
\end{equation}
where $ \mathbf{y}_{\mathrm{dr},m} $ is the $ m $-th row of $ \mathbf{Y}_{\mathrm{dr}} $.

By substituting \eqref{eq:soltoxtn} into \eqref{eq:hacrfcal}, the solution to the problem $ \mathcal{P}_{1} $ can be given in the following proposition.

\begin{prop}[Solutions to the problem $ \mathcal{P}_{1} $]\label{prop:solutot1u1}
	By assuming that the first receive digital RF chain is set as the reference, i.e., $ u_{1,1}=c_{\mathrm{dr}} $, the solutions to $ \mathbf{t}_1 $ and $ \mathbf{u}_1 $ can be given by
	\begin{align}
		&\hat{\mathbf{u}}_1=c_{\mathrm{dr}}\bigl[1,\tilde{\mathbf{y}}_{\mathrm{dr}}^H\check{\mathbf{Y}}_{\mathrm{dr}}(\check{\mathbf{Y}}_{\mathrm{dr}}^H\check{\mathbf{Y}}_{\mathrm{dr}})^{-1}\bigr]^H,\label{eq:solu1}\\
		&\hat{\mathbf{t}}_1=\frac{1}{c_{\mathrm{dr}}P_{\mathrm{dr}}}\Bigl[\mathbf{1}_{P_{\mathrm{dr}}}^T\otimes\bigl(\mathbf{X}_{\mathrm{dr}}^*\mathbf{X}_{\mathrm{dr}}^T\bigr)^{-1}\mathbf{X}_{\mathrm{dr}}^*\Bigr]\mathbf{y}_{\mathrm{dt}},
		\label{eq:solt1}
	\end{align}
	where $ \tilde{\mathbf{y}}_{\mathrm{dr}}=[\mathrm{vec}(\tilde{\mathbf{Y}}_{\mathrm{dr},1})^T,\cdots,\mathrm{vec}(\tilde{\mathbf{Y}}_{\mathrm{dr},P_{\mathrm{dr}}})^T]^T \in \mathbb{C}^{P_{\mathrm{dr}}Q_{\mathrm{dr}}(M_{\mathrm{r}}-1)}$,$ \tilde{\mathbf{Y}}_{\mathrm{dr},p} $ consists of the second to the last row of $ \bar{\mathbf{Y}}_{\mathrm{dr},p} $, $ \check{\mathbf{Y}}_{\mathrm{dr}}=[(\mathbf{y}_{\mathrm{dr},1}\otimes \mathbf{I}_{M_{\mathrm{r}}-1}),\cdots,(\mathbf{y}_{\mathrm{dr},(P_{\mathrm{dr}}-1)M_{\mathrm{r}}+1}\otimes \mathbf{I}_{M_{\mathrm{r}}-1})]^T\in \mathbb{C}^{P_{\mathrm{dr}}Q_{\mathrm{dr}}(M_{\mathrm{r}}-1)\times (M_{\mathrm{r}}-1)} $, and $ \mathbf{y}_{\mathrm{dt}}=[\mathbf{y}_{\mathrm{dr},1},\cdots,\mathbf{y}_{\mathrm{dr},(P_{\mathrm{dr}}-1)M_{\mathrm{r}}+1}]^T $.
\end{prop}

\begin{IEEEproof}
	See Appendix \ref{proof:solutot1u1}.
\end{IEEEproof}

\begin{rem}[The special solution to $ \mathcal{P}_1 $]
	Equations \eqref{eq:solu1} and \eqref{eq:solt1} give the general solutions to $ \mathcal{P}_1 $ and are dependent on the value of $ c_{\mathrm{dr}} $. In practice, it is difficult to determine the value of $ c_{\mathrm{dr}} $. To avoid this issue, the mismatch coefficient of the reference can be set to $ 1 $, i.e., $ c_{\mathrm{dr}}=1 $. In this case, equations \eqref{eq:solu1} and \eqref{eq:solt1} degenerate to a special solution to $ \mathcal{P}_1 $. Since the vectors parallel to $ \mathbf{t}_1 $ and $ \mathbf{u}_1 $ can be applied to the calibration, the special solution to $ \mathcal{P}_1 $ still works for the reciprocity calibration.
\end{rem}

Then, the mismatch coefficients of analog RF chains can be estimated by solving $\mathcal{P}_{2}$. By exploiting the geometry channel model of mmWave, the calibration problem of analog chains can be further written as
\begin{equation}
	\mathcal{P}_{2.1}:\min_{\mathbf{U}_2,\mathbf{T}_2,\boldsymbol{\Theta},\boldsymbol{\Phi},\mathbf{H}_{\alpha}}\quad\left\|\mathbf{Y}_{\mathrm{da}}-\bar{\mathbf{B}}_{\mathrm{da}}^T\mathbf{U}_2\mathbf{A}_{\mathrm{r}}\mathbf{H}_{\alpha}\mathbf{A}_{\mathrm{t}}^T\mathbf{T}_2\tilde{\mathbf{X}}_{\mathrm{da}}\right\|_{\mathrm{F}}^2,
\end{equation}
where $ \tilde{\mathbf{X}}_{\mathrm{da}}=\bar{\mathbf{F}}_{\mathrm{da}}\mathbf{X}_{\mathrm{da}} $, $\mathbf{A}_{\mathrm{t}}=[\mathbf{a}_{\mathrm{t}}(\theta_{1}),\cdots,\mathbf{a}_{\mathrm{t}}(\theta_{K})]\in \mathbb{C}^{N_{\mathrm{t}}\times K}$, $\mathbf{A}_{\mathrm{r}}=[\mathbf{a}_{\mathrm{r}}(\phi_{1}),\cdots,\mathbf{a}_{\mathrm{r}}(\phi_{K})]\in \mathbb{C}^{N_{\mathrm{r}}\times K}$, and $\mathbf{H}_{\alpha}=\mathrm{diag}(\alpha_{1},\cdots,\alpha_{K})\sqrt{N_{\mathrm{t}}N_{\mathrm{r}}}/{\sqrt{K}}$. As the variables are correlated with each other, this problem is nonconvex and there is no tractable solution to the problem. To solve $\mathcal{P}_{2.1}$ efficiently, inspired by \cite{Bezdek2003Convergence}, we propose an alternating optimization algorithm to solve a locally optimal solution.  

During the $ l_{\mathrm{ao}} $-th iteration,  we apply the least square algorithm to estimate the diagonal matrices $ \mathbf{U}_2 $, $ \mathbf{T}_2 $, $ \mathbf{H}_{\alpha} $, then, propose an algorithm to estimate the AoA and AoD matrices $\boldsymbol{\Theta} $, $ \boldsymbol{\Phi} $.

\begin{lem}[Solution to the diagonal matrices]\label{lem:soltot2u2h}
	During the $ l_{\mathrm{ao}}$-th iteration, when $ \mathbf{T}_2^{l_{\mathrm{ao}}-1},\mathbf{U}_2^{l_{\mathrm{ao}}-1}$, $\boldsymbol{\Theta}^{l_{\mathrm{ao}}-1}$, and $\boldsymbol{\Phi}^{l_{\mathrm{ao}}-1} $ are known, the diagonal elements of $ \mathbf{H}_{\alpha}^{l_{\mathrm{ao}}} $ can be estimated by
	\if\thecol1
	\begin{equation}
			{\mathbf{h}}_{\alpha}^{l_{\mathrm{ao}}}=\mathrm{arg}\min_{\mathbf{h}_{\alpha}}\bar{g}(\mathbf{T}_2^{l_{\mathrm{ao}}-1},\mathbf{U}_2^{l_{\mathrm{ao}}-1},\mathbf{H}_{\alpha},\boldsymbol{\Theta}^{l_{\mathrm{ao}}-1},\boldsymbol{\Phi}^{l_{\mathrm{ao}}-1})
			=(\boldsymbol{\Gamma}_{\mathrm{h}}^H\boldsymbol{\Gamma}_{\mathrm{h}})^{-1}\boldsymbol{\Gamma}_{\mathrm{h}}^H\mathrm{vec}\left\lbrace \mathbf{Y}_{\mathrm{da}} \right\rbrace,
		\label{eq:estchgain}
	\end{equation}
	\else
	\begin{equation}
		\begin{split}
			{\mathbf{h}}_{\alpha}^{l_{\mathrm{ao}}}&=\mathrm{arg}\min_{\mathbf{h}_{\alpha}}\bar{g}(\mathbf{T}_2^{l_{\mathrm{ao}}-1},\mathbf{U}_2^{l_{\mathrm{ao}}-1},\mathbf{H}_{\alpha},\boldsymbol{\Theta}^{l_{\mathrm{ao}}-1},\boldsymbol{\Phi}^{l_{\mathrm{ao}}-1})\\
			&=(\boldsymbol{\Gamma}_{\mathrm{h}}^H\boldsymbol{\Gamma}_{\mathrm{h}})^{-1}\boldsymbol{\Gamma}_{\mathrm{h}}^H\mathrm{vec}\left\lbrace \mathbf{Y}_{\mathrm{da}} \right\rbrace,
		\end{split}
		\label{eq:estchgain}
	\end{equation}
	\fi
	where $\boldsymbol{\Gamma}_{\mathrm{h}}=(\tilde{\mathbf{X}}_{\mathrm{da}}^T\mathbf{T}_2^{l_{\mathrm{ao}}-1}\mathbf{A}_{\mathrm{t}}^{l_{\mathrm{ao}}-1}\odot\bar{\mathbf{B}}_{\mathrm{r}}\mathbf{U}_2^{l_{\mathrm{ao}}-1}\mathbf{A}_{\mathrm{r}}^{l_{\mathrm{ao}}-1})$. When $ \mathbf{T}_2^{l_{\mathrm{ao}}-1},\mathbf{H}_{\alpha}^{l_{\mathrm{ao}}}$, and $\boldsymbol{\Theta}^{l_{\mathrm{ao}}-1}$, $\boldsymbol{\Phi}^{l_{\mathrm{ao}}-1} $ are known, the diagonal elements of $ \mathbf{U}_2^{l_{\mathrm{ao}}} $ can be estimated by
	\if\thecol1
	\begin{equation}
			\mathbf{u}_2^{l_{\mathrm{ao}}}=\mathrm{arg}\min_{\mathbf{u}_{2}}\bar{g}(\mathbf{T}_2^{l_{\mathrm{ao}}-1},\mathbf{U}_2,\mathbf{H}_{\alpha}^{l_{\mathrm{ao}}},\boldsymbol{\Theta}^{l_{\mathrm{ao}}-1},\boldsymbol{\Phi}^{l_{\mathrm{ao}}-1})
			=(\boldsymbol{\Gamma}_{\mathrm{u}}^H\boldsymbol{\Gamma}_{\mathrm{u}})^{-1}\boldsymbol{\Gamma}_{\mathrm{u}}^H\mathrm{vec}\left\lbrace \mathbf{Y}_{\mathrm{da}} \right\rbrace,
		\label{eq:estu2}
	\end{equation}
	\else
	\begin{equation}
		\begin{split}
			\mathbf{u}_2^{l_{\mathrm{ao}}}&=\mathrm{arg}\min_{\mathbf{u}_{2}}\bar{g}(\mathbf{T}_2^{l_{\mathrm{ao}}-1},\mathbf{U}_2,\mathbf{H}_{\alpha}^{l_{\mathrm{ao}}},\boldsymbol{\Theta}^{l_{\mathrm{ao}}-1},\boldsymbol{\Phi}^{l_{\mathrm{ao}}-1})\\
			&=(\boldsymbol{\Gamma}_{\mathrm{u}}^H\boldsymbol{\Gamma}_{\mathrm{u}})^{-1}\boldsymbol{\Gamma}_{\mathrm{u}}^H\mathrm{vec}\left\lbrace \mathbf{Y}_{\mathrm{da}} \right\rbrace,
		\end{split}
		\label{eq:estu2}
	\end{equation}
	\fi
	where $\boldsymbol{\Gamma}_{\mathrm{u}}=(\tilde{\mathbf{X}}_{\mathrm{da}}^T\mathbf{T}_2^{l_{\mathrm{ao}}-1}\mathbf{A}_{\mathrm{t}}^{l_{\mathrm{ao}}-1}\mathbf{H}_{\alpha}^{l_{\mathrm{ao}}}(\mathbf{A}_{\mathrm{r}}^{l_{\mathrm{ao}}-1})^T\odot\bar{\mathbf{B}}_{\mathrm{r}})$. Similarly, by giving $ \mathbf{U}_2^{l_{\mathrm{ao}}},\mathbf{H}_{\alpha}^{l_{\mathrm{ao}}}$, $\boldsymbol{\Theta}^{l_{\mathrm{ao}}-1}$, and $\boldsymbol{\Phi}^{l_{\mathrm{ao}}-1} $, the diagonal entries of $ \mathbf{T}_2 $ can be given by
	\if\thecol1
	\begin{equation}
			\mathbf{t}_2^{l_{\mathrm{ao}}}=\mathrm{arg}\min_{\mathbf{t}_{2}}\bar{g}(\mathbf{T}_2,\mathbf{U}_2^{l_{\mathrm{ao}}},\mathbf{H}_{\alpha}^{l_{\mathrm{ao}}},\boldsymbol{\Theta}^{l_{\mathrm{ao}}-1},\boldsymbol{\Phi}^{l_{\mathrm{ao}}-1})
			=(\boldsymbol{\Gamma}_{\mathrm{t}}^H\boldsymbol{\Gamma}_{\mathrm{t}})^{-1}\boldsymbol{\Gamma}_{\mathrm{t}}^H\mathrm{vec}\left\lbrace \mathbf{Y}_{\mathrm{da}} \right\rbrace,
		\label{eq:estt2}
	\end{equation}
	\else
	\begin{equation}
		\begin{split}
			\mathbf{t}_2^{l_{\mathrm{ao}}}&=\mathrm{arg}\min_{\mathbf{t}_{2}}\bar{g}(\mathbf{T}_2,\mathbf{U}_2^{l_{\mathrm{ao}}},\mathbf{H}_{\alpha}^{l_{\mathrm{ao}}},\boldsymbol{\Theta}^{l_{\mathrm{ao}}-1},\boldsymbol{\Phi}^{l_{\mathrm{ao}}-1})\\
			&=(\boldsymbol{\Gamma}_{\mathrm{t}}^H\boldsymbol{\Gamma}_{\mathrm{t}})^{-1}\boldsymbol{\Gamma}_{\mathrm{t}}^H\mathrm{vec}\left\lbrace \mathbf{Y}_{\mathrm{da}} \right\rbrace,
		\end{split}
		\label{eq:estt2}
	\end{equation}
\fi
	where $\boldsymbol{\Gamma}_{\mathrm{t}}=(\tilde{\mathbf{X}}_{\mathrm{da}}^T\odot\bar{\mathbf{B}}_{\mathrm{r}}\mathbf{U}_2^{l_{\mathrm{ao}}}\mathbf{A}_{\mathrm{r}}^{l_{\mathrm{ao}}-1}\mathbf{H}_{\alpha}^{l_{\mathrm{ao}}}(\mathbf{A}_{\mathrm{t}}^{l_{\mathrm{ao}}-1})^T)$.
\end{lem}

\begin{algorithm}[t]
	\caption{The AoAs/AoDs updating}
	\label{alg:updirections}
	\begin{algorithmic}[1]
		\REQUIRE $ \mathbf{U}_2^{l_{\mathrm{ao}}} $, $ \mathbf{T}_2^{l_{\mathrm{ao}}} $, $ \mathbf{H}_{\alpha}^{l_{\mathrm{ao}}} $, $ \boldsymbol{\Theta}^{l_{\mathrm{ao}}-1} $, $ \boldsymbol{\Phi}^{l_{\mathrm{ao}}-1} $, and the convergence condition $ \epsilon_{\mathrm{an}} $.
		\STATE Initialize $ l_{\mathrm{an}}=1 $, and $ \bar{\boldsymbol{\Phi}}^{l_{\mathrm{an}}-1}=\boldsymbol{\Phi}^{l_{\mathrm{ao}}-1} $.
		\REPEAT
		\STATE Compute the array steering matrix $ \mathbf{A}_{\mathrm{r}}(\bar{\boldsymbol{\Phi}}^{l_{\mathrm{an}}-1}) $ and its gradient matrix $ \bar{\mathbf{A}}_{\mathrm{r}}(\bar{\boldsymbol{\Phi}}^{l_{\mathrm{an}}-1}) $;
		\STATE Compute the equivalent receive signal matrix $ \mathbf{Y}_{\mathrm{dar}}=\mathbf{Y}_{\mathrm{da}}-\bar{\mathbf{B}}_{\mathrm{da}}^T\mathbf{U}_2^{l_{\mathrm{ao}}}\mathbf{A}_{\mathrm{r}}(\bar{\boldsymbol{\Phi}}^{l_{\mathrm{an}}-1})\mathbf{H}_{\mathrm{r}}^{l_{\mathrm{ao}}} $;
		\STATE Compute the increase of direction angles $ \boldsymbol{\xi} = \Re\{(\boldsymbol{\Gamma}_{\xi}^H\boldsymbol{\Gamma}_{\xi})\}^{-1}\Re\{\boldsymbol{\Gamma}_{\xi}^H\mathrm{vec}(\mathbf{Y}_{\mathrm{dar}})\} $;
		\STATE Upgrade the direction angles $\bar{\boldsymbol{\Phi}}^{l_{\mathrm{an}}}=\bar{\boldsymbol{\Phi}}^{l_{\mathrm{an}}-1}+\boldsymbol{\xi}$, and set $ l_{\mathrm{an}}= l_{\mathrm{an}}+1 $;
		\UNTIL{$\|\bar{\boldsymbol{\Phi}}^{l_{\mathrm{an}}}-\bar{\boldsymbol{\Phi}}^{l_{\mathrm{an}}-1}\|_{\mathrm{F}}^2<\epsilon_{\mathrm{an}} $}
		\ENSURE The updated $ \boldsymbol{\Phi}^{l_{\mathrm{ao}}} $.
	\end{algorithmic}
\end{algorithm}

\begin{IEEEproof}
	The complete proof is presented in Appendix C of Supplementary Material.
\end{IEEEproof}	

\textcolor{black}{
Finally, we propose a AoAs and AoDs updating algorithm\footnote{The update method of AoA/AoD is not restricted to the algorithm proposed in this paper. Some existing methods, such as those presented in \cite{Zhang2022MMVBased}, can be employed.} when $ \mathbf{U}_2^{l_{\mathrm{ao}}} $, $ \mathbf{T}_2^{l_{\mathrm{ao}}} $, $ \mathbf{H}_{\alpha}^{l_{\mathrm{ao}}} $, $ \boldsymbol{\Theta}^{l_{\mathrm{ao}}-1} $, and $ \boldsymbol{\Phi}^{l_{\mathrm{ao}}-1} $ are given. 
Since the AoAs and AoDs can be estimated using the same approaches, we introduce the updating algorithm by taking updating the AoDs $ \boldsymbol{\Phi} $ as an example. Since the problem for estimating AoDs is nonlinear, it is difficult to solve the AoDs directly. 
To address this issue, as presented in \cite{Yang2013OffGrid}, the nonlinear problem is transformed into a series of linear problems by using the first-order Taylor expansion to approximate the array steering vector. Let $ \xi_{k} \ (k\in[1:K])$ denotes the differences between the estimated AoDs $ \bar{\boldsymbol{\Phi}} $ and the real AoDs $ \boldsymbol{\Phi} $.  By assuming that the differences are small, the array steering vector can be approximated by the first-order Taylor expansion denoted as $ \mathbf{a}_{\mathrm{r}}(\phi_k)=\mathbf{a}_{\mathrm{r}}(\bar{\phi}_k)+\bar{\mathbf{a}}_{\mathrm{r}}(\bar{\phi}_k)\xi_k $, where $ \bar{\mathbf{a}}_{\mathrm{r}}(\bar{\phi}_k)=\partial \mathbf{a}_{\mathrm{r}}(\phi_k)/\partial \phi_k|_{\phi_k=\bar{\phi}_k}= \mathbf{a}_{\mathrm{r}}(\bar{\phi}_k)\circ[0,-j\frac{2\pi d}{\lambda}\cos\bar{\phi}_k,\cdots,-j\frac{2\pi d}{\lambda}(N_{\mathrm{r}}-1)\cos\bar{\phi}_k]^T  $.
}
Then, the problem $ \mathcal{P}_{2.1} $ can be further formulated as 
\begin{equation}
	\mathcal{P}_{2.2}:\min_{\{\xi_{k}\}_{k\in[1:K]}}\quad \left\|\mathbf{Y}_{\mathrm{dar}}^{l_{\mathrm{ao}}}-\bar{\mathbf{B}}_{\mathrm{da}}^T\mathbf{U}_2^{l_{\mathrm{ao}}}\bar{\mathbf{A}}_{\mathrm{r}}(\bar{\boldsymbol{\Phi}})\boldsymbol{\Lambda}\mathbf{H}_{\mathrm{r}}^{l_{\mathrm{ao}}}\right\|_{\mathrm{F}}^2,
\end{equation}
where  $ \mathbf{H}_{\mathrm{r}}^{l_{\mathrm{ao}}} = \mathbf{H}_{\alpha}^{l_{\mathrm{ao}}}\mathbf{A}_{\mathrm{t}}^T\mathbf{T}_2^{l_{\mathrm{ao}}}\tilde{\mathbf{X}}_{\mathrm{da}} $, $ \mathbf{Y}_{\mathrm{dar}}=\mathbf{Y}_{\mathrm{da}}-\bar{\mathbf{B}}_{\mathrm{da}}^T\mathbf{U}_2^{l_{\mathrm{ao}}}\mathbf{A}_{\mathrm{r}}(\bar{\boldsymbol{\Phi}})\mathbf{H}_{\mathrm{r}}^{l_{\mathrm{ao}}} $, $ \boldsymbol{\Lambda}=\mathrm{diag}(\xi_1,\cdots,\xi_K) $, and $ \bar{\mathbf{A}}_{\mathrm{r}}(\bar{\boldsymbol{\Phi}})=[\bar{\mathbf{a}}_{\mathrm{r}}(\bar{\phi}_1),\cdots,\bar{\mathbf{a}}_{\mathrm{r}}(\bar{\phi}_K)] $. The solution can be given by
\begin{equation}
	\boldsymbol{\xi} = \Re\{(\boldsymbol{\Gamma}_{\xi}^H\boldsymbol{\Gamma}_{\xi})\}^{-1}\Re\{\boldsymbol{\Gamma}_{\xi}^H\mathrm{vec}(\mathbf{Y}_{\mathrm{dar}})\},
\end{equation}
$ \boldsymbol{\Gamma}_{\xi}=(\mathbf{H}_{\mathrm{r}}^{l_{\mathrm{ao}}})^T\odot \bar{\mathbf{B}}_{\mathrm{da}}^T\mathbf{U}_2^{l_{\mathrm{ao}}}\bar{\mathbf{A}}_{\mathrm{r}}(\bar{\boldsymbol{\Phi}}) $. Since $ \xi_k $ is assumed to be small, the updating requires several iterations, and the iterative updating algorithm is summarized as Algorithm \ref{alg:updirections}.

It is worth noting that the initial values of AoAs $ \boldsymbol{\Theta} $ and AoDs $ \boldsymbol{\Phi} $ can be roughly calculated by direction finding methods, e.g., the modified MUSIC algorithm in \cite{Qi2019OffGrid}.

Finally, based on Lemma \ref{lem:soltot2u2h} and Algorithm \ref{alg:updirections}, the problem $ \mathcal{P}_{2.1} $ can be solved by an alternating optimization algorithm, which is summarized as Algorithm \ref{alg:ADMM}. 
\begin{algorithm}[t]
	\caption{Alternating Optimization for solving $ \mathcal{P}_{2.1} $}
	\label{alg:ADMM} 
	\begin{algorithmic}[1]
		\REQUIRE The received signals $ \mathbf{Y}_{\mathrm{da}} $, and the convergence threshold $ \epsilon $.
		\STATE Set $ l_{\mathrm{ao}}=1 $; initialize $ \mathbf{U}_2^{l_{\mathrm{ao}}-1} $, $ \mathbf{T}_2^{l_{\mathrm{ao}}-1} $, and $ \mathbf{H}_{\alpha}^{l_{\mathrm{ao}}-1} $ randomly; initialize AoAs $ \boldsymbol{\Theta}^{l_{\mathrm{ao}}-1} $ and AoDs $ \boldsymbol{\Phi}^{l_{\mathrm{ao}}-1}$ by the modified MUSIC algorithm in \cite{Qi2019OffGrid}; 
		\REPEAT
		\STATE Estimate the channel gain $ \mathbf{H}_{\alpha}^{l_{\mathrm{ao}}} $ by using \eqref{eq:estchgain};
		\STATE Estimate mismatch coefficients $ \mathbf{T}_2^{l_{\mathrm{ao}}} $ by using \eqref{eq:estt2};
		\STATE Estimated mismatch coefficients $ \mathbf{U}_2^{l_{\mathrm{ao}}} $ by using \eqref{eq:estu2};
		\STATE Upgrade the AoAs $ \boldsymbol{\Theta}^{l_{\mathrm{ao}}} $ and AoDs $ \boldsymbol{\Phi}^{l_{\mathrm{ao}}}$ with Algorithm \ref{alg:updirections}; set $ l_{\mathrm{ao}}=l_{\mathrm{ao}}+1 $;
		\UNTIL{$|\bar{g}(\mathbf{T}_2^{l_{\mathrm{ao}}-1},\mathbf{U}_2^{l_{\mathrm{ao}}-1},\mathbf{H}_{\alpha}^{l_{\mathrm{ao}}-1},\boldsymbol{\Theta}^{l_{\mathrm{ao}}-1},\boldsymbol{\Phi}^{l_{\mathrm{ao}}-1})-\bar{g}(\mathbf{T}_2^{l_{\mathrm{ao}}},\mathbf{U}_2^{l_{\mathrm{ao}}},\mathbf{H}_{\alpha}^{l_{\mathrm{ao}}},\boldsymbol{\Theta}^{l_{\mathrm{ao}}},\boldsymbol{\Phi}^{l_{\mathrm{ao}}})| <\epsilon $}
		\ENSURE The mismatch coefficients $ \mathbf{U}_2^{l_{\mathrm{ao}}} $ and $ \mathbf{T}_2^{l_{\mathrm{ao}}} $.
	\end{algorithmic}
\end{algorithm}

\begin{rem}[Convergence analysis]
	In Algorithm \ref{alg:updirections}, each iteration can minimize the objective of $ \mathcal{P}_{2.1} $, i.e.,
	\if\thecol1
	\begin{equation}
		\left\|\mathbf{Y}_{\mathrm{dar}}^{l_{\mathrm{ao}}}-\bar{\mathbf{B}}_{\mathrm{da}}^T\mathbf{U}_2^{l_{\mathrm{ao}}}\bar{\mathbf{A}}_{\mathrm{r}}(\bar{\boldsymbol{\Phi}}^{l_{\mathrm{an}}-1})\boldsymbol{\Lambda}\mathbf{H}_{\mathrm{r}}^{l_{\mathrm{ao}}}\right\|_{\mathrm{F}}^2\ge \left\|\mathbf{Y}_{\mathrm{dar}}^{l_{\mathrm{ao}}}-\bar{\mathbf{B}}_{\mathrm{da}}^T\mathbf{U}_2^{l_{\mathrm{ao}}}\bar{\mathbf{A}}_{\mathrm{r}}(\bar{\boldsymbol{\Phi}}^{l_{\mathrm{an}}})\boldsymbol{\Lambda}\mathbf{H}_{\mathrm{r}}^{l_{\mathrm{ao}}}\right\|_{\mathrm{F}}^2.
	\end{equation}
	\else
	\begin{equation}
		\begin{split}
			&\left\|\mathbf{Y}_{\mathrm{dar}}^{l_{\mathrm{ao}}}-\bar{\mathbf{B}}_{\mathrm{da}}^T\mathbf{U}_2^{l_{\mathrm{ao}}}\bar{\mathbf{A}}_{\mathrm{r}}(\bar{\boldsymbol{\Phi}}^{l_{\mathrm{an}}-1})\boldsymbol{\Lambda}\mathbf{H}_{\mathrm{r}}^{l_{\mathrm{ao}}}\right\|_{\mathrm{F}}^2\ge\\
			&\left\|\mathbf{Y}_{\mathrm{dar}}^{l_{\mathrm{ao}}}-\bar{\mathbf{B}}_{\mathrm{da}}^T\mathbf{U}_2^{l_{\mathrm{ao}}}\bar{\mathbf{A}}_{\mathrm{r}}(\bar{\boldsymbol{\Phi}}^{l_{\mathrm{an}}})\boldsymbol{\Lambda}\mathbf{H}_{\mathrm{r}}^{l_{\mathrm{ao}}}\right\|_{\mathrm{F}}^2.
		\end{split}
	\end{equation}
	\fi
	Thus, Algorithm \ref{alg:updirections} can achieve a local convergence. For Algorithm \ref{alg:ADMM}, each alternating optimization can minimize the objective $ \bar{g}(\mathbf{T}_2,\mathbf{U}_2,\mathbf{H}_{\alpha},\boldsymbol{\Theta},\boldsymbol{\Phi}) $. In other words, we have
	\if\thecol1
	\begin{equation}
		\begin{split}
			&\bar{g}(\mathbf{T}_2^{l_{\mathrm{ao}}-1},\mathbf{U}_2^{l_{\mathrm{ao}}-1},\mathbf{H}_{\alpha}^{l_{\mathrm{ao}}-1},\boldsymbol{\Theta}^{l_{\mathrm{ao}}-1},\boldsymbol{\Phi}^{l_{\mathrm{ao}}-1})\ge\bar{g}(\mathbf{T}_2^{l_{\mathrm{ao}}-1},\mathbf{U}_2^{l_{\mathrm{ao}}-1},\mathbf{H}_{\alpha}^{l_{\mathrm{ao}}},\boldsymbol{\Theta}^{l_{\mathrm{ao}}-1},\boldsymbol{\Phi}^{l_{\mathrm{ao}}-1})\\
			&\ge\bar{g}(\mathbf{T}_2^{l_{\mathrm{ao}}-1},\mathbf{U}_2^{l_{\mathrm{ao}}},\mathbf{H}_{\alpha}^{l_{\mathrm{ao}}},\boldsymbol{\Theta}^{l_{\mathrm{ao}}-1},\boldsymbol{\Phi}^{l_{\mathrm{ao}}-1})\ge\cdots\ge\bar{g}(\mathbf{T}_2^{l_{\mathrm{ao}}},\mathbf{U}_2^{l_{\mathrm{ao}}},\mathbf{H}_{\alpha}^{l_{\mathrm{ao}}},\boldsymbol{\Theta}^{l_{\mathrm{ao}}},\boldsymbol{\Phi}^{l_{\mathrm{ao}}}),
		\end{split}		
	\end{equation}
	\else
	\begin{equation}
		\begin{split}
			&\bar{g}(\mathbf{T}_2^{l_{\mathrm{ao}}-1},\mathbf{U}_2^{l_{\mathrm{ao}}-1},\mathbf{H}_{\alpha}^{l_{\mathrm{ao}}-1},\boldsymbol{\Theta}^{l_{\mathrm{ao}}-1},\boldsymbol{\Phi}^{l_{\mathrm{ao}}-1})\\
			\ge&\bar{g}(\mathbf{T}_2^{l_{\mathrm{ao}}-1},\mathbf{U}_2^{l_{\mathrm{ao}}-1},\mathbf{H}_{\alpha}^{l_{\mathrm{ao}}},\boldsymbol{\Theta}^{l_{\mathrm{ao}}-1},\boldsymbol{\Phi}^{l_{\mathrm{ao}}-1})\ge\cdots\\
			\ge&\bar{g}(\mathbf{T}_2^{l_{\mathrm{ao}}},\mathbf{U}_2^{l_{\mathrm{ao}}},\mathbf{H}_{\alpha}^{l_{\mathrm{ao}}},\boldsymbol{\Theta}^{l_{\mathrm{ao}}},\boldsymbol{\Phi}^{l_{\mathrm{ao}}}),
		\end{split}		
	\end{equation}
	\fi
	and thus, Algorithm \ref{alg:ADMM} converges to a minimum.
\end{rem}

	\textcolor{black}{
	Similarly, the mismatch coefficients of the receive RF chains of the BS and the transmit RF chains of the UE can be estimated by the uplink calibration. During the uplink calibration, the UE transmits pilots to the BS, and the BS estimates the mismatch coefficients by Proposition 3 and Algorithm 2. With estimated mismatch coefficients, the reciprocity mismatch can be compensated in the digital domain. Thus, the overall procedure of HAC can be summarized as follows.
	\begin{enumerate}[label=\textbf{Step \arabic*},leftmargin=3.5em]
		\item \emph{(Downlink calibration)} The BS sends downlink pilots to the UE. After receiving the downlink pilots, the UE jointly estimates the mismatch coefficients of the transmit RF chains of BS and the receive RF chains of UE by Algorithm 2;
		\item \emph{(Uplink calibration)} The UE transmits uplink pilots to the BS. Based on the received uplink pilots, the BS calculates the mismatch coefficients of the receive RF chains of BS and the transmit RF chains of UE with Algorithm 2;
		\item \emph{(Mismatch coefficients feedback)} The UE feeds back the estimated mismatch coefficients to the BS, and the BS sends the mismatch coefficients to the UE;
		\item \emph{(Reciprocity mismatch compensation)}  During data transmission phases, CSI $ \mathbf{H} $ can be estimated by utilizing the knowledge of mismatch coefficients and some existing approaches , e.g., the methods presented in \cite{Guo2017MillimeterWave,Venugopal2017Channel,Zhang2022MMVBased}. Then, the equivalent downlink CSI can be formulated as $ \mathbf{U}_2\mathbf{H}^T\mathbf{T}_2 $, and the precoding/combining and beamforming matrices can be designed by some existing approaches, e.g. the methods proposed in \cite{Sohrabi2016Hybrid,Yu2016Alternating,Lin2019Hybrida}. Finally, the mismatch of digital RF chains can be compensated by multiplying the inverse of mismatch coefficients matrices of digital chains on the precoding/combining matrices.
	\end{enumerate}
	}

\textcolor{black}{
\subsection{Extend HAC to the multi-user scenario}
Although the HAC is introduced in a point-to-point HBF system, it can be also applied in multi-user HBF systems, where a single BS serves $ G $ UEs simultaneously. Each UE is equipped with an HBF transceiver. Same as the single-UE case, the HAC consists of the downlink calibration and the uplink calibration. For both downlink and uplink calibrations, the pilots and beamforming designs remain consistent to the single-UE case.
}

\textcolor{black}{
\emph{Downlink calibration:} The BS sends training pilots to the UEs, and the $ g $-th UE received signal can be denoted as
\begin{equation}
	\mathbf{y}_{\mathrm{d},g,l}=\mathbf{D}_{\mathrm{r},g,l}^T\mathbf{U}_{g,1}\mathbf{B}_{\mathrm{r},g,l}^T\mathbf{U}_{g,2}\mathbf{H}_{g}^T\mathbf{T}_2\mathbf{F}_{\mathrm{t},l}\mathbf{T}_1\mathbf{W}_{\mathrm{t},l}\mathbf{x}_{\mathrm{d},l}+\tilde{\mathbf{n}}_{\mathrm{d},g,l},
	\label{eq:murecsig}
\end{equation}
where $ \mathbf{H}_{g} $ denotes the wireless channel between the $ g $-th UE and the BS, $ \mathbf{U}_{g,1} $ and $ \mathbf{U}_{g,2} $ are the mismatch coefficient matrices of the receive digital and analog chains of the $ g $-th UE. Since \eqref{eq:murecsig} of each UE is same as \eqref{eq:generalsgianl}, it can be solved by the same approach, i.e., Proposition \ref{prop:HACproblemdcp}, Proposition \ref{prop:solutot1u1}, and Algorithm \ref{alg:ADMM}.
}

\textcolor{black}{
\emph{Uplink calibration:} The UEs sends training pilots to the BS, and the received signal can be given by
\begin{equation}
		\mathbf{y}_{\mathrm{u},l}=\mathbf{W}_{\mathrm{r},l}^T\mathbf{R}_1\mathbf{F}_{\mathrm{r},l}^T\mathbf{R}_2\mathbf{H}_{\mathrm{mu}}\bar{\mathbf{V}}_{2}\mathbf{B}_{\mathrm{t},l}\bar{\mathbf{V}}_{1}\mathbf{D}_{\mathrm{t},l}\mathbf{x}_{\mathrm{u},l}+\tilde{\mathbf{n}}_{\mathrm{u},l},
		\label{eq:muulrecsig}
\end{equation}
where $ \mathbf{H}_{\mathrm{mu}}=[\mathbf{H}_1,\cdots,\mathbf{H}_G] $, $ \bar{\mathbf{V}}_2=\mathrm{blkdiag}(\mathbf{V}_{1,2},\cdots,\mathbf{V}_{G,2}) $, $ \mathbf{B}_{\mathrm{t},l}=\mathrm{blkdiag}(\mathbf{B}_{\mathrm{t},1,l},\cdots,\mathbf{B}_{\mathrm{t},G,l}) $, $ \bar{\mathbf{V}}_{1}=\mathrm{blkdiag}(\mathbf{V}_{1,1},\cdots,\mathbf{V}_{G,1}) $, $ \mathbf{x}_{\mathrm{u},l}=[\mathbf{x}_{\mathrm{u},1,l}^T\mathbf{D}_{\mathrm{t},1,l}^T,\cdots,\mathbf{x}_{\mathrm{u},G,l}^T\mathbf{D}_{\mathrm{t},G,l}^T]^T $, $ \mathbf{V}_{g,1} $ and $ \mathbf{V}_{g,2} $ represent the mismatch matrices of the transmit digital and analog chains of the $ g $-th UE. The uplink calibration problem can be formulated as
\begin{equation}
	\min_{\bar{\mathbf{V}}_i,\mathbf{R}_i,\mathbf{H}_{\mathrm{mu}}}\  \sum_{l=1}^{L_{\mathrm{u}}}\left\|\mathbf{y}_{\mathrm{u},l}-\mathbf{W}_{\mathrm{r},l}^T\mathbf{R}_1\mathbf{F}_{\mathrm{r},l}^T\mathbf{R}_2\mathbf{H}_{\mathrm{mu}}\bar{\mathbf{V}}_{2}\mathbf{B}_{\mathrm{t},l}\bar{\mathbf{V}}_{1}\mathbf{x}_{\mathrm{u},l}\right\|_{\mathrm{F}}^2,
\end{equation}
which can be decomposed into the calibration problem of digital chains and analog chains by Proposition \ref{prop:HACproblemdcp}. The uplink calibration problem of digital chains can be solved by Proposition \ref{prop:solutot1u1}, whereas the problem of analog chains can not solved by Algorithm \ref{alg:ADMM} due to the different structure of $ \mathbf{H}_{\mathrm{mu}} $. Fortunately, if UEs feed back AoAs and AoDs estimated in the downlink calibration to the BS, the BS only estimates the mismatch coefficients $ \bar{\mathbf{V}}_2 $ and $ \mathbf{R}_2 $ which can be solved by alternating optimization approach similar to Algorithm \ref{alg:ADMM}.
}

\section{Performance Analysis of Reciprocity Calibration}
In this section, we will analyze the performance of the proposed HAC. The minimum length of the calibration pilots will be first derived, followed by the overhead and computational complexity analysis. To measure the performance of the proposed calibration approach, the Cram\'er-Rao lower bound will be derived as the benchmark of the calibration performance.

\subsection{Overhead and Complexity of HAC}
Based on the calibration signal design and estimation approaches, we can derive the requirements of the length of calibration pilots.

\begin{prop}[Length of downlink pilots]\label{prop:pilotlength}
	The proposed downlink training and estimation approaches require that the length of pilots meets the following conditions
	\begin{equation}
		\begin{cases}
			Q_{\mathrm{dr}}\geq M_{\mathrm{t}},\\
			P_{\mathrm{dr}}\geq 1,\\
			Q_{\mathrm{da}}\geq N_{\mathrm{t}}-K+1,\\
			P_{\mathrm{da}}\geq N_{\mathrm{r}}-K+1.
		\end{cases}
		\label{eq:lengthofpilot}
	\end{equation}
\end{prop}

\begin{IEEEproof}
The complete proof is presented in Appendix D of Supplementary Material.
\end{IEEEproof}

Based on the above pilot requirements and the proposed calibration algorithms, the overhead and computational complexity of HAC can be given in the following lemma.

\textcolor{black}{
	\begin{rem}[Overhead and complexity of the proposed HAC]
		By considering the length of pilots exactly meets the requirements denoted as \eqref{eq:lengthofpilot}, the overhead of downlink training is proportional to $ M_{\mathrm{t}}+N_{\mathrm{r}}N_{\mathrm{r}} $.  In each iteration , the computational complexity is mainly caused by computing the inverse of matrices. Let $ \mathcal{O}(L_{\mathrm{an}}K^3) $ denote the total iteration number of updating AoA/AoD and $ L_{\mathrm{ao}} $ represent the iteration number of Algorithm \ref{alg:ADMM}. The complexity of solve problem $ \mathcal{P}_{2.1} $ can be given by $ \mathcal{O}[L_{\mathrm{ao}}(M_{\mathrm{r}}^3+M_{\mathrm{t}}^3+N_{\mathrm{r}}^3+N_{\mathrm{t}}^3+L_{\mathrm{ao}}K^3)] $. 
		The comparisons between the overhead and complexity of HAC and CRC are shown in Table \ref{tab:comhaccrc}, which indicates that the proposed HAC requires requires less overhead. According to experiments, both Algorithm \ref{alg:updirections} and Algorithm \ref{alg:ADMM} converge after several iterations, and the complexity of the HAC is also lower than the CRC.
	\end{rem}
	\begin{table}[t]
		\centering
		\caption{Comparison of HAC and CRC}
		\label{tab:comhaccrc}
		\begin{tabular}{ccc}
			\toprule
			&Overhead&Complexity\\
			\midrule
			CRC &$ N_{\mathrm{t}}M_{\mathrm{t}}N_{\mathrm{r}} $&$ \mathcal{O}(N_{\mathrm{t}}^3M_{\mathrm{t}}^3N_{\mathrm{r}}^3M_{\mathrm{r}}^3) $\\
			HAC &$ M_{\mathrm{t}}+N_{\mathrm{t}}N_{\mathrm{r}} $&$ \mathcal{O}[L_{\mathrm{ao}}(M_{\mathrm{r}}^3+M_{\mathrm{t}}^3+N_{\mathrm{r}}^3+N_{\mathrm{t}}^3+L_{\mathrm{ao}}K^3)] $\\
			\bottomrule
		\end{tabular}
	\end{table}
	}

\subsection{CRLB of Calibration Coefficients}
To verify the performance of proposed joint estimation approaches, we derive the CRLB of $\tilde{\mathbf{u}}_1$, $\mathbf{t}_1$, $ \mathbf{u}_2 $, and $ \mathbf{t}_2 $ to be the performance benchmark, where $ \tilde{\mathbf{u}}_1=[u_{1,2},\cdots, u_{1,M_{\mathrm{r}}}]^T $. We first define the variable vectors as
\begin{equation}
	\begin{split}
		\boldsymbol{\eta}=&[\Re\{\tilde{\mathbf{u}}_1^T\},\Im\{\tilde{\mathbf{u}}_1^T\},\Re\{\mathbf{t}_1^T\},\Im\{\mathbf{t}_1^T\},[\Re\{\mathbf{u}_2^T\},\Im\{\mathbf{u}_2^T\},\\
		&\Re\{\mathbf{t}_2^T\},\Im\{\mathbf{t}_2^T\},\Re\{\mathbf{h}_{\alpha}^T\},\Im\{\mathbf{h}_{\alpha}^T\},\boldsymbol{\Theta}^T,\boldsymbol{\Phi}^T]^T,\\
	\end{split}
\end{equation}
\begin{equation}
	\boldsymbol{\eta}_{\mathrm{ut}}=[\tilde{\mathbf{u}}_1^T, \mathbf{t}_1^T, \mathbf{u}_2^T, \mathbf{t}_2^T]^T,
\end{equation}
and the transformation function vector $ \mathbf{g}(\boldsymbol{\eta}) $  as
\if\thecol1
\begin{equation}
	\begin{split}
		\boldsymbol{\eta}_{\mathrm{ut}}=\mathbf{g}(\boldsymbol{\eta})=&\bigl[\Re\{\tilde{\mathbf{u}}_1^T\}+j\Im\{\tilde{\mathbf{u}}_1^T\},\Re\{\mathbf{t}_1^T\}+j\Im\{\mathbf{t}_1^T\},\\
		&\Re\{\mathbf{u}_2^T\}+j\Im\{\mathbf{u}_2^T\},\Re\{\mathbf{t}_2^T\}+j\Im\{\mathbf{t}_2^T\}\bigr]^T.
	\end{split}
\end{equation}
\else
\begin{equation}
	\begin{split}
		\boldsymbol{\eta}_{\mathrm{ut}}&=\mathbf{g}(\boldsymbol{\eta})\\
		&=\bigl[\Re\{\tilde{\mathbf{u}}_1^T\}+j\Im\{\tilde{\mathbf{u}}_1^T\},\Re\{\mathbf{t}_1^T\}+j\Im\{\mathbf{t}_1^T\},\\
		&\quad\ \Re\{\mathbf{u}_2^T\}+j\Im\{\mathbf{u}_2^T\},\Re\{\mathbf{t}_2^T\}+j\Im\{\mathbf{t}_2^T\}\bigr]^T.
	\end{split}
\end{equation}
\fi
Based on this definition, the CRLB of the equivalent mismatch coefficients $\boldsymbol{\eta}_{\mathrm{ut}}$ can be defined as follows.

\begin{deff}[CRLB of ${\boldsymbol{\eta}}_{\mathrm{ut}}$]\label{def:CLRB}
	According to the transformation relation in \cite{Kay1993Fundamentalsa}, the CRLB of $\boldsymbol{\eta}_{\mathrm{ut}}$ can be given by
	\begin{equation}
		\mathrm{CRLB}({\eta}_{\mathrm{ut},i}) =\left[\frac{\partial \mathbf{g}(\boldsymbol{\eta})}{\partial \boldsymbol{\eta}^T}\boldsymbol{\mathcal{I}}(\boldsymbol{\eta})^{-1}\left(\frac{\partial \mathbf{g}(\boldsymbol{\eta})}{\partial \boldsymbol{\eta}^T}\right)^H\right]_{i,i},
		\label{eq:defCRLB}
	\end{equation}
	where $\boldsymbol{\mathcal{I}}(\boldsymbol{\eta})$ denotes the Fisher information matrix of $\boldsymbol{\eta}$, and $ {\eta}_{\mathrm{ut},i} $ is the $ i $-th entry of $\boldsymbol{\eta}_{\mathrm{ut}}$.
\end{deff}	

\begin{lem}[Transformation of the Fisher information matrix]\label{lem:fimtrn}
	\if\thecol1
	By dividing the variables into two parts and defining the vectors $ \boldsymbol{\eta}_1=[\Re\{\tilde{\mathbf{u}}_1^T\},\Im\{\tilde{\mathbf{u}}_1^T\},\Re\{\mathbf{t}_1^T\},\Im\{\mathbf{t}_1^T\}]^T $ and $ \boldsymbol{\eta}_2=[\Re\{\mathbf{u}_2^T\},\Im\{\mathbf{u}_2^T\},\\\Re\{\mathbf{t}_2^T\},\Im\{\mathbf{t}_2^T\},\Re\{\mathbf{h}_{\alpha}^T\},\Im\{\mathbf{h}_{\alpha}^T\},\boldsymbol{\Theta}^T,\boldsymbol{\Phi}^T]^T $, the Fisher information matrix $\boldsymbol{\mathcal{I}}(\boldsymbol{\eta})$ can be further denoted as
	\begin{equation}
		\boldsymbol{\mathcal{I}}(\boldsymbol{\eta})=\mathrm{blkdiag}[\boldsymbol{\mathcal{I}}(\boldsymbol{\eta}_1),\boldsymbol{\mathcal{I}}(\boldsymbol{\eta}_2)],
	\end{equation}
	where $\boldsymbol{\mathcal{I}}(\boldsymbol{\eta}_{i})$ denotes the Fisher information matrix of $\boldsymbol{\eta}_{i}$, $ \forall i\in\{1,2\} $.
	\else
	By dividing the variables into two parts and defining the vectors $ \boldsymbol{\eta}_1=[\Re\{\tilde{\mathbf{u}}_1^T\},\Im\{\tilde{\mathbf{u}}_1^T\},\Re\{\mathbf{t}_1^T\},\Im\{\mathbf{t}_1^T\}]^T $ and $ \boldsymbol{\eta}_2=[\Re\{\mathbf{u}_2^T\},\Im\{\mathbf{u}_2^T\},\Re\{\mathbf{t}_2^T\},\Im\{\mathbf{t}_2^T\},\Re\{\mathbf{h}_{\alpha}^T\},\Im\{\mathbf{h}_{\alpha}^T\},\boldsymbol{\Theta}^T,\boldsymbol{\Phi}^T]^T $, the Fisher information matrix $\boldsymbol{\mathcal{I}}(\boldsymbol{\eta})$ can be further denoted as
	\begin{equation}
		\boldsymbol{\mathcal{I}}(\boldsymbol{\eta})=\mathrm{blkdiag}[\boldsymbol{\mathcal{I}}(\boldsymbol{\eta}_1),\boldsymbol{\mathcal{I}}(\boldsymbol{\eta}_2)],
	\end{equation}
	where $\boldsymbol{\mathcal{I}}(\boldsymbol{\eta}_{i})$ denotes the Fisher information matrix of $\boldsymbol{\eta}_{i}$, $ \forall i\in\{1,2\} $.
	\fi
\end{lem}

\begin{IEEEproof}
	The complete proof is presented in Appendix E of Supplementary Material.
\end{IEEEproof}

Thus, to derive the closed-form expressions of the CRLB of $ \boldsymbol{\eta}_{\mathrm{ut}} $, we first derive the closed-form expression of $ \mathcal{I}(\boldsymbol{\eta}_{1}) $ denoted as follows.

\begin{lem}[Closed-form expression of $ \mathcal{I}(\boldsymbol{\eta}_1) $ ]\label{lem:fisherinfomatu1t1}
	The closed-form expression can be given by
	\begin{equation}
		\boldsymbol{\mathcal{I}}(\boldsymbol{\eta}_1)=\mathrm{blkdiag}(\boldsymbol{\mathcal{I}}(\boldsymbol{\eta}_{1,1}),\boldsymbol{\mathcal{I}}(\boldsymbol{\eta}_{1,2})),
		\label{eq:FIMt1u1}
	\end{equation}
	where $ \boldsymbol{\mathcal{I}}(\boldsymbol{\eta}_{1,1})=\lim_{\gamma\rightarrow 0}2\gamma^{-1}\sum_{p=1}^{P_{\mathrm{dr}}}\|\mathbf{x}_{\mathrm{tn},p}\|^2\mathbf{I}_{2M_{\mathrm{r}}-2} $, $ \boldsymbol{\mathcal{I}}(\boldsymbol{\eta}_{1,2})=2\rho_{\mathrm{c}}|\beta_{\mathrm{d}}|^2L_{\mathrm{dr}}\sigma_{\mathrm{n}}^{-2}\mathbf{I}_{2M_{\mathrm{t}}} $, $ \boldsymbol{\eta}_{1,1}=[\Re\{\tilde{\mathbf{u}}_1^T\},\Im\{\tilde{\mathbf{u}}_1^T\}]^T $, and $ \boldsymbol{\eta}_{1,2}=[\Re\{\mathbf{t}_1^T\},\Im\{\mathbf{t}_1^T\}]^T $.
\end{lem}

\begin{IEEEproof}
	The complete proof is presented in Appendix F of Supplementary Material.
\end{IEEEproof}

Similarly, for deriving the closed-form expressions of $ \boldsymbol{\eta}_{\mathrm{ut}} $, we derive $ \mathcal{I}(\boldsymbol{\eta}_{2}) $ denoted in the following lemma.

\begin{lem}[Closed-form expression of $ \boldsymbol{I}(\boldsymbol{\eta}_2) $]\label{lem:fisherinfomatu2t2}
	The closed-form expression of $ \boldsymbol{I}(\boldsymbol{\eta}_2) $ can be given by
	\begin{equation}
		\boldsymbol{I}(\boldsymbol{\eta}_2)=\frac{2}{\sigma_{\mathrm{n}}^2}\Re \left\lbrace \boldsymbol{\Upsilon}_{\mathrm{\eta}}^H\boldsymbol{\Upsilon}_{\mathrm{\eta}} \right\rbrace,
		\label{eq:FIMt2u2}
	\end{equation}
	\if\thecol1
	where $ \boldsymbol{\Upsilon}_{\mathrm{\eta}}=[\mathbf{\Gamma}_{\mathrm{t}},j\mathbf{\Gamma}_{\mathrm{t}},\mathbf{\Gamma}_{\mathrm{u}},j\mathbf{\Gamma}_{\mathrm{u}},\mathbf{\Gamma}_{\mathrm{h}},j\mathbf{\Gamma}_{\mathrm{h}},\mathbf{\Gamma}_{\mathrm{\theta}},\mathbf{\Gamma}_{\mathrm{\Phi}}] $, $ \mathbf{\Gamma_{\mathrm{h}}} $, $ \mathbf{\Gamma}_{\mathrm{u}} $, and $ \mathbf{\Gamma_{\mathrm{t}}} $ are defined in Lemma \ref{lem:soltot2u2h}, $ \mathbf{\Gamma_{\mathrm{\theta}}}=(\tilde{\mathbf{X}}_{\mathrm{da}}^T\mathbf{T}_2\otimes\bar{\mathbf{B}}_{\mathrm{r}}\mathbf{U}_2\mathbf{A}_{\mathrm{r}}\mathbf{H}_{\alpha})\mathbf{E}_{\mathrm{x},N_{\mathrm{t}}K}\mathrm{blkdiag}(\bar{\mathbf{a}}_{\mathrm{t}}(\theta_1),\cdots,\bar{\mathbf{a}}_{\mathrm{t}}(\theta_K)) $, $ \bar{\mathbf{a}}_{\mathrm{t}}(\theta_k) = \mathbf{a}_{\mathrm{t}}(\theta_k)\circ[0,-j\frac{2\pi d}{\lambda}\cos\theta_k,\cdots,\\-j\frac{2\pi d}{\lambda}(N_{\mathrm{t}}-1)\cos\theta_k]^T $, and $ \mathbf{E}_{\mathrm{x},N_{\mathrm{t}},K}=\sum_{k=1}^{K}(\mathbf{e}_k^T\otimes \mathbf{I}_{N_{\mathrm{t}}}\otimes \mathbf{e}_k) $, $ \mathbf{e}_k $ is the $ k $-the column of $ \mathbf{I}_{K} $, $ \mathbf{\Gamma_{\mathrm{\Phi}}} =  (\tilde{\mathbf{X}}_{\mathrm{da}}^T\mathbf{T}_2\mathbf{A}_{\mathrm{t}}\mathbf{H}_{\alpha}\otimes\bar{\mathbf{B}}_{\mathrm{r}}\mathbf{U}_2)\mathrm{blkdiag}(\bar{\mathbf{a}}_{\mathrm{r}}(\phi_1),\cdots,\bar{\mathbf{a}}_{\mathrm{r}}(\phi_K)) $, and $ \bar{\mathbf{a}}_{\mathrm{r}}(\phi_k) = \mathbf{a}_{\mathrm{r}}(\phi_k)\circ[0,-j\frac{2\pi d}{\lambda}\cos\phi_k,\cdots,\\-j\frac{2\pi d}{\lambda}(N_{\mathrm{r}}-1)\cos\phi_k]^T $.
	\else
	where $ \boldsymbol{\Upsilon}_{\mathrm{\eta}}=[\mathbf{\Gamma}_{\mathrm{t}},j\mathbf{\Gamma}_{\mathrm{t}},\mathbf{\Gamma}_{\mathrm{u}},j\mathbf{\Gamma}_{\mathrm{u}},\mathbf{\Gamma}_{\mathrm{h}},j\mathbf{\Gamma}_{\mathrm{h}},\mathbf{\Gamma}_{\mathrm{\theta}},\mathbf{\Gamma}_{\mathrm{\Phi}}] $, $ \mathbf{\Gamma_{\mathrm{h}}} $, $ \mathbf{\Gamma}_{\mathrm{u}} $, and $ \mathbf{\Gamma_{\mathrm{t}}} $ are defined in Lemma \ref{lem:soltot2u2h}, $ \mathbf{\Gamma_{\mathrm{\theta}}}=(\tilde{\mathbf{X}}_{\mathrm{da}}^T\mathbf{T}_2\otimes\bar{\mathbf{B}}_{\mathrm{r}}\mathbf{U}_2\mathbf{A}_{\mathrm{r}}\mathbf{H}_{\alpha})\mathbf{E}_{\mathrm{x},N_{\mathrm{t}}K}\mathrm{blkdiag}(\bar{\mathbf{a}}_{\mathrm{t}}(\theta_1),\cdots,\bar{\mathbf{a}}_{\mathrm{t}}(\theta_K)) $, $ \bar{\mathbf{a}}_{\mathrm{t}}(\theta_k) = \mathbf{a}_{\mathrm{t}}(\theta_k)\circ[0,-j\frac{2\pi d}{\lambda}\cos\theta_k,\cdots,-j\frac{2\pi d}{\lambda}(N_{\mathrm{t}}-1)\cos\theta_k]^T $, and $ \mathbf{E}_{\mathrm{x},N_{\mathrm{t}},K}=\sum_{k=1}^{K}(\mathbf{e}_k^T\otimes \mathbf{I}_{N_{\mathrm{t}}}\otimes \mathbf{e}_k) $, $ \mathbf{e}_k $ is the $ k $-the column of $ \mathbf{I}_{K} $, $ \mathbf{\Gamma_{\mathrm{\Phi}}} =  (\tilde{\mathbf{X}}_{\mathrm{da}}^T\mathbf{T}_2\mathbf{A}_{\mathrm{t}}\mathbf{H}_{\alpha}\otimes\bar{\mathbf{B}}_{\mathrm{r}}\mathbf{U}_2)\mathrm{blkdiag}(\bar{\mathbf{a}}_{\mathrm{r}}(\phi_1),\cdots,\bar{\mathbf{a}}_{\mathrm{r}}(\phi_K)) $, and $ \bar{\mathbf{a}}_{\mathrm{r}}(\phi_k) = \mathbf{a}_{\mathrm{r}}(\phi_k)\circ[0,-j\frac{2\pi d}{\lambda}\cos\phi_k,\cdots,-j\frac{2\pi d}{\lambda}(N_{\mathrm{r}}-1)\cos\phi_k]^T $.
	\fi
\end{lem}
\begin{IEEEproof}
	The complete proof is presented in Appendix G of Supplementary Material.
\end{IEEEproof}

Based on Lemma \ref{lem:fisherinfomatu1t1} and Lemma \ref{lem:fisherinfomatu2t2}, the CRLB of $ \boldsymbol{\eta}_{\mathrm{ut}} $ can be given in the following proposition.
\begin{prop}[CRLB of $ \boldsymbol{\eta}_{\mathrm{ut}} $]\label{prop:CRLBDL}
	Based on the Definition  \ref{def:CLRB}, Lemma \ref{lem:fisherinfomatu1t1}, and Lemma \ref{lem:fisherinfomatu2t2}, the closed-form expression of the CRLB of $ \boldsymbol{\eta}_{\mathrm{ut}} $ can be given by
	\begin{equation}
		\mathrm{CRLB}(\eta_{\mathrm{ut},i})=\left[\boldsymbol{\mathcal{J}}(\boldsymbol{\eta}_{\mathrm{ut}})\right]_{i,i},
		\label{eq:CRLBut12}
	\end{equation}
	with
	\if\thecol1
	\begin{equation}
		\boldsymbol{\mathcal{J}}(\boldsymbol{\eta}_{\mathrm{ut}})=\mathrm{blkdiag}\left(\mathbf{0}_{M_{\mathrm{r}}-1},\frac{\sigma_{\mathrm{n}}^{2}}{\rho_{\mathrm{c}}|\beta_{\mathrm{d}}|^2L_{\mathrm{dr}}}\mathbf{I}_{M_\mathrm{t}},\frac{\sigma_{\mathrm{n}}^2}{2}\boldsymbol{\Pi}\left(\Re\left\lbrace \boldsymbol{\Upsilon}_{\eta}^H\boldsymbol{\Upsilon}_{\eta} \right\rbrace\right)^{-1}\boldsymbol{\Pi}^H\right),
	\end{equation}
	\else
	\begin{equation}
		\begin{split}
			\boldsymbol{\mathcal{J}}(\boldsymbol{\eta}_{\mathrm{ut}})=\mathrm{blkdiag}\Big(\mathbf{0}_{M_{\mathrm{r}}-1},\frac{\sigma_{\mathrm{n}}^{2}}{\rho_{\mathrm{c}}|\beta_{\mathrm{d}}|^2L_{\mathrm{dr}}}\mathbf{I}_{M_\mathrm{t}},\\\frac{\sigma_{\mathrm{n}}^2}{2}\boldsymbol{\Pi}\left(\Re\left\lbrace \boldsymbol{\Upsilon}_{\eta}^H\boldsymbol{\Upsilon}_{\eta} \right\rbrace\right)^{-1}\boldsymbol{\Pi}^H\Big),
		\end{split}
	\end{equation}
	\fi
	where $ \boldsymbol{\Pi}=[\mathrm{blkdiag}([\mathbf{I}_{N_{\mathrm{r}}},j\mathbf{I}_{N_{\mathrm{r}}}],[\mathbf{I}_{N_{\mathrm{t}}},j\mathbf{I}_{N_{\mathrm{t}}}]),\mathbf{0}_{N_{\mathrm{r}}+N_{\mathrm{t}},4K}] $.
\end{prop}

\begin{IEEEproof}
	The complete proof is presented in Appendix H of Supplementary Material.
\end{IEEEproof}

\begin{rem}[CRLB analysis]\label{rem:CRLBan}
	From \eqref{eq:CRLBut12}, the CRLB of $ u_{1,m} \ (m\in[1:M_{\mathrm{r}}])$ is equal to zeros. This result is because $ u_{1,m} $ is estimated from the deterministic signals. The CRLB of $ t_{1,m}\ (m\in[1:M_{\mathrm{t}}]) $ can be given by $ \sigma_{\mathrm{n}}^2/(\rho_{\mathrm{c}}|\beta_{\mathrm{d}}|^2L_{\mathrm{dr}}) $, which indicates that increasing the pilots can improve the accuracy of $ t_{1,m} $.
\end{rem}



\section{Simulation Results and Discussions}
In this section, we will provide simulation results to evaluate the performance of the proposed reciprocity calibration approach for the mmWave-HBF system.

The system parameters are set as follows. Analog RF chains of the BS and UE, and the number of data streams are different in each simulation, while the number of digital RF chains equals to a quarter of the number of analog RF chains, i.e., $ M_{\mathrm{t}}=N_{\mathrm{t}}/4 $ and $ M_{\mathrm{r}}=N_{\mathrm{r}}/4 $. The path number $ K $ of the mmWave channel is set to $ 4 $. The variance $ \sigma_{\alpha}^2 $ of the channel gain $ \alpha_k $ is set to $ 1 $, and the AoAs and DoAs obey the uniform distribution, i.e., $ \{\theta_{k},\phi_k\}\sim\mathcal{U}(-\pi/2,\pi/2) $. Then, the amplitudes of reciprocity mismatch coefficients obey the log-normal distribution, i.e., $ \{\ln|t_{i,m}|,\ln|r_{i,m}|,\ln|u_{i,m}|,\ln|v_{i,m}|\}\sim\mathcal{CN}(0,0.01) $, and the phases of reciprocity mismatch coefficients follow the uniform distribution, i.e., $ \{\angle t_{i,m},\angle r_{i,m},\angle u_{i,m},\angle v_{i,m}\}\sim\mathcal{U}(-\pi/6,\pi/6) $. Further, the length of pilots is set to $ Q_{\mathrm{dr}}=M_{\mathrm{r}} $, $ P_{\mathrm{dr}}=1 $, and $ Q_{\mathrm{da}}=P_{\mathrm{da}}=125 $. Finally, the variance of the AWGN is $ 1 $, and $ \bar{\rho}_{\mathrm{c}}=\rho_{\mathrm{c}}/\sigma_{\mathrm{n}}^2 $ denotes the average SNR of received calibration signals during the simulation.

\subsection{NMSE of Estimated Mismatch Parameters}
To illustrate the performance of the proposed HAC calibration approach, we compare the normalized mean square error (NMSE) of the mismatch coefficients with the CRLB. The NMSE of the mismatch coefficients is defined as $ \mathrm{NMSE}(\boldsymbol{\eta}_{\mathrm{ut}})=\mathbb{E}\left\lbrace   \|\boldsymbol{\eta}_{\mathrm{ut}}-\hat{\boldsymbol{\eta}}_{\mathrm{ut}}\|_{\mathrm{F}}^2/\|\boldsymbol{\eta}_{\mathrm{ut}}\|_{\mathrm{F}}^2  \right\rbrace $. It is worth noting that the Oracle HAC represents the reciprocity calibration with the knowledge of AoAs and AODs, which is a performance benchmark of the proposed HAC.

\if\thecol1
\begin{figure}
	\centering
	\vspace{-2em}
	\begin{minipage}{0.45\linewidth}
		\includegraphics[width=\linewidth]{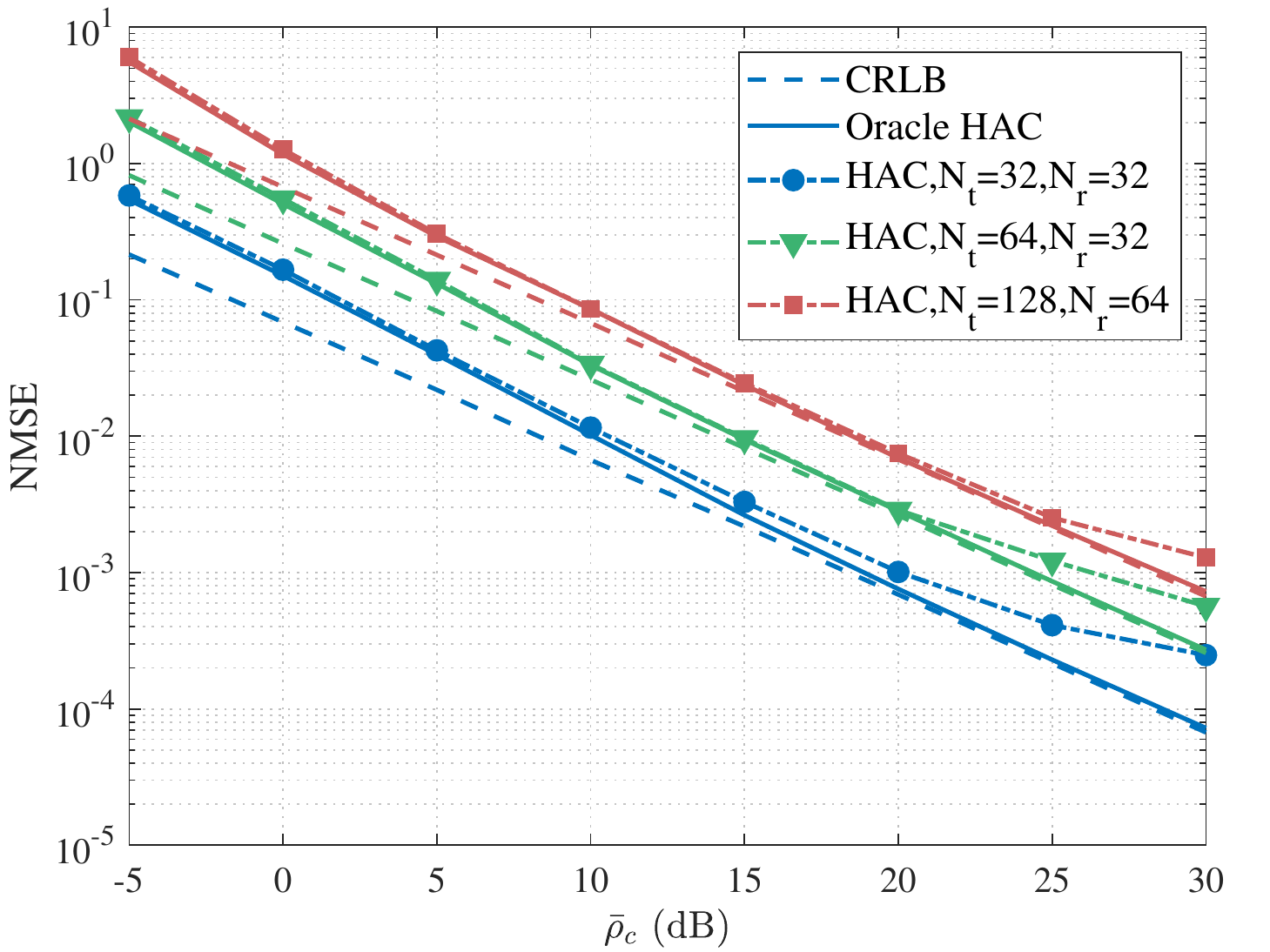}
		\vspace{-3em}
		\caption{The NMSE of the estimated mismatch coefficients versus the SNR of calibration signals.}
		\label{fig:NMSEVSSNR}
	\end{minipage}
	\hspace{2em}
	\begin{minipage}{0.45\linewidth}
		\includegraphics[width=\linewidth]{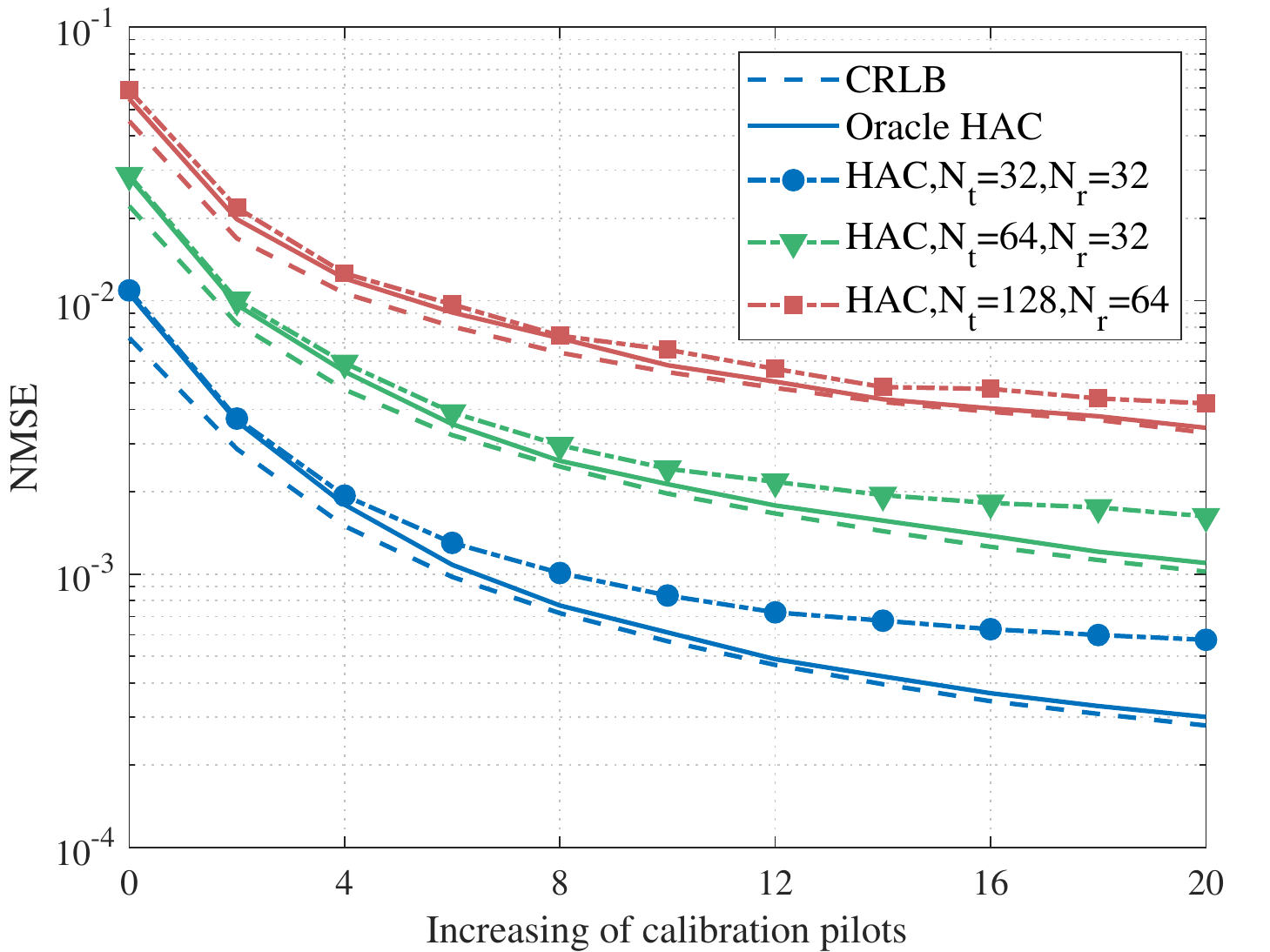}
		\vspace{-3em}
		\caption{The NMSE of the estimated mismatch coefficients versus the length of calibration pilots.}
		\label{fig:NMSEVSPilotsLength}
	\end{minipage}
	\vspace{-2em}
\end{figure}
\else
\begin{figure}
	\centering
	\includegraphics[width=0.8\linewidth]{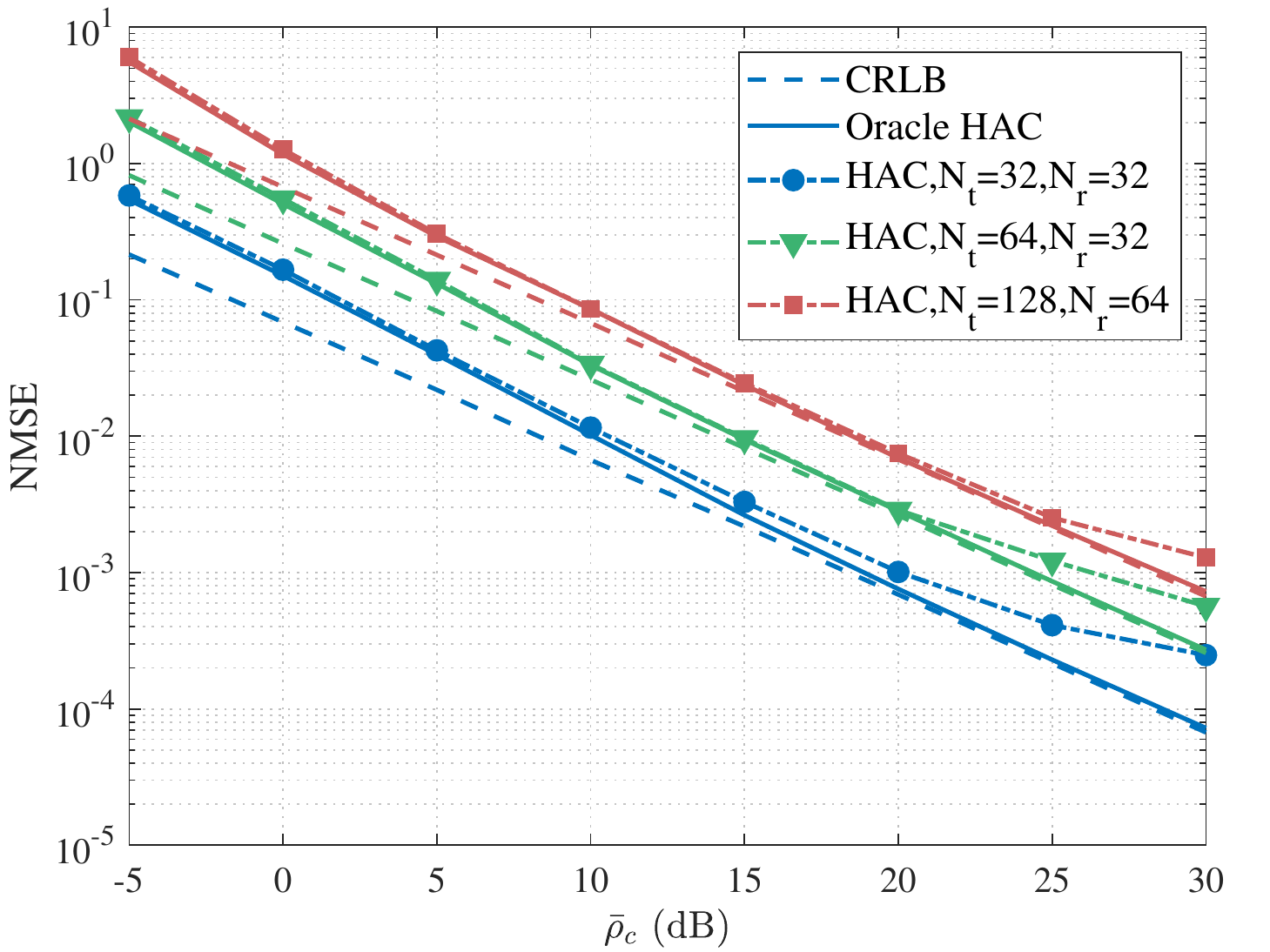}
	\caption{The NMSE of the estimated mismatch coefficients versus the SNR of calibration signals.}
	\label{fig:NMSEVSSNR}
\end{figure}

\begin{figure}
	\centering
	\includegraphics[width=0.8\linewidth]{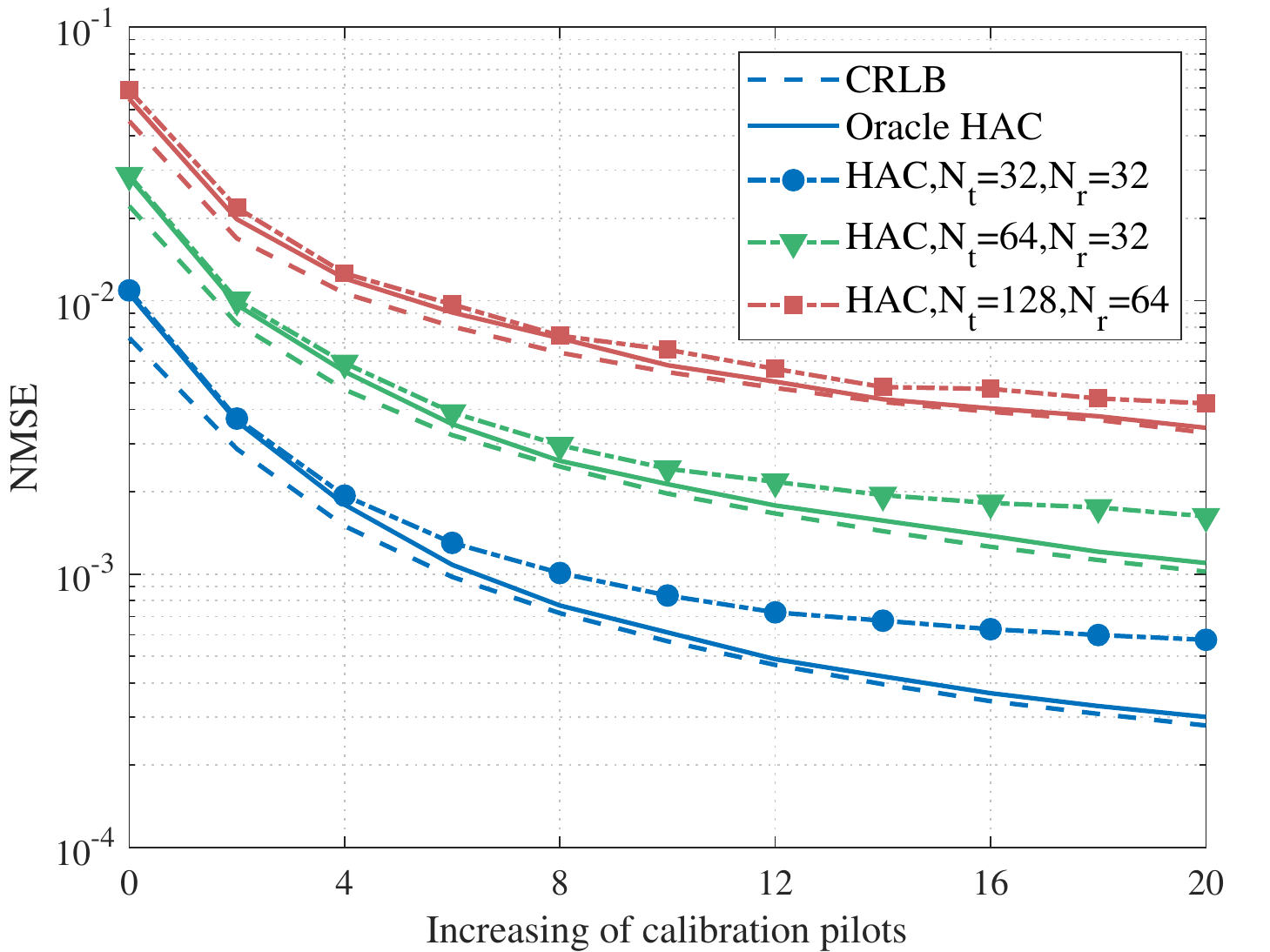}
	\caption{The NMSE of the estimated mismatch coefficients versus the length of calibration pilots.}
	\label{fig:NMSEVSPilotsLength}
\end{figure}
\fi

Fig. \ref{fig:NMSEVSSNR} demonstrates the NMSE of the mismatch coefficients versus the SNR $ \bar{\rho}_{\mathrm{c}} $ of the calibration signals, where the antenna numbers of the BS and the UE are set to three different sets of parameters given by $ (N_{\mathrm{t}},N_{\mathrm{r}})\in\{(32,32),(64,32),(128,64)\} $. It can be seen that the NMSE of the proposed HAC gradually achieves a floor with the increase of calibration SNR. This is because the solution to the nonconvex problem gets stuck in local optima. Further, the figure also shows that the floor effect can be alleviated when the antenna number increases, which is because the independence between array steering vectors increases with the increase of antenna number. Besides, we find that the NMSE increases with the antenna number. This result indicates that the system with more antennas requires higher calibration SNR to guarantee the same calibration performance as the system possessing fewer antennas.

Then, the NMSE of the mismatch coefficients versus the length of calibration pilots is illustrated in Fig. \ref{fig:NMSEVSPilotsLength} with the SNR of calibration signals set to $ \bar{\rho}_{\mathrm{c}}=10 $ dB. From the figure, it can be found that the NMSE and CRLB decrease with the increase of calibration pilots, which is consistent with the theoretical results shown in Proposition \ref{prop:CRLBDL}. This result indicates that better performance of the proposed HAC can be achieved at the cost of overhead or power. Also, the curves of the proposed HAC gradually approach floors when the length of pilots increases, while the curves of the Oracle HAC gradually converge to the CRLB. Increasing the antenna number can reduce the floor effect.

\subsection{NMSE of Channel Estimation}
To examine the efficacy of the reciprocity calibration in mmWave-HBF systems, we study the NMSE of the uplink channel estimation by using the two-dimension MUSIC algorithm proposed in \cite{Guo2017MillimeterWave}, and the pilot block is set to $ 40 $.The NSME of the estimated channel is defined as
\begin{equation}
	\mathrm{NMSE}(\mathbf{H}_{\mathrm{UL}})=\mathbb{E}\left\lbrace \|\hat{\mathbf{H}}_{\mathrm{UL}}-\mathbf{H}_{\mathrm{UL}}\|_{\mathrm{F}}^2/\|\mathbf{H}_{\mathrm{UL}}\|_{\mathrm{F}}^2\right\rbrace,
\end{equation}
where $ \hat{\mathbf{H}}_{\mathrm{UL}} $ denotes the estimated channel from the uplink pilots. It is worth noting that "Perfect Cal." denotes the mismatch coefficients $ \mathbf{U}_1 $, $ \mathbf{T}_1 $, $ \mathbf{U}_2 $, and $ \mathbf{T}_2 $ are known perfectly, and "Without Cal." represents that the mismatch coefficients are completely unknown.

\if\thecol1
\begin{figure}
	\centering
	\vspace{-2em}
	\begin{minipage}{0.45\linewidth}
		\includegraphics[width=\linewidth]{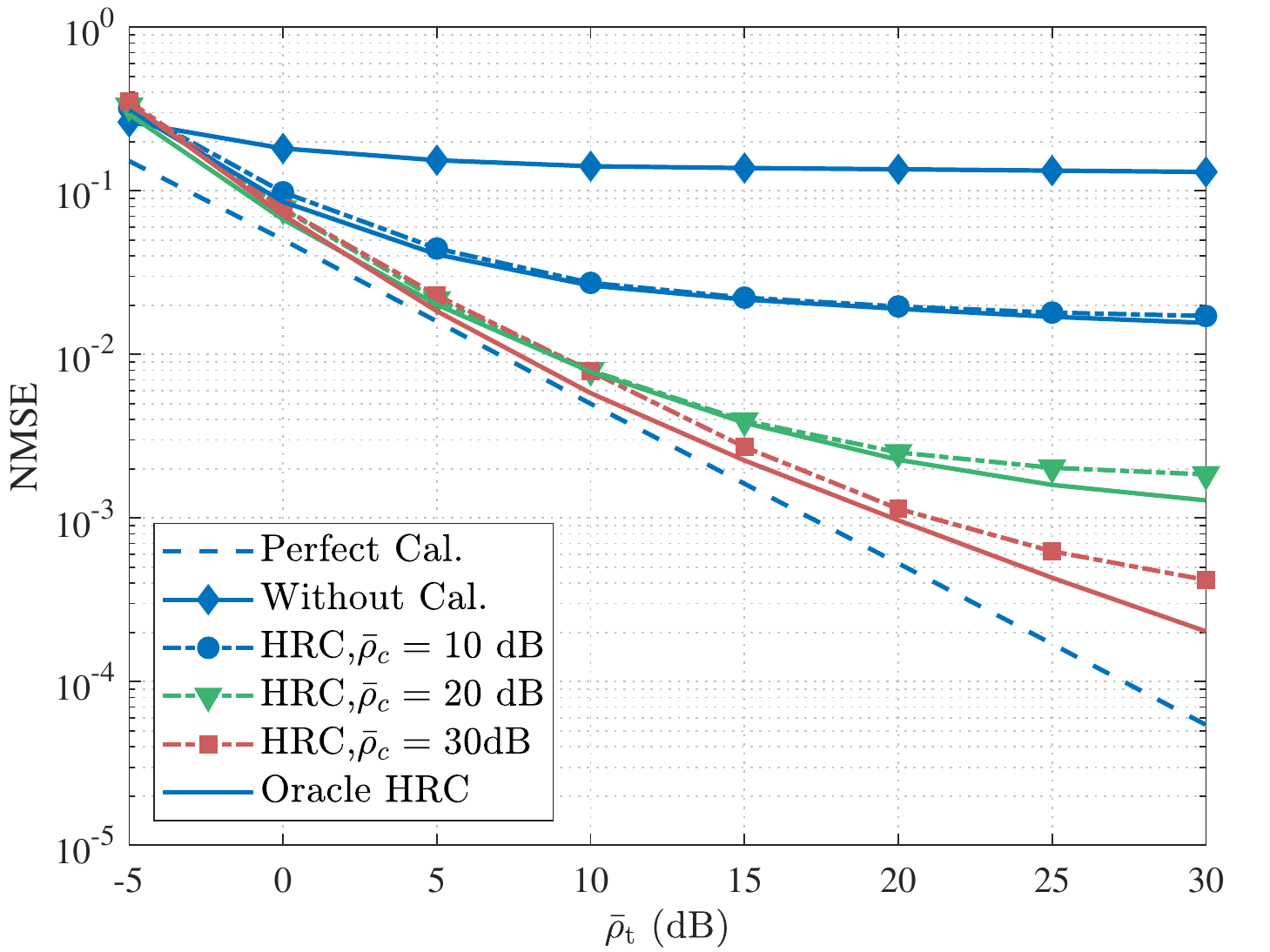}
		\vspace{-3em}
		\caption{The NMSE of the uplink channel estimation versus the SNR of training signals.}
		\label{fig:Channel_NMSEVSSNR}
	\end{minipage}
	\hspace{2em}
	\begin{minipage}{0.45\linewidth}
		\includegraphics[width=\linewidth]{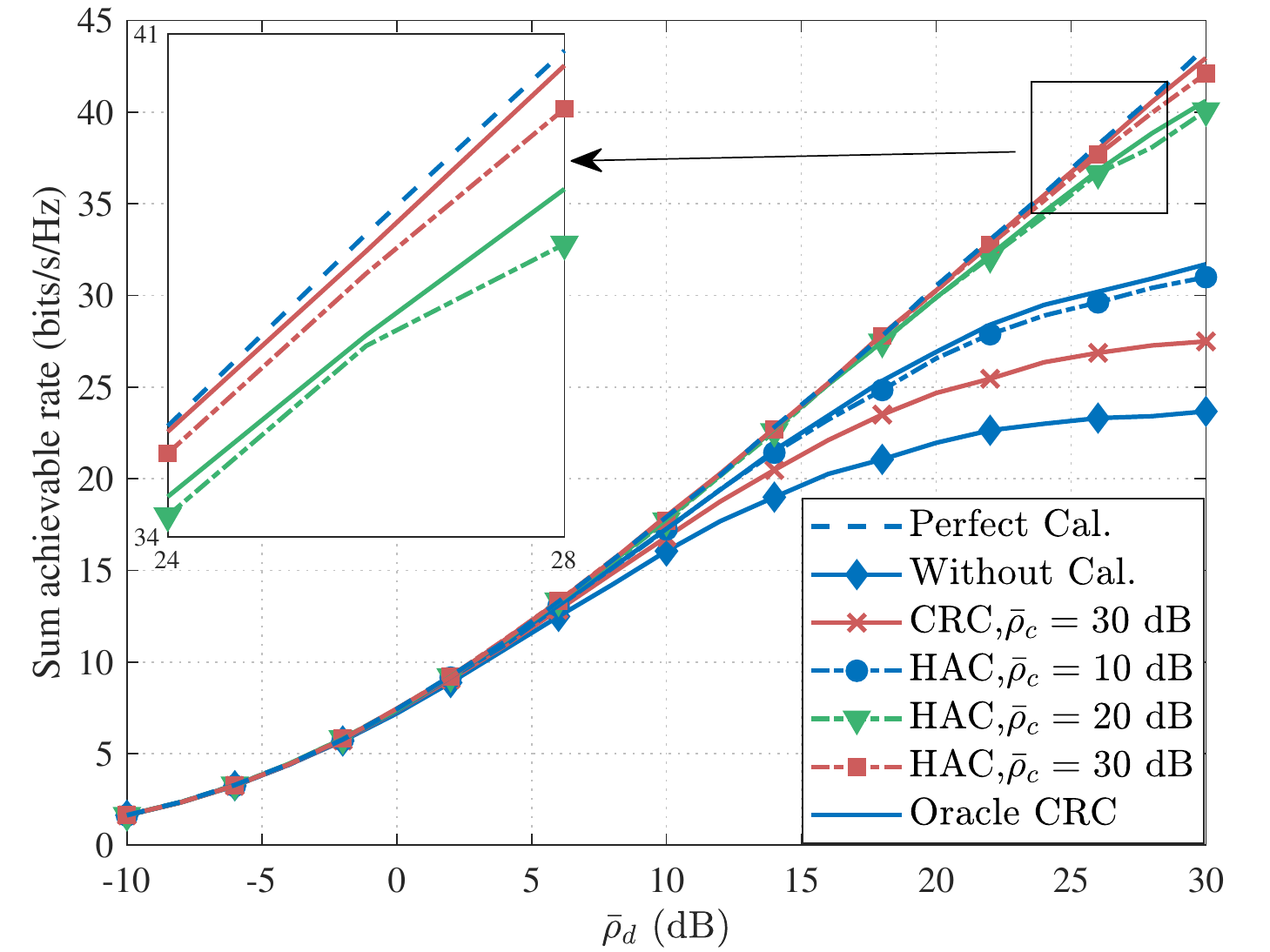}
		\vspace{-3em}
		\caption{The sum achievable rate of downlink transmission versus the SNR of transmit signals.}
		\label{fig:rateVSSNR}
	\end{minipage}
	\vspace{-2em}
\end{figure}
\else
\begin{figure}
	\centering
	\includegraphics[width=0.8\linewidth]{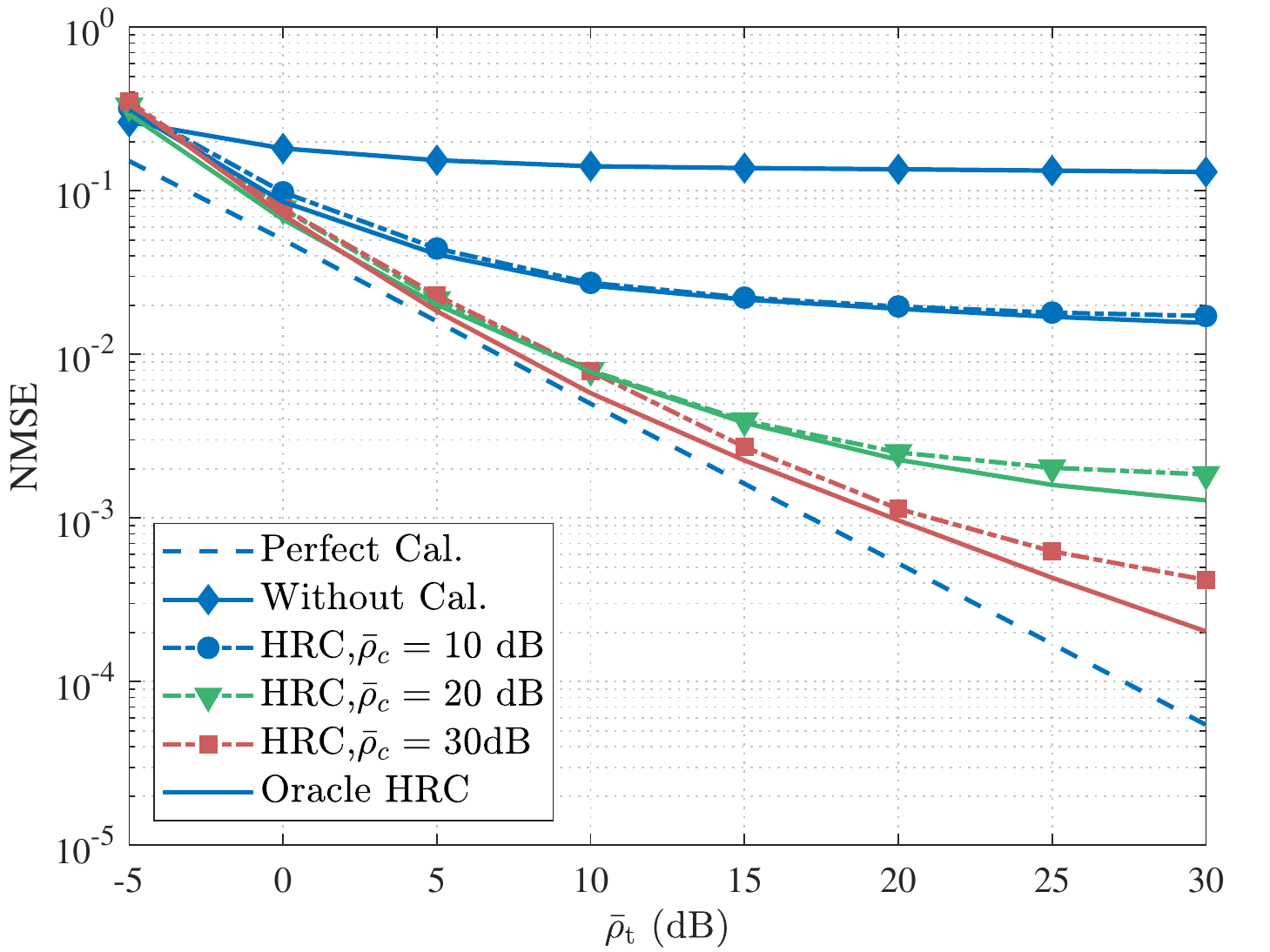}
	\caption{The NMSE of the uplink channel estimation versus the SNR of training signals.}
	\label{fig:Channel_NMSEVSSNR}
\end{figure}
\fi

Fig. \ref{fig:Channel_NMSEVSSNR} demonstrates the NMSE of estimated uplink channel versus the SNR $ \bar{\rho}_{\mathrm{t}} $ of the training signals for the channel estimation, where the SNR $ \bar{\rho}_{\mathrm{c}}$ of calibration signals is set to $ 10 $ dB, $ 20 $ dB, and $ 30 $ dB. It can be seen that the NMSE of the perfect calibration decreases with the increase of the training SNR, while the NMSE of the uncalibrated case almost remains constant. The NMSE of the channel estimation with the proposed HAC also decreases and gradually achieves floors with the training SNR increasing. The floor effect is caused by the estimation error of mismatch coefficients. When the SNR of calibration signals increases, the curves of the proposed HAC can approach the curve of the perfect calibration, which is because the estimation error of mismatch coefficients decreases. Further, since the NMSE of HAC is much less than the NMSE of the uncalibrated case, the proposed HAC can improve the system performance of the mmWave-HBF system, significantly.

\subsection{Achievable Rate of Downlink Transmission}
To further examine the efficacy of the reciprocity calibration in mmWave-HBF systems, we study the achievable rate of the downlink transmission. During the downlink transmission, the transmit analog beamforming $ \mathbf{F}_\mathrm{t} $ is set to $ \mathbf{F}_\mathrm{t}=\angle \bar{\mathbf{V}}_{\mathrm{a}} $, and the receive analog beamforming $ \mathbf{B}_{\mathrm{r}} $ is equal to $ \mathbf{B}_{\mathrm{r}}=\angle \bar{\mathbf{U}}_{\mathrm{a}}^* $, where $ \bar{\mathbf{V}}_{\mathrm{a}} $ is the first $ M_{\mathrm{t}} $ columns of $ \mathbf{V}_{\mathrm{a}} $, $ \bar{\mathbf{U}}_{\mathrm{a}} $ is the first $ M_{\mathrm{r}} $ columns of $ \mathbf{U}_{\mathrm{a}} $, $ \mathbf{V}_{\mathrm{a}} $ and $ \mathbf{U}_{\mathrm{a}} $ are obtained from the SVD of $ \hat{\mathbf{H}}_{\mathrm{DL}}= \hat{\mathbf{U}}_2\mathbf{H}\hat{\mathbf{T}}_2 $, i.e., $ \hat{\mathbf{H}}_{\mathrm{DL}}= \mathbf{U}_{\mathrm{d}}\boldsymbol{\Sigma}_{\mathrm{a}}\mathbf{V}_{\mathrm{a}}^H$. The digital precoding and the digital receiver are set to $ \mathbf{W}_{\mathrm{t}} = \bar{\mathbf{V}}_{\mathrm{d}} $ and $ \mathbf{D}_{\mathrm{r}}=\bar{\mathbf{U}}_{\mathrm{d}} $, where $  \bar{\mathbf{V}}_{\mathrm{d}} $ and $ \bar{\mathbf{U}}_{\mathrm{d}} $ consist of the first $ N_{\mathrm{s}} $ columns of $  \mathbf{V}_{\mathrm{d}} $ and $ \mathbf{U}_{\mathrm{d}} $, $  \mathbf{V}_{\mathrm{d}} $ and $ \mathbf{U}_{\mathrm{d}} $ can be obtained from the SVD of the equivalent downlink channel $ \mathbf{H}_{\mathrm{eq}}=\mathbf{U}_1\mathbf{B}_{\mathrm{r}}^T\bar{\mathbf{H}}_{\mathrm{DL}}\mathbf{F}_{\mathrm{t}}\mathbf{T}_1 $. Thus, based on the downlink transmission model \eqref{eq:downlinktrans}, the sum achievable rate can be denoted as
\begin{equation}
	R_{\mathrm{DL}} = \sum_{n_{\mathrm{s}}=1}^{N_{\mathrm{s}}}\mathbb{E}\left\lbrace \log\left(1+\frac{\rho_{\mathrm{d}}|\bar{h}_{n_{\mathrm{s}},n_{\mathrm{s}}}|^2}{\rho_{\mathrm{d}}\sum_{i\neq n_{\mathrm{s}}}^{N_{\mathrm{s}}}|\bar{h}_{\mathrm{eq},n_{\mathrm{s}},i}|^2+\bar{\sigma}_{\mathrm{n}}^2}\right) \right\rbrace,
\end{equation}
where $ N_{\mathrm{s}} $ denotes the data stream number, and $ \bar{\sigma}_{\mathrm{n}}^2= \|\mathbf{d}_{\mathrm{r},n_{\mathrm{s}}}^T\mathbf{U}_1\mathbf{B}_{\mathrm{r}}^T\|_{\mathrm{F}}^2\sigma_{\mathrm{n}}^2 $, $ \bar{h}_{\mathrm{eq},n_{\mathrm{s}},i} $ denotes the $ i $-th entry of $ \bar{\mathbf{h}}_{\mathrm{eq},n_{\mathrm{s}}} $, and $ \bar{\mathbf{h}}_{\mathrm{eq},n_{\mathrm{s}}} $ is the $ n_{\mathrm{s}} $-row of $ \mathbf{D}_{\mathrm{r}}^T\mathbf{U}_1\mathbf{B}_{\mathrm{r}}^T\mathbf{H}_{\mathrm{DL}}\mathbf{F}_{\mathrm{t}}\mathbf{T}_2\mathbf{W}_{\mathrm{t}}\mathbf{T}_1  $.

\if\thecol2
\begin{figure}
	\centering
	\includegraphics[width=0.8\linewidth]{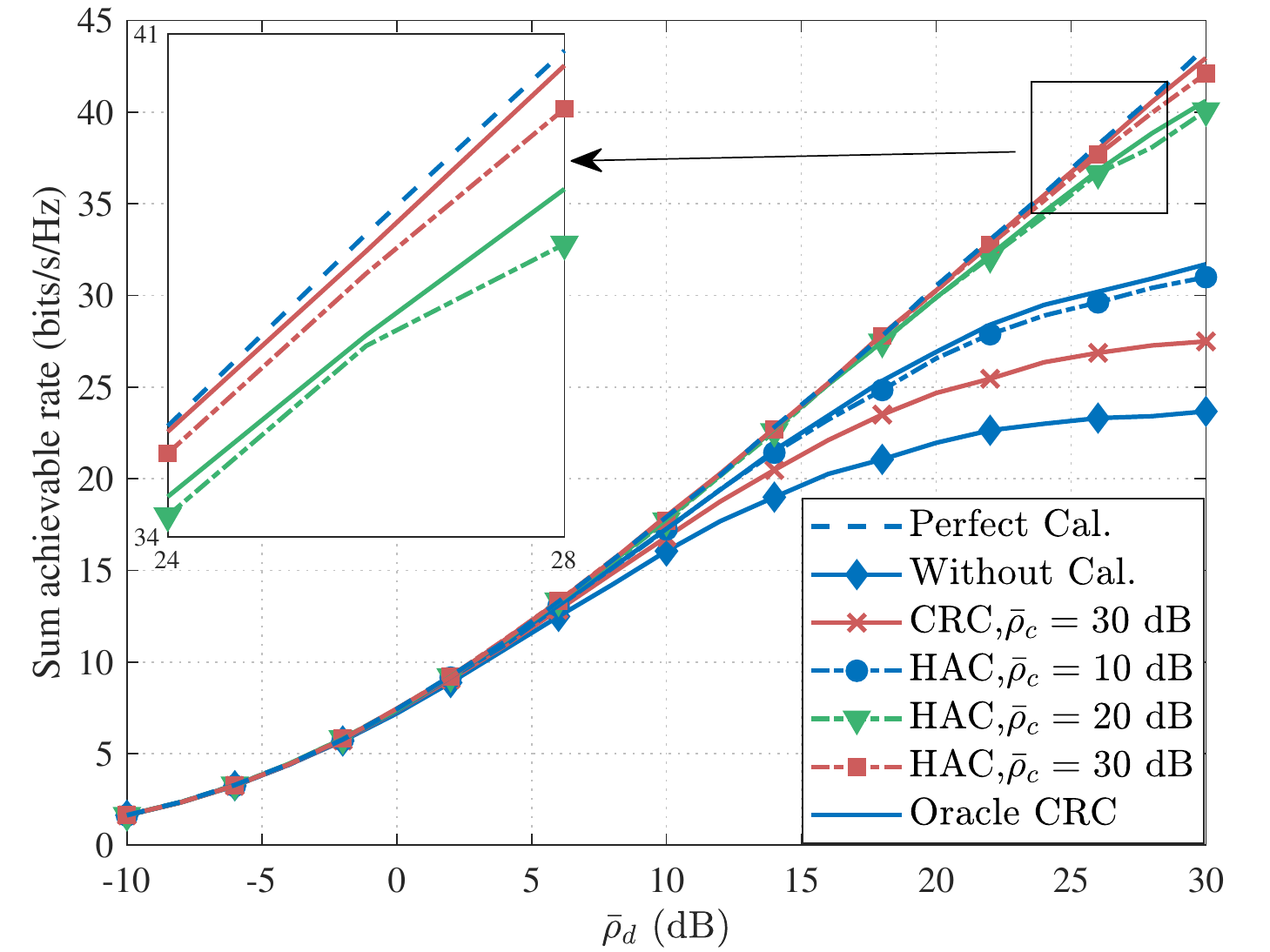}
	\caption{The sum achievable rate of downlink transmission versus the SNR of transmit signals.}
	\label{fig:rateVSSNR}
\end{figure}
\fi

The sum achievable rate of downlink transmission with the reciprocity calibration versus the downlink transmission SNR $ \bar{\rho}_{\mathrm{d}} $ is illustrated in Fig. \ref{fig:rateVSSNR}. From the figure, it can be observed that the curves of the systems with the perfect calibration, HAC, CRC, and without calibration 
achieve the same performance in the low SNR regime. This is because the impact of the noise is much larger than the reciprocity mismatch when the transmission SNR is small. In the high SNR regime, the achievable rate of the uncalibrated system converges to an upper bound, which is caused by multi-stream interference. Also, the achievable rate of the system applying HAC achieves the upper limit when the transmit SNR increases. For the perfectly calibrated system, the achievable rate linearly increases with the log function of the transmit SNR increasing. This is because the multi-stream interference can be completely mitigated by the HBF beamforming with perfect knowledge of mismatch coefficients. Further, we can also find that the achievable rate of HAC with $ \bar{\rho}_{c}=30 $ dB is almost twice larger than that of the uncalibrated case. This result implies that the reciprocity calibration can significantly improve the system performance as expected. Besides, the achievable rate of the system using HAC is larger than that using CRC, which indicates the proposed HAC outperforms the CRC.

Fig. \ref{fig:rateVSSNRCal} demonstrates the sum achievable rate versus the SNR of the calibration signals, where the number $ N_{\mathrm{s}} $ of data streams is set to $ 2 $ and $ 4 $, and the SNR $ \bar{\rho}_{\mathrm{d}} $ of the downlink transmission signals is set to $ 30 $ dB. It can be found that the sum achievable rate increases with the SNR of calibration signals increasing. This is because the estimation error decreases with the increase of the calibration SNR and the power of the interference decreases with the decrease of the estimation error of the mismatch coefficients. The gaps between the rates of the system using the proposed HAC and those of the perfectly calibrated systems are much smaller than the gaps between the rates of the system using CRC and those of the perfectly calibrated system. Further, when $ N_{\mathrm{s}}=2 $, the curve of HAC approaches the curve of the perfect calibration more quickly. This result indicates that  the system transmitting more data streams requires higher calibration SNR to have the same performance loss.

\if\thecol1
\begin{figure}
	\centering
	\vspace{-2em}
	\begin{minipage}{0.45\linewidth}
		\includegraphics[width=\linewidth]{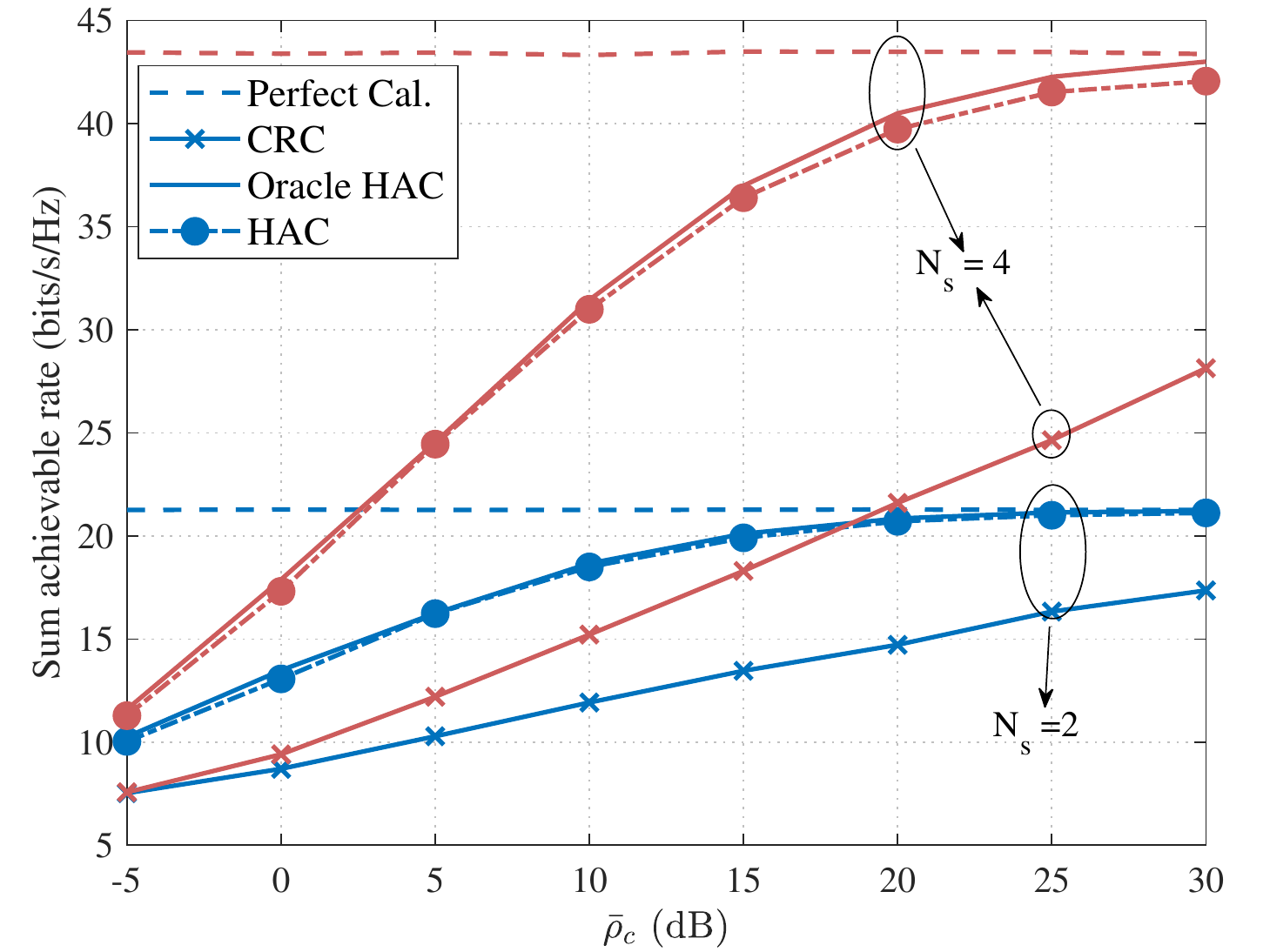}
		\vspace{-3em}
		\caption{The sum achievable rate of the downlink transmission versus the SNR of calibration signals.}
		\label{fig:rateVSSNRCal}
	\end{minipage}
	\hspace{2em}
	\begin{minipage}{0.45\linewidth}
		\includegraphics[width=\linewidth]{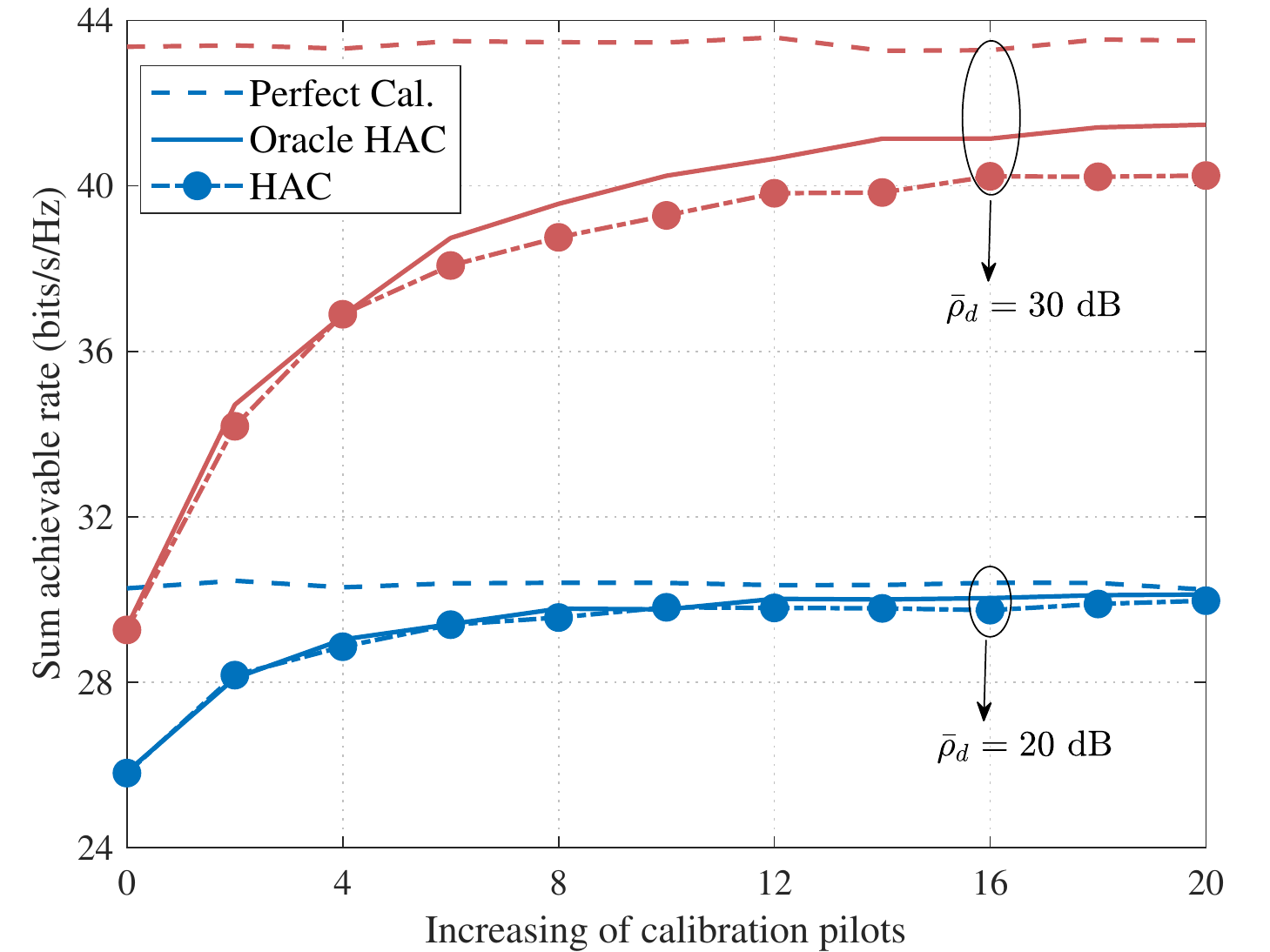}
		\vspace{-3em}
		\caption{The sum achievable rate of the downlink transmission versus the length of calibration pilots.}
		\label{fig:rateVSCalPilots}
	\end{minipage}
	\vspace{-2em}
\end{figure}
\else
\begin{figure}
	\centering
	\includegraphics[width=0.8\linewidth]{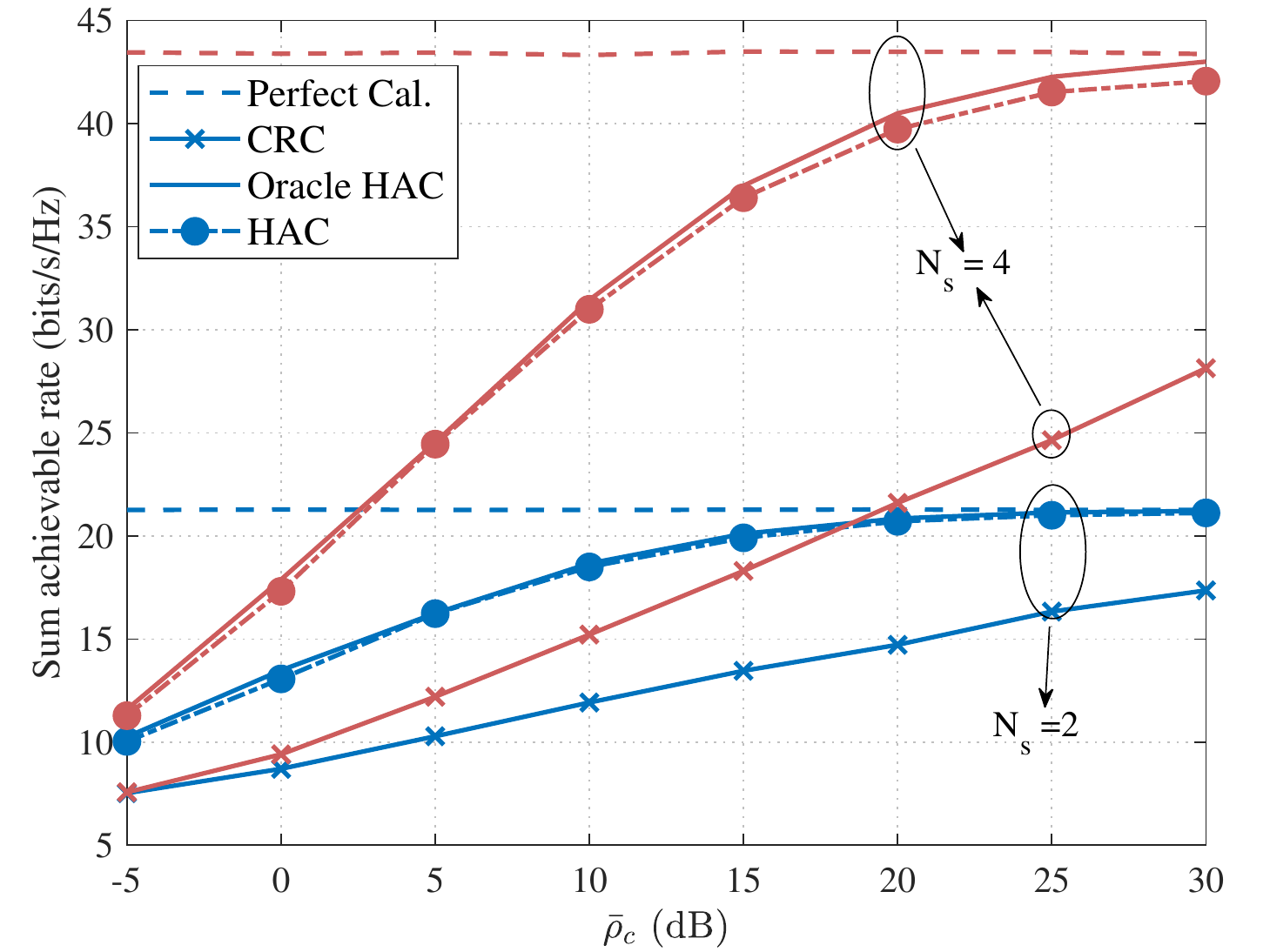}
	\caption{The sum achievable rate of the downlink transmission versus the SNR of calibration signals.}
	\label{fig:rateVSSNRCal}
\end{figure}
\fi

Finally, we examine the sum achievable rate versus the length of calibration pilots illustrated in Fig. \ref{fig:rateVSCalPilots}, where the SNR $ \bar{\rho}_{\mathrm{d}} $ of the downlink transmission signals is set to $ 20 $ dB and $ 30 $ dB, and the SNR of calibrations signals is set to $ 10 $ dB. From the figure, it can be observed that the achievable rate increases with the length of calibration pilots increasing, which is because the estimation error of the mismatch coefficients decreases with the increase of the calibration pilots. Further, the system with higher transmit SNR requires smaller calibration errors to guarantee the same performance loss.

\section{Conclusion}
In this paper, we have proposed a hierarchical-absolute reciprocity calibration for the mmWave-HBF system with the fully-connected phase shifter network. By proposing a specific beamforming design, the reciprocity calibration of the HBF system has been decoupled into the reciprocity calibration of digital RF chains and analog RF chains. Based on the decoupling, the entire reciprocity calibration problem of the HBF system has been equivalently decomposed into two subproblems corresponding to the reciprocity calibrations of digital and analog RF chains. Theoretical analysis has revealed that the overhead and computational complexity of the proposed HAC is much smaller than the conventional reciprocity calibration of HBF systems due to the decoupling. Further, based on the proposed calibration approach, we have derived the CRLB of the mismatch coefficients, which indicated that the estimation errors of the mismatch coefficients of digital and analog RF chains were independent, and the mismatch coefficients of receive digital chains could be estimated perfectly. Finally, simulation results have demonstrated that the proposed HAC significantly improved the system performance and outperformed the conventional calibration.

\if\thecol2
\begin{figure}
	\centering
	\includegraphics[width=0.8\linewidth]{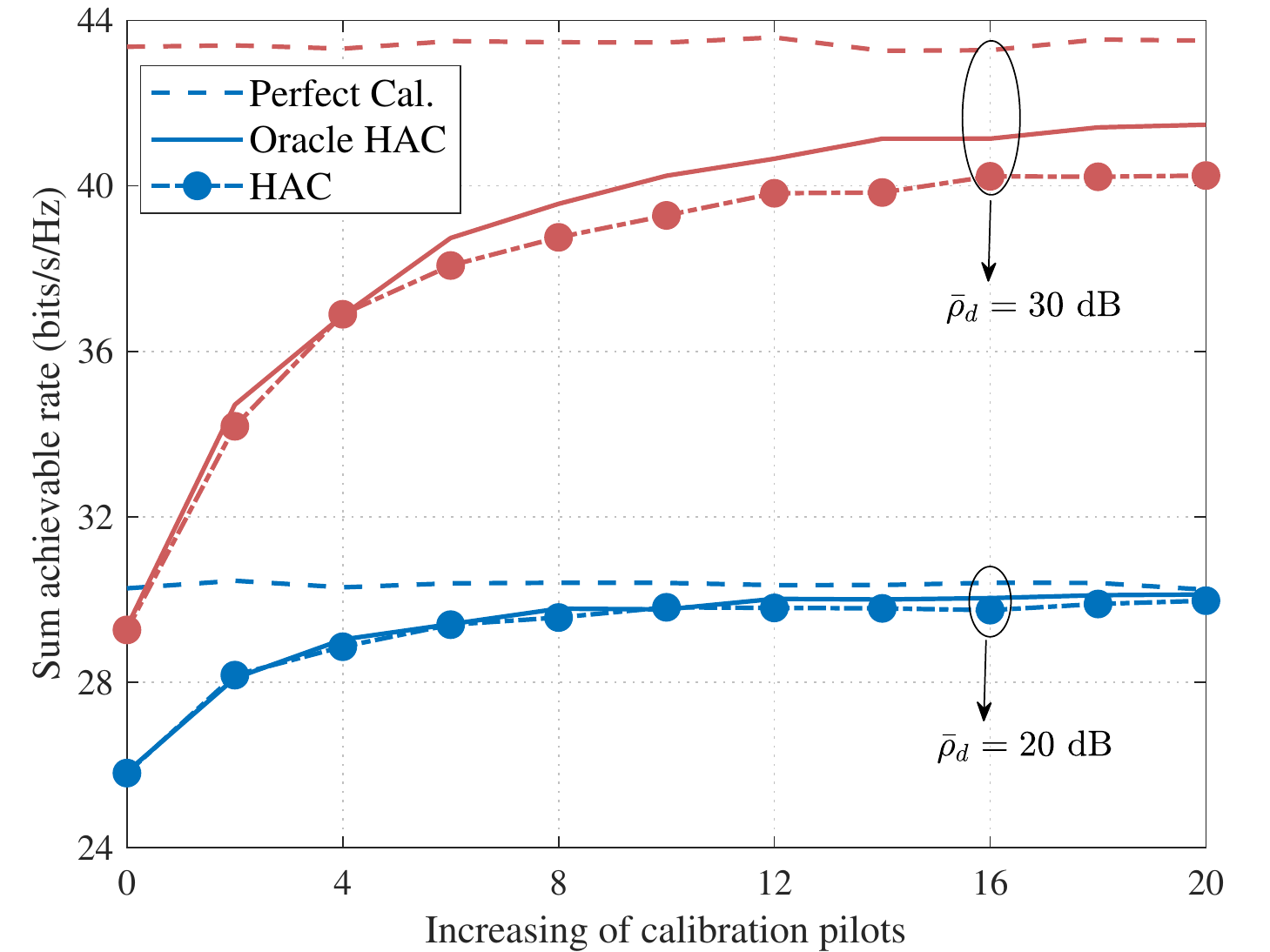}
	\caption{The sum achievable rate of the downlink transmission versus the length of calibration pilots.}
	\label{fig:rateVSCalPilots}
\end{figure}
\fi

\appendices
\section{Proof of Proposition \ref{prop:HACproblemdcp}} \label{proof:prodecople}
Based on the specific design of the digital/analog beamforming matrices, when $ l\le L_{\mathrm{dr}} $, the received signal in \eqref{eq:generalsgianl} can be rewritten as
\if\thecol1
\begin{equation}
	\mathbf{y}_{\mathrm{d},l}=\frac{1}{\sqrt{M_{\mathrm{t}}}}\mathbf{u}_1\mathbf{b}_{\mathrm{dr}}^T\mathbf{H}_{\mathrm{DL}}\mathbf{f}_{\mathrm{dr}}\mathbf{t}_1^T\mathbf{x}_{q}+\mathbf{u}_1\mathbf{b}_{\mathrm{dr}}^T\mathbf{n}_{\mathrm{d},l}\\
	=\beta_{\mathrm{d}}\mathbf{u}_1\mathbf{t}_1^T\mathbf{x}_{q}+\mathbf{u}_1\mathbf{b}_{\mathrm{dr}}^T\mathbf{n}_{\mathrm{d},l},
	\label{eq:recsignaldrpri}
\end{equation}
\else
\begin{equation}
	\begin{split}
		\mathbf{y}_{\mathrm{d},l}&=\frac{1}{\sqrt{M_{\mathrm{t}}}}\mathbf{u}_1\mathbf{b}_{\mathrm{dr}}^T\mathbf{H}_{\mathrm{DL}}\mathbf{f}_{\mathrm{dr}}\mathbf{t}_1^T\mathbf{x}_{q}+\mathbf{u}_1\mathbf{b}_{\mathrm{dr}}^T\mathbf{n}_{\mathrm{d},l}\\
		&=\beta_{\mathrm{d}}\mathbf{u}_1\mathbf{t}_1^T\mathbf{x}_{q}+\mathbf{u}_1\mathbf{b}_{\mathrm{dr}}^T\mathbf{n}_{\mathrm{d},l},
	\end{split}
	\label{eq:recsignaldrpri}
\end{equation}
\fi
where $ \beta_{\mathrm{d}}=\frac{1}{\sqrt{M_{\mathrm{t}}}}\mathbf{b}_{\mathrm{dr}}^T\mathbf{H}_{\mathrm{DL}}\mathbf{f}_{\mathrm{dr}} $, $\mathbf{u}_1$ consists of the diagonal entries of $\mathbf{U}_1$, and $\mathbf{t}_1$ is composed of the diagonal entries of $\mathbf{T}_1$. By stacking all $ L_{\mathrm{dr}} $-length signals $ \mathbf{y}_{\mathrm{d},l} $ into the matrix form, the received signals can be further denoted as
\begin{equation}
	\mathbf{Y}_{\mathrm{dr}}=\beta_{\mathrm{d}}(\mathbf{1}_{P_{\mathrm{dr}}}\otimes\mathbf{u}_1)\mathbf{t}_1^T\mathbf{X}_{\mathrm{dr}}+(\mathbf{I}_{P_{\mathrm{dr}}}\otimes\mathbf{u}_1\mathbf{b}_{\mathrm{dr}}^T)\mathbf{N}_{\mathrm{dr}},
	\label{eq:recsignaldr}
\end{equation}
where $ \mathbf{Y}_{\mathrm{dr}}=[\bar{\mathbf{Y}}_{\mathrm{dr},1}^T,\cdots,\bar{\mathbf{Y}}_{\mathrm{dr},P_{\mathrm{dr}}}^T]^T $, $ \bar{\mathbf{Y}}_{\mathrm{dr},p}=[\mathbf{y}_{\mathrm{d},(p-1)Q_{\mathrm{dr}}+1},\cdots,\mathbf{y}_{\mathrm{d},pQ_{\mathrm{dr}}}] $, $\mathbf{X}_{\mathrm{dr}}=[\mathbf{x}_{1},\cdots,\mathbf{x}_{Q_{\mathrm{dr}}}]$, $ \mathbf{N}_{\mathrm{dr}} = [\bar{\mathbf{N}}_{\mathrm{dr},1}^T,\cdots,\bar{\mathbf{N}}_{\mathrm{dr},P_{\mathrm{dr}}}^T]^T $, $ \bar{\mathbf{N}}_{\mathrm{dr},p}=[\mathbf{n}_{\mathrm{d},(p-1)Q_{\mathrm{dr}}+1},\cdots,\mathbf{n}_{\mathrm{d},pQ_{\mathrm{dr}}}] $.

Similarly, when $ l> L_{\mathrm{dr}} $, by substituting the designed beamforming matrices into \eqref{eq:generalsgianl}, the received signal can be rewritten as
\if\thecol1
\begin{equation}
	\mathbf{y}_{\mathrm{d},l}=\mathbf{D}_{\mathrm{da},p}^T\mathbf{U}_1\mathbf{B}_{\mathrm{da},p}^T\mathbf{H}_{\mathrm{DL}}\mathbf{F}_{\mathrm{da},q}\mathbf{T}_1\mathbf{W}_{\mathrm{da},q}\mathbf{x}_{q}+\tilde{\mathbf{n}}_{\mathrm{d},l}
	\overset{\equlabel}{=}\left[\begin{matrix}
		u_{1,1}t_{1,1}\mathbf{b}_{\mathrm{da},p,1}^T\mathbf{H}_{\mathrm{DL}}\mathbf{f}_{\mathrm{da},q,1}x_{q,1}+\tilde{n}_{\mathrm{d},l}\\
		\mathbf{0}_{M_{\mathrm{r}}-1}
	\end{matrix}
	\right],
\end{equation}
\else
\begin{equation}
	\begin{split}
		\mathbf{y}_{\mathrm{d},l}&=\mathbf{D}_{\mathrm{da},p}^T\mathbf{U}_1\mathbf{B}_{\mathrm{da},p}^T\mathbf{H}_{\mathrm{DL}}\mathbf{F}_{\mathrm{da},q}\mathbf{T}_1\mathbf{W}_{\mathrm{da},q}\mathbf{x}_{q}+\tilde{\mathbf{n}}_{\mathrm{d},l}\\
		&\overset{\equlabel}{=}\left[\begin{matrix}
			u_{1,1}t_{1,1}\mathbf{b}_{\mathrm{da},p,1}^T\mathbf{H}_{\mathrm{DL}}\mathbf{f}_{\mathrm{da},q,1}x_{q,1}+\tilde{n}_{\mathrm{d},l}\\
			\mathbf{0}_{M_{\mathrm{r}}-1}
		\end{matrix}
		\right],
	\end{split}
\end{equation}
\fi
which indicates that signal $ y_{\mathrm{d},l,1} $ received by the first receive digital RF chain is valid.
By stacking all $ L_{\mathrm{da}} $-length signals $ y_{\mathrm{d},l,1} $ into matrix form, the received signal can be further denoted as
\begin{equation}
	\mathbf{Y}_{\mathrm{da}}=u_{1,1}t_{1,1}\bar{\mathbf{B}}_{\mathrm{da}}^T\mathbf{H}_{\mathrm{DL}}\bar{\mathbf{F}}_{\mathrm{da}}\mathbf{X}_{\mathrm{da}}+\mathbf{N}_{\mathrm{da}}
	\label{eq:recsignalda}
\end{equation}
\if\thecol1
where $ \mathbf{Y}_{\mathrm{da}}=[\mathbf{y}_{\mathrm{da},1}^T,\cdots,\mathbf{y}_{\mathrm{da},P_{\mathrm{da}}}^T]^T $, $ \mathbf{y}_{\mathrm{da},p}=[y_{\mathrm{d},(p-1)Q_{\mathrm{da}}+1,1},\cdots,y_{\mathrm{d},pQ_{\mathrm{da}},1}] $, $ \bar{\mathbf{B}}_{\mathrm{da}}=[\mathbf{b}_{\mathrm{da},1,1},\cdots,\mathbf{b}_{\mathrm{da},P_{\mathrm{da}},1}] $, $ \bar{\mathbf{F}}_{\mathrm{da}}=[\mathbf{f}_{\mathrm{da},1,1},\cdots,\mathbf{f}_{\mathrm{da},Q_{\mathrm{da}},1}] $, $ \mathbf{X}_{\mathrm{da}}=\mathrm{diag}(x_{1,1},\cdots,x_{Q_{\mathrm{da}},1}) $, $ \mathbf{N}_{\mathrm{da}}=\mathrm{blkdiag}(\mathbf{b}_{\mathrm{da},1,1}^T,\cdots,\mathbf{b}_{\mathrm{da},P_{\mathrm{da}},1}^T)[\bar{\mathbf{N}}_{\mathrm{da},1}^T,\\\cdots, \bar{\mathbf{N}}_{\mathrm{da},P_{\mathrm{da}}}^T]^T$, and $ \bar{\mathbf{N}}_{\mathrm{da},p}=[\mathbf{n}_{\mathrm{d},(p-1)Q_{\mathrm{da}}+1},\cdots,\mathbf{n}_{\mathrm{d},pQ_{\mathrm{da}}}] $.
\else
where $ \mathbf{Y}_{\mathrm{da}}=[\mathbf{y}_{\mathrm{da},1}^T,\cdots,\mathbf{y}_{\mathrm{da},P_{\mathrm{da}}}^T]^T $, $ \mathbf{y}_{\mathrm{da},p}=[y_{\mathrm{d},(p-1)Q_{\mathrm{da}}+1,1},\cdots,y_{\mathrm{d},pQ_{\mathrm{da}},1}] $, $ \bar{\mathbf{B}}_{\mathrm{da}}=[\mathbf{b}_{\mathrm{da},1,1},\cdots,\mathbf{b}_{\mathrm{da},P_{\mathrm{da}},1}] $, $ \bar{\mathbf{F}}_{\mathrm{da}}=[\mathbf{f}_{\mathrm{da},1,1},\cdots,\mathbf{f}_{\mathrm{da},Q_{\mathrm{da}},1}] $, $ \mathbf{X}_{\mathrm{da}}=\mathrm{diag}(x_{1,1},\cdots,x_{Q_{\mathrm{da}},1}) $, $ \mathbf{N}_{\mathrm{da}}=\mathrm{blkdiag}(\mathbf{b}_{\mathrm{da},1,1}^T,\cdots,\mathbf{b}_{\mathrm{da},P_{\mathrm{da}},1}^T)[\bar{\mathbf{N}}_{\mathrm{da},1}^T,\cdots, \bar{\mathbf{N}}_{\mathrm{da},P_{\mathrm{da}}}^T]^T$, and $ \bar{\mathbf{N}}_{\mathrm{da},p}=[\mathbf{n}_{\mathrm{d},(p-1)Q_{\mathrm{da}}+1},\cdots,\mathbf{n}_{\mathrm{d},pQ_{\mathrm{da}}}] $.
\fi

By substituting \eqref{eq:recsignaldr} and \eqref{eq:recsignalda} into \eqref{eq:priprobelm}, the optimization problem can be rewritten as
\if\thecol1
\begin{equation}
	\min_{\mathbf{u}_{\mathrm{1}},\mathbf{t}_{\mathrm{1}},\mathbf{U}_2,\mathbf{T}_2,\mathbf{H}}\quad \underbrace{\left\|\mathbf{Y}_{\mathrm{dr}}-\beta_{\mathrm{d}}(\mathbf{1}_{P_{\mathrm{dr}}}\otimes\mathbf{u}_1)\mathbf{t}_1^T\mathbf{X}_{\mathrm{dr}}\right\|_{\mathrm{F}}^2}_{f(\beta_{\mathrm{d}},\mathbf{u}_1,\mathbf{t}_1)}+\underbrace{\left\|\mathbf{Y}_{\mathrm{da}}-u_{1,1}t_{1,1}\bar{\mathbf{B}}_{\mathrm{da}}^T\mathbf{H}_{\mathrm{DL}}\bar{\mathbf{F}}_{\mathrm{da}}\mathbf{X}_{\mathrm{da}}\right\|_{\mathrm{F}}^2}_{g(u_{1,1},t_{1,1},\mathbf{U}_2,\mathbf{T}_2,\mathbf{H})}.
	\label{eq:priproblemmatrix}
\end{equation}
\else
\begin{equation}
	\begin{split}
		\min_{\mathbf{u}_{\mathrm{1}},\mathbf{t}_{\mathrm{1}},\mathbf{U}_2,\mathbf{T}_2,\mathbf{H}}\quad &\underbrace{\left\|\mathbf{Y}_{\mathrm{dr}}-\beta_{\mathrm{d}}(\mathbf{1}_{P_{\mathrm{dr}}}\otimes\mathbf{u}_1)\mathbf{t}_1^T\mathbf{X}_{\mathrm{dr}}\right\|_{\mathrm{F}}^2}_{f(\beta_{\mathrm{d}},\mathbf{u}_1,\mathbf{t}_1)}+\\
		&\underbrace{\left\|\mathbf{Y}_{\mathrm{da}}-u_{1,1}t_{1,1}\bar{\mathbf{B}}_{\mathrm{da}}^T\mathbf{H}_{\mathrm{DL}}\bar{\mathbf{F}}_{\mathrm{da}}\mathbf{X}_{\mathrm{da}}\right\|_{\mathrm{F}}^2}_{g(u_{1,1},t_{1,1},\mathbf{U}_2,\mathbf{T}_2,\mathbf{H})}.
	\end{split}
	\label{eq:priproblemmatrix}
\end{equation}
\fi
From \eqref{eq:priproblemmatrix}, it can be observed that the functions $ f(\beta_{\mathrm{d}},\mathbf{u}_1,\mathbf{t}_1) $ and $ g(u_{1,1},t_{1,1},\mathbf{U}_2,\mathbf{T}_2,\mathbf{H}) $ are coupled to each other through the variables $ \beta_{\mathrm{d}} $, $ u_{1,1} $, and $ t_{1,1} $.  Fortunately, any vector $ \hat{\mathbf{t}}_1 $ parallel to the mismatch coefficient vector $ \mathbf{t}_1 $ can be exploited to calibrate the reciprocity mismatch of digital RF chains. This fact indicates that the unknown variable $ \beta_{\mathrm{d}} $ in the function $ f(\beta_{\mathrm{d}},\mathbf{u}_1,\mathbf{t}_1) $ can be regarded as a scale factor of $ \mathbf{t}_1 $ or $ \mathbf{u}_1 $. Similarly, since any vector $ \hat{\mathbf{u}}_2 $ parallel to the mismatch coefficient vector $ \mathbf{u}_2 $ can be used to calibrate the reciprocity mismatch of analog RF chains, the unknown variable $ u_{1,1}t_{1,1} $ in the function $ g(u_{1,1},t_{1,1},\mathbf{U}_2,\mathbf{T}_2,\mathbf{H}) $ can be also treated as a scale factor of $ \mathbf{U}_2 $ or $ \mathbf{T}_2 $.  Therefore, \eqref{eq:priproblemmatrix} can be equivalently written as
\if\thecol1
\begin{equation}
	\min_{\mathbf{u}_{\mathrm{1}},\mathbf{t}_{\mathrm{1}},\mathbf{U}_2,\mathbf{T}_2,\mathbf{H}}\quad \underbrace{\left\|\mathbf{Y}_{\mathrm{dr}}-(\mathbf{1}_{P_{\mathrm{dr}}}\otimes\mathbf{u}_1)\mathbf{t}_1^T\mathbf{X}_{\mathrm{dr}}\right\|_{\mathrm{F}}^2}_{\bar{f}(\mathbf{u}_1,\mathbf{t}_1)}+\underbrace{\left\|\mathbf{Y}_{\mathrm{da}}-\bar{\mathbf{B}}_{\mathrm{da}}^T\mathbf{U}_2\mathbf{H}^T\mathbf{T}_2\bar{\mathbf{F}}_{\mathrm{da}}\mathbf{X}_{\mathrm{da}}\right\|_{\mathrm{F}}^2}_{\bar{g}(\mathbf{U}_2,\mathbf{T}_2,\mathbf{H})}.
	\label{eq:priproelmmatrixdef}
\end{equation}
\else
\begin{equation}
	\begin{split}
		\min_{\mathbf{u}_{\mathrm{1}},\mathbf{t}_{\mathrm{1}},\mathbf{U}_2,\mathbf{T}_2,\mathbf{H}}\quad &\underbrace{\left\|\mathbf{Y}_{\mathrm{dr}}-(\mathbf{1}_{P_{\mathrm{dr}}}\otimes\mathbf{u}_1)\mathbf{t}_1^T\mathbf{X}_{\mathrm{dr}}\right\|_{\mathrm{F}}^2}_{\bar{f}(\mathbf{u}_1,\mathbf{t}_1)}+\\
		&\underbrace{\left\|\mathbf{Y}_{\mathrm{da}}-\bar{\mathbf{B}}_{\mathrm{da}}^T\mathbf{U}_2\mathbf{H}^T\mathbf{T}_2\bar{\mathbf{F}}_{\mathrm{da}}\mathbf{X}_{\mathrm{da}}\right\|_{\mathrm{F}}^2}_{\bar{g}(\mathbf{U}_2,\mathbf{T}_2,\mathbf{H})}.
	\end{split}
	\label{eq:priproelmmatrixdef}
\end{equation}
\fi
It is worth noting that the solutions to the problem \eqref{eq:priproelmmatrixdef} are also the solutions to the problem \eqref{eq:priproblemmatrix}. Further, since $ \bar{f}(\mathbf{u}_1,\mathbf{t}_1) $ and $ \bar{g}(\mathbf{U}_2,\mathbf{T}_2,\mathbf{H}) $ are independent of each other, the minimum sum of these two functions is equal to the sum of the minimum of each function, i.e., $ \min\{ \bar{f}(\mathbf{u}_1,\mathbf{t}_1)+\bar{g}(\mathbf{U}_2,\mathbf{T}_2,\mathbf{H})\}=\min\{\bar{f}(\mathbf{u}_1,\mathbf{t}_1)\}+\min\{\bar{g}(\mathbf{U}_2,\mathbf{T}_2,\mathbf{H})\} $. Consequently, the problem \eqref{eq:priproelmmatrixdef} can be decoupled into two independent subproblems denoted as
\begin{equation}
	\begin{split}
		&\min_{\ \ \mathbf{u}_{\mathrm{1}},\mathbf{t}_{\mathrm{1}}\ }\quad\bar{f}(\mathbf{u}_1,\mathbf{t}_1),\\
		&\min_{\mathbf{T}_2,\mathbf{U}_2,\mathbf{H}}\quad\bar{g}(\mathbf{U}_2,\mathbf{T}_2,\mathbf{H}).
	\end{split}
\end{equation}
Thus, Proposition \ref{prop:HACproblemdcp} holds.

\section{Proof of Proposition \ref{prop:solutot1u1}}\label{proof:solutot1u1}
By substituting $\mathbf{x}_{\mathrm{dt},p}$ in \eqref{eq:soltoxtn} and $ u_{1,1}=1 $ into \eqref{eq:hacrfcal}, the objective of the problem $\mathcal{P}_{1}$ can be further denoted as
\if\thecol1
\begin{equation}
	\begin{split}
		\bar{f}(\mathbf{u}_1,\mathbf{t}_1)
		=&\sum_{p=1}^{P_{\mathrm{dr}}}\Bigl[\left\|\mathbf{y}_{\mathrm{dr},(p-1)M_{\mathrm{r}}+1}^T- c_{\mathrm{dr}}\mathbf{X}_{\mathrm{dr}}^T\mathbf{t}_1\right\|_{\mathrm{F}}^2+\left\|\tilde{\mathbf{Y}}_{\mathrm{dr},p}-\frac{1}{c_{\mathrm{dr}}}\tilde{\mathbf{u}}_1\mathbf{y}_{\mathrm{dr},(p-1)M_{\mathrm{r}}+1}\right\|_{\mathrm{F}}^2\Bigr]\\
		&=\check{f}(\tilde{\mathbf{u}}_1,\mathbf{t}_1)
	\end{split}
\end{equation}
\else
\begin{equation}
	\begin{split}
		\bar{f}(\mathbf{u}_1,\mathbf{t}_1)
		=&\sum_{p=1}^{P_{\mathrm{dr}}}\Bigl[\left\|\mathbf{y}_{\mathrm{dr},(p-1)M_{\mathrm{r}}+1}^T- c_{\mathrm{dr}}\mathbf{X}_{\mathrm{dr}}^T\mathbf{t}_1\right\|_{\mathrm{F}}^2\\
		&+\left\|\tilde{\mathbf{Y}}_{\mathrm{dr},p}-\frac{1}{c_{\mathrm{dr}}}\tilde{\mathbf{u}}_1\mathbf{y}_{\mathrm{dr},(p-1)M_{\mathrm{r}}+1}\right\|_{\mathrm{F}}^2\Bigr]\\
		=&\check{f}(\tilde{\mathbf{u}}_1,\mathbf{t}_1)
	\end{split}
\end{equation}
\fi
where $ \tilde{\mathbf{Y}}_{\mathrm{dr},p} $ consists of the second to the last row of $ \bar{\mathbf{Y}}_{\mathrm{dr},p} $, and $ \tilde{\mathbf{u}}_1=[u_{1,2},\cdots,u_{1,M_{\mathrm{r}}}]^T $. By defining $ \mathbf{y}_{\mathrm{dt}}=[\mathbf{y}_{\mathrm{dr},1},\cdots,\mathbf{y}_{\mathrm{dr},(P_{\mathrm{dr}}-1)M_{\mathrm{r}}+1}]^T $, $ \tilde{\mathbf{y}}_{\mathrm{dr}}=[\mathrm{vec}(\tilde{\mathbf{Y}}_{\mathrm{dr},1})^T,\cdots,\mathrm{vec}(\tilde{\mathbf{Y}}_{\mathrm{dr},P_{\mathrm{dr}}})^T]^T $, and $ \check{\mathbf{Y}}_{\mathrm{dr}}=[(\mathbf{y}_{\mathrm{dr},1}\otimes \mathbf{I}_{M_{\mathrm{r}}-1}),\cdots,(\mathbf{y}_{\mathrm{dr},(P_{\mathrm{dr}}-1)M_{\mathrm{r}}+1}\otimes \mathbf{I}_{M_{\mathrm{r}}-1})]^T $, $ \check{f}(\tilde{\mathbf{u}}_1,\mathbf{t}_1) $ can be rewritten into the matrix form denoted as
\if\thecol1
\begin{equation}
	\check{f}(\tilde{\mathbf{u}}_1,\mathbf{t}_1)=\left\|\mathbf{y}_{\mathrm{dt}}-c_{\mathrm{dr}}(\mathbf{1}_{P_\mathrm{\mathrm{dr}}}\otimes\mathbf{X}_{\mathrm{dr}}^T)\mathbf{t}_1\right\|_{\mathrm{F}}^2
	+\left\|\tilde{\mathbf{y}}_{\mathrm{dr}}-\frac{1}{c_{\mathrm{dr}}}\check{\mathbf{Y}}_{\mathrm{dr}}\tilde{\mathbf{u}}_1\right\|_{\mathrm{F}}^2.
	\label{eq:objecttrans2}
\end{equation} 
\else
\begin{equation}
	\begin{split}
		\check{f}(\tilde{\mathbf{u}}_1,\mathbf{t}_1)=&\left\|\mathbf{y}_{\mathrm{dt}}-c_{\mathrm{dr}}(\mathbf{1}_{P_\mathrm{\mathrm{dr}}}\otimes\mathbf{X}_{\mathrm{dr}}^T)\mathbf{t}_1\right\|_{\mathrm{F}}^2\\
		&+\left\|\tilde{\mathbf{y}}_{\mathrm{dr}}-\frac{1}{c_{\mathrm{dr}}}\check{\mathbf{Y}}_{\mathrm{dr}}\tilde{\mathbf{u}}_1\right\|_{\mathrm{F}}^2.
	\end{split}
	\label{eq:objecttrans2}
\end{equation} 
\fi
Then, $ \mathcal{P}_1 $ can be equivalently transformed into
\begin{equation}
	\mathcal{P}_{1,2}:\min_{\tilde{\mathbf{u}}_1,\mathbf{t}_1}\quad\check{f}(\tilde{\mathbf{u}}_1,\mathbf{t}_1).
\end{equation}

Since the objective $ \check{f}(\tilde{\mathbf{u}}_1,\mathbf{t}_1) $  of $ \mathcal{P}_{1.2} $ is the sum of two convex functions, $ \mathcal{P}_{1.2} $ is a convex problem without constraints. Thus, to solve $ \tilde{\mathbf{u}}_1 $, we take the partial derivative of $ \check{f}(\tilde{\mathbf{u}}_1,\mathbf{t}_1) $ with respect to $ \tilde{\mathbf{u}}_1 $  as follows
\if\thecol1
\begin{equation}
	\frac{\partial \check{f}(\tilde{\mathbf{u}}_1,\mathbf{t}_1)}{\partial \tilde{\mathbf{u}}_1}=\frac{\partial \left\|\tilde{\mathbf{y}}_{\mathrm{dr}}-\frac{1}{c_{\mathrm{dr}}}\check{\mathbf{Y}}_{\mathrm{dr}}\tilde{\mathbf{u}}_1\right\|_{\mathrm{F}}^2}{\partial \tilde{\mathbf{u}}_1}=-\check{\mathbf{Y}}_{\mathrm{dr}}^T(\tilde{\mathbf{y}}_{\mathrm{dr}}-\frac{1}{c_{\mathrm{dr}}}\check{\mathbf{Y}}_{\mathrm{dr}}\tilde{\mathbf{u}}_1)^*.
\end{equation}
\else
\begin{equation}
	\begin{split}
		\frac{\partial \check{f}(\tilde{\mathbf{u}}_1,\mathbf{t}_1)}{\partial \tilde{\mathbf{u}}_1}
		&=\frac{\partial \left\|\tilde{\mathbf{y}}_{\mathrm{dr}}-\frac{1}{c_{\mathrm{dr}}}\check{\mathbf{Y}}_{\mathrm{dr}}\tilde{\mathbf{u}}_1\right\|_{\mathrm{F}}^2}{\partial \tilde{\mathbf{u}}_1}\\
		&=-\check{\mathbf{Y}}_{\mathrm{dr}}^T(\tilde{\mathbf{y}}_{\mathrm{dr}}-\frac{1}{c_{\mathrm{dr}}}\check{\mathbf{Y}}_{\mathrm{dr}}\tilde{\mathbf{u}}_1)^*.
	\end{split}
\end{equation}
\fi
By setting the derivative equal to zero, the solution to $\mathbf{u}_1$ can be given by
\begin{equation}
	\tilde{\mathbf{u}}_1\overset{(b)}{=}c_{\mathrm{dr}}(\check{\mathbf{Y}}_{\mathrm{dr}}^H\check{\mathbf{Y}}_{\mathrm{dr}})^{-1}\check{\mathbf{Y}}_{\mathrm{dr}}^H\tilde{\mathbf{y}}_{\mathrm{dr}},
\end{equation}
where the condition $ (b) $ holds when $ \check{\mathbf{Y}}_{\mathrm{dr}} $ is a full column rank matrix.

Similarly, by taking the partial derivative of $ \check{f}(\tilde{\mathbf{u}}_1,\mathbf{t}_1) $ with respect to $ \mathbf{t}_1 $ and setting the partial derivative to zero, we can obtain the solution to $ \mathbf{t}_1 $ denoted as follows
\if\thecol1
\begin{equation}
	\begin{split}
		\hat{\mathbf{t}}_1&\overset{(c)}{=}\frac{1}{c_{\mathrm{dr}}}\bigl[(\mathbf{1}_{P_{\mathrm{dr}}}\otimes\mathbf{X}_{\mathrm{dr}}^T)^H(\mathbf{1}_{P_{\mathrm{dr}}}\otimes\mathbf{X}_{\mathrm{dr}}^T)\bigr]^{-1} (\mathbf{1}_{P_{\mathrm{dr}}}\otimes\mathbf{X}_{\mathrm{dr}}^T)^H\mathbf{y}_{\mathrm{dt}}\\
		&=\frac{1}{c_{\mathrm{dr}}P_{\mathrm{dr}}}\Bigl[\mathbf{1}_{P_{\mathrm{dr}}}^T\otimes\bigl(\mathbf{X}_{\mathrm{dr}}^*\mathbf{X}_{\mathrm{dr}}^T\bigr)^{-1}\mathbf{X}_{\mathrm{dr}}^*\Bigr]\mathbf{y}_{\mathrm{dt}},
	\end{split}
\end{equation}
\else
\begin{equation}
	\begin{split}
		\hat{\mathbf{t}}_1&\overset{(c)}{=}\frac{1}{c_{\mathrm{dr}}}\bigl[(\mathbf{1}_{P_{\mathrm{dr}}}\otimes\mathbf{X}_{\mathrm{dr}}^T)^H(\mathbf{1}_{P_{\mathrm{dr}}}\otimes\mathbf{X}_{\mathrm{dr}}^T)\bigr]^{-1}\cdot\\
		&\quad\quad (\mathbf{1}_{P_{\mathrm{dr}}}\otimes\mathbf{X}_{\mathrm{dr}}^T)^H\mathbf{y}_{\mathrm{dt}}\\
		&=\frac{1}{c_{\mathrm{dr}}P_{\mathrm{dr}}}\Bigl[\mathbf{1}_{P_{\mathrm{dr}}}^T\otimes\bigl(\mathbf{X}_{\mathrm{dr}}^*\mathbf{X}_{\mathrm{dr}}^T\bigr)^{-1}\mathbf{X}_{\mathrm{dr}}^*\Bigr]\mathbf{y}_{\mathrm{dt}},
	\end{split}
\end{equation}
\fi
where the condition $ (c) $ holds when $ (\mathbf{1}_{P_{\mathrm{dr}}}\otimes\mathbf{X}_{\mathrm{dr}}^T) $ is full column rank. Thus, Proposition \ref{prop:solutot1u1} holds.

\AtNextBibliography{\small}
\printbibliography
\end{refsection}


 \newpage
 \begin{refsection}
 \DeclareFieldFormat{labelnumber}{R#1} 
 \setcounter{page}{1}

 \begin{center}
 	{\huge Supplementary Material for ``Hierarchical-Absolute Reciprocity Calibration for Millimeter-wave Hybrid Beamforming Systems"}
	
 	 Li Chen,~Rongjiang Nie,~Yunfei Chen,~and~Weidong Wang
 \end{center}

 Some proofs are omitted in the main paper for readability, and we provide the missing content in this  supplementary material for completeness.

 \appendices
 \setcounter{section}{2}
 \section{Proof of Lemma \ref{lem:soltot2u2h}}\label{proof:soltot2u2h}
 To estimate the diagonal matrix $ \mathbf{H}_{\alpha} $, the objective $ \bar{g}(\mathbf{T}_2,\mathbf{U}_2,\mathbf{H}_{\alpha},\boldsymbol{\Theta},\boldsymbol{\Phi}) $  of $ \mathcal{P}_{2.1} $ can be further denoted as
 \if\thecol1
 \begin{equation}
 	\bar{g}(\mathbf{T}_2,\mathbf{U}_2,\mathbf{H}_{\alpha},\boldsymbol{\Theta},\boldsymbol{\Phi})
 	=\left\|\mathrm{vec}\left\lbrace \mathbf{Y}_{\mathrm{da}} \right\rbrace-(\tilde{\mathbf{X}}_{\mathrm{da}}^T\mathbf{T}_2\mathbf{A}_{\mathrm{t}}\odot\bar{\mathbf{B}}_{\mathrm{r}}\mathbf{U}_2\mathbf{A}_{\mathrm{r}})\mathbf{h}_{\alpha}\right\|_{\mathrm{F}}^2.
 	\label{eq:gammah}
 \end{equation}
 \else
 \begin{equation}
 	\begin{split}
 		&\bar{g}(\mathbf{T}_2,\mathbf{U}_2,\mathbf{H}_{\alpha},\boldsymbol{\Theta},\boldsymbol{\Phi})\\
 		&\quad\quad=\left\|\mathrm{vec}\left\lbrace \mathbf{Y}_{\mathrm{da}} \right\rbrace-(\tilde{\mathbf{X}}_{\mathrm{da}}^T\mathbf{T}_2\mathbf{A}_{\mathrm{t}}\odot\bar{\mathbf{B}}_{\mathrm{r}}\mathbf{U}_2\mathbf{A}_{\mathrm{r}})\mathbf{h}_{\alpha}\right\|_{\mathrm{F}}^2.
 	\end{split}
 	\label{eq:gammah}
 \end{equation}
 \fi
 When $\mathbf{U}_2,\mathbf{T}_2,\boldsymbol{\Theta},\boldsymbol{\Phi}$ are known during the $ l_{\mathrm{ao}} $-iteration, $ \bar{g}(\mathbf{T}_2,\mathbf{U}_2,\mathbf{H}_{\alpha},\boldsymbol{\Theta},\boldsymbol{\Phi}) $ is a convex function corresponding to $ \mathbf{h}_{\alpha} $. Thus, $ \mathbf{h}_{\alpha} $ can be estimated by taking the derivative of $ \bar{g}(\mathbf{T}_2,\mathbf{U}_2,\mathbf{H}_{\alpha},\boldsymbol{\Theta},\boldsymbol{\Phi}) $, and the solution to $ \mathbf{h}_{\alpha} $ is given by
 \begin{equation}
 	\mathbf{h}_{\alpha}=(\boldsymbol{\Gamma}_{\mathrm{h}}^H\boldsymbol{\Gamma}_{\mathrm{h}})^{-1}\boldsymbol{\Gamma}_{\mathrm{h}}^H\mathrm{vec}\left\lbrace \mathbf{Y}_{\mathrm{da}} \right\rbrace,
 \end{equation}
 where $\boldsymbol{\Gamma}_{\mathrm{h}}=(\tilde{\mathbf{X}}_{\mathrm{da}}^T\mathbf{T}_2\mathbf{A}_{\mathrm{t}}\odot\bar{\mathbf{B}}_{\mathrm{r}}\mathbf{U}_2\mathbf{A}_{\mathrm{r}})$.

 Similarly, the diagonal entries of $ \mathbf{U}_2 $ and $ \mathbf{T}_2 $ can be estimated as
 \if\thecol1
 \begin{align}
 	&\mathbf{u}_2=\mathrm{arg}\min_{\mathbf{u}_2}\|\mathrm{vec}\left\lbrace \mathbf{Y}_{\mathrm{da}} \right\rbrace-\boldsymbol{\Gamma}_{\mathrm{u}}\mathbf{u}_2\|_{\mathrm{F}}^2=(\boldsymbol{\Gamma}_{\mathrm{u}}^H\boldsymbol{\Gamma}_{\mathrm{u}})^{-1}\boldsymbol{\Gamma}_{\mathrm{u}}^H\mathrm{vec}\left\lbrace \mathbf{Y}_{\mathrm{da}} \right\rbrace,\label{eq:estu2pro}\\
 	&\mathbf{t}_2=\mathrm{arg}\min_{\mathbf{t}_{2}}\left\| \mathrm{vec}\{\mathbf{Y}_{\mathrm{da}}\}-\boldsymbol{\Gamma}_{\mathrm{t}}\mathbf{t}_2\right\|_{\mathrm{F}}^2
 	=(\boldsymbol{\Gamma}_{\mathrm{t}}^H\boldsymbol{\Gamma}_{\mathrm{t}})^{-1}\boldsymbol{\Gamma}_{\mathrm{t}}^H\mathrm{vec}\left\lbrace \mathbf{Y}_{\mathrm{da}} \right\rbrace,\label{eq:gammat}
 \end{align}
 \else
 \begin{align}
 	\mathbf{u}_2&=\mathrm{arg}\min_{\mathbf{u}_2}\|\mathrm{vec}\left\lbrace \mathbf{Y}_{\mathrm{da}} \right\rbrace-\boldsymbol{\Gamma}_{\mathrm{u}}\mathbf{u}_2\|_{\mathrm{F}}^2\nonumber\\
 	&=(\boldsymbol{\Gamma}_{\mathrm{u}}^H\boldsymbol{\Gamma}_{\mathrm{u}})^{-1}\boldsymbol{\Gamma}_{\mathrm{u}}^H\mathrm{vec}\left\lbrace \mathbf{Y}_{\mathrm{da}} \right\rbrace,\label{eq:estu2pro}\\
 	\mathbf{t}_2&=\mathrm{arg}\min_{\mathbf{t}_{2}}\left\| \mathrm{vec}\{\mathbf{Y}_{\mathrm{da}}\}-\boldsymbol{\Gamma}_{\mathrm{t}}\mathbf{t}_2\right\|_{\mathrm{F}}^2\nonumber\\
 	&=(\boldsymbol{\Gamma}_{\mathrm{t}}^H\boldsymbol{\Gamma}_{\mathrm{t}})^{-1}\boldsymbol{\Gamma}_{\mathrm{t}}^H\mathrm{vec}\left\lbrace \mathbf{Y}_{\mathrm{da}} \right\rbrace,\label{eq:gammat}
 \end{align}
 \fi
 where $\boldsymbol{\Gamma}_{\mathrm{u}}=(\tilde{\mathbf{X}}_{\mathrm{da}}^T\mathbf{T}_2\mathbf{A}_{\mathrm{t}}\mathbf{H}_{\alpha}(\mathbf{A}_{\mathrm{r}})^T\odot\bar{\mathbf{B}}_{\mathrm{r}})$, and $\boldsymbol{\Gamma}_{\mathrm{t}}=(\tilde{\mathbf{X}}_{\mathrm{da}}^T\odot\bar{\mathbf{B}}_{\mathrm{r}}\mathbf{U}_2\mathbf{A}_{\mathrm{r}}\mathbf{H}_{\alpha}(\mathbf{A}_{\mathrm{t}})^T)$. Thus, Lemma \ref{lem:soltot2u2h} holds.

 \section{Proof of Proposition \ref{prop:pilotlength}}\label{proof:pilotlength}
 We first derive the pilot requirement for calibrating the digital RF chains. According to \eqref{eq:solu1}, to guarantee a unique solution to $\mathbf{u}_1$, the matrix $\check{\mathbf{Y}}_{\mathrm{dr}}$ should have column rank, which is always satisfied when $ P_{\mathrm{dr}}\geq 1 $ and $ Q_{\mathrm{dr}}\geq 1 $. Similarly, based on \eqref{eq:solt1}, for computing $ \mathbf{t}_1 $, the matrix $ \mathbf{1}_{P_\mathrm{dr}}\otimes\mathbf{X}_{\mathrm{dr}}^T $ should have column rank, i.e., $ \mathrm{rank}\{\mathbf{1}_{P_\mathrm{dr}}\otimes\mathbf{X}_{\mathrm{dr}}^T\}=\mathrm{rank}\{\mathbf{X}_{\mathrm{dr}}^T\}=\min\{M_{\mathrm{t}},Q_{\mathrm{dr}}\}\ge M_{\mathrm{t}} $. Thus, the pilots of digital RF chain calibration should satisfy
 \begin{equation}
 	Q_{\mathrm{dr}}\geq M_{\mathrm{t}},\text{ and }
 	P_{\mathrm{dr}}\geq 1.
 \end{equation}

 Then, we derive the pilot requirement for calibrating the analog RF chains. To guarantee the unique solution to $ \mathbf{h}_{\alpha} $, $\boldsymbol{\Gamma}_{\mathrm{h}}$ must be full column rank, i.e.,
 \begin{equation}
 	\mathrm{krank}(\tilde{\mathbf{X}}_{\mathrm{da}}^T\mathbf{T}_2\mathbf{A}_{\mathrm{t}})+\mathrm{krank}(\bar{\mathbf{B}}_{\mathrm{r}}\mathbf{U}_2\mathbf{A}_{\mathrm{r}})\geq K+1.
 \end{equation}
 Since it is difficult to determine the krank of any matrices, we derive a sufficient condition to guarantee the above inequality to hold, which is denoted as
 \begin{equation}
 	Q_{\mathrm{da}}\geq K, \text{and } 
 	P_{\mathrm{da}}\geq K.
 	\label{eq:lengthofh}
 \end{equation}
 To guarantee the unique solution to $ \mathbf{t}_2 $ during the iteration, $\boldsymbol{\Gamma}_{\mathrm{t}}^H\boldsymbol{\Gamma}_{\mathrm{t}}$ must be full rank, i.e., $ \mathrm{rank}(\boldsymbol{\Gamma}_{\mathrm{t}}^H\boldsymbol{\Gamma}_{\mathrm{t}})=N_{\mathrm{t}} $. $ \boldsymbol{\Gamma}_{\mathrm{t}}^H\boldsymbol{\Gamma}_{\mathrm{t}} $ can be further expressed by the Hadamard product denoted as
 \begin{equation}
 	\boldsymbol{\Gamma}_{\mathrm{t}}^H\boldsymbol{\Gamma}_{\mathrm{t}}
 	=(\tilde{\mathbf{X}}_{\mathrm{da}}^*\tilde{\mathbf{X}}_{\mathrm{da}}^T)\circ(\underbrace{\mathbf{A}_{\mathrm{t}}^*\mathbf{H}_{\alpha}^H\mathbf{A}_{\mathrm{r}}^H\mathbf{U}_2^*\bar{\mathbf{B}}_{\mathrm{r}}^H\bar{\mathbf{B}}_{\mathrm{r}}\mathbf{U}_2\mathbf{A}_{\mathrm{r}}\mathbf{H}_{\alpha}\mathbf{A}_{\mathrm{t}}^T)}_{\tilde{\boldsymbol{\Gamma}}_{\mathrm{t}}}.
 \end{equation}
 According to \cite[(10)]{Horn2020Rank}, the rank of the Hadamard product is given by
 \begin{equation}
 	\mathrm{rank}(\boldsymbol{\Gamma}_{\mathrm{t}}^H\boldsymbol{\Gamma}_{\mathrm{t}})\geq \min\{\mathrm{krank}(\tilde{\mathbf{X}}_{\mathrm{da}}^*\tilde{\mathbf{X}}_{\mathrm{da}}^T)+\mathrm{rank}(\tilde{\boldsymbol{\Gamma}}_{\mathrm{t}})-1,N_{\mathrm{t}}\}.
 \end{equation}
 As $ \tilde{\mathbf{X}}_{\mathrm{da}} $ is artificially designed, we assume that $ \mathrm{krank}(\tilde{\mathbf{X}}_{\mathrm{da}}^*\tilde{\mathbf{X}}_{\mathrm{da}}^T)=\mathrm{rank}(\tilde{\mathbf{X}}_{\mathrm{da}}^*\tilde{\mathbf{X}}_{\mathrm{da}})=\min(Q_{\mathrm{da}},N_{\mathrm{t}}) $. Further, because of $ K\ll N_{\mathrm{t}} $, $ \mathrm{rank}(\tilde{\boldsymbol{\Gamma}}_{\mathrm{t}})=\min(P_{\mathrm{da}},K) $. Accordingly, we can obtain the inequality denoted as
 \begin{equation}
 	\min(Q_{\mathrm{da}},N_{\mathrm{t}})+\min(P_{\mathrm{da}},K)\geq N_{\mathrm{t}}+1.
 	\label{eq:lengthoft}
 \end{equation}
 Similarly, to guarantee the unique solution to $ \mathbf{u}_2 $ during the iteration, the rank of $\boldsymbol{\Gamma}_{\mathrm{u}}^H\boldsymbol{\Gamma}_{\mathrm{u}}$ must be $ N_{\mathrm{r}} $. Since $ \bar{\mathbf{B}}_{\mathrm{r}} $ is artificially designed, we can obtain the following inequality denoted as
 \begin{equation}
 	\min(Q_{\mathrm{da}},K)+\min(P_{\mathrm{da}},N_{\mathrm{r}})\geq N_{\mathrm{r}}+1.
 	\label{eq:lengthofu}
 \end{equation}
 The solution to the inequalities consisting of \eqref{eq:lengthofh}, \eqref{eq:lengthoft}, and \eqref{eq:lengthofu} is given by
 \begin{equation}
 	Q_{\mathrm{da}}\geq N_{\mathrm{t}}-K+1,\text{ and }
 	P_{\mathrm{da}}\geq N_{\mathrm{r}}-K+1.
 \end{equation}
 Therefore, Proposition \ref{prop:pilotlength} holds.

 \section{Proof of Lemma \ref{lem:fimtrn}}\label{proof:fimtrn}
 We use $\mathbf{y}_{\mathrm{d}}$ to denote the received signal vector for estimating $ \boldsymbol{\eta} $. The corresponding probability density function of $\mathbf{y}_{\mathrm{d}}$ is defined as $p(\mathbf{y}_{\mathrm{d}};\boldsymbol{\eta})$. Then, the Fisher information matrix $\boldsymbol{\mathcal{I}}(\boldsymbol{\eta})$ can be defined as \cite{Kay1993Fundamentalsa}
 \if\thecol1
 \begin{equation}
 	\boldsymbol{\mathcal{I}}(\boldsymbol{\eta})=\mathbb{E}\left\lbrace\left[\frac{\partial \ln p(\mathbf{y}_{\mathrm{d}};\boldsymbol{\eta})}{\partial\boldsymbol{\eta}}\right] \left[\frac{\partial \ln p(\mathbf{y}_{\mathrm{d}};\boldsymbol{\eta})}{\partial\boldsymbol{\eta}}\right]^T\right\rbrace\\
 	=-\mathbb{E}\left\lbrace \frac{\partial}{\partial \boldsymbol{\eta}}\left[\frac{\partial}{\partial \boldsymbol{\eta}}\ln p(\mathbf{y}_{\mathrm{d}};\boldsymbol{\eta})\right]^T\right\rbrace.
 	\label{eq:fimdef}
 \end{equation}
 \else
 \begin{equation}
 	\begin{split}
 		\boldsymbol{\mathcal{I}}(\boldsymbol{\eta})&=\mathbb{E}\left\lbrace\left[\frac{\partial \ln p(\mathbf{y}_{\mathrm{d}};\boldsymbol{\eta})}{\partial\boldsymbol{\eta}}\right] \left[\frac{\partial \ln p(\mathbf{y}_{\mathrm{d}};\boldsymbol{\eta})}{\partial\boldsymbol{\eta}}\right]^T\right\rbrace\\
 		&=-\mathbb{E}\left\lbrace \frac{\partial}{\partial \boldsymbol{\eta}}\left[\frac{\partial}{\partial \boldsymbol{\eta}}\ln p(\mathbf{y}_{\mathrm{d}};\boldsymbol{\eta})\right]^T\right\rbrace.
 	\end{split}
 	\label{eq:fimdef}
 \end{equation}
 \fi

 \if\thecol1
 In Section \ref{sec:proforanddecom}, the received training signal $ \mathbf{y}_{\mathrm{d}} $ consists of two independent parts $ \mathbf{y}_{\mathrm{dr}} $ and $ \mathbf{y}_{\mathrm{da}} $, i.e., $ \mathbf{y}_{\mathrm{d}}=[\mathbf{y}_{\mathrm{dr}}^T,\mathbf{y}_{\mathrm{da}}^T]^T $, where $ \mathbf{y}_{\mathrm{dr}} $ is utilized to estimate $ \tilde{\mathbf{u}}_1$ and $ \mathbf{t}_1 $, and $ \mathbf{y}_{\mathrm{da}} $ is used to estimate $ \mathbf{u}_2 $ and $ \mathbf{t}_2 $. Thus, by dividing the vector $\boldsymbol{\eta}$ into two independent parts, i.e., $\boldsymbol{\eta}=[\boldsymbol{\eta}_{1}^T,\boldsymbol{\eta}_2^T]^T$, the corresponding probability density function of $ \mathbf{y}_{\mathrm{d}} $ can be further denoted as $p(\mathbf{y}_{\mathrm{d}};\boldsymbol{\eta})=p_1(\mathbf{y}_{\mathrm{dr}};\boldsymbol{\eta}_1)p_2(\mathbf{y}_{\mathrm{da}};\boldsymbol{\eta}_2)$, where $\boldsymbol{\eta}_{1}=[\Re\{\tilde{\mathbf{u}}_1^T\},\Im\{\tilde{\mathbf{u}}_1^T\},\Re\{\mathbf{t}_1^T\},\Im\{\mathbf{t}_1^T\}]^T$, $\boldsymbol{\eta}_2=[\Re\{\mathbf{u}_2^T\},\Im\{\mathbf{u}_2^T\},\\\Re\{\mathbf{t}_2^T\},\Im\{\mathbf{t}_2^T\}, \Re\{\mathbf{h}_{\alpha}^T\},\Im\{\mathbf{h}_{\alpha}^T\},\boldsymbol{\Theta}^T,\boldsymbol{\Phi}^T]^T$, $ p_1(\mathbf{y}_{\mathrm{dr}};\boldsymbol{\eta}_1) $ and $ p_2(\mathbf{y}_{\mathrm{da}};\boldsymbol{\eta}_2)$  denote the corresponding probability density functions of $ \mathbf{y}_{\mathrm{dr}} $ and $ \mathbf{y}_{\mathrm{da}} $. Then, based on \eqref{eq:fimdef}, the Fisher information matrix can be further given by
 \else
 In Section \ref{sec:proforanddecom}, the received training signal $ \mathbf{y}_{\mathrm{d}} $ consists of two independent parts $ \mathbf{y}_{\mathrm{dr}} $ and $ \mathbf{y}_{\mathrm{da}} $, i.e., $ \mathbf{y}_{\mathrm{d}}=[\mathbf{y}_{\mathrm{dr}}^T,\mathbf{y}_{\mathrm{da}}^T]^T $, where $ \mathbf{y}_{\mathrm{dr}} $ is utilized to estimate $ \tilde{\mathbf{u}}_1$ and $ \mathbf{t}_1 $, and $ \mathbf{y}_{\mathrm{da}} $ is used to estimate $ \mathbf{u}_2 $ and $ \mathbf{t}_2 $. Thus, by dividing the vector $\boldsymbol{\eta}$ into two independent parts, i.e., $\boldsymbol{\eta}=[\boldsymbol{\eta}_{1}^T,\boldsymbol{\eta}_2^T]^T$, the corresponding probability density function of $ \mathbf{y}_{\mathrm{d}} $ can be further denoted as $p(\mathbf{y}_{\mathrm{d}};\boldsymbol{\eta})=p_1(\mathbf{y}_{\mathrm{dr}};\boldsymbol{\eta}_1)p_2(\mathbf{y}_{\mathrm{da}};\boldsymbol{\eta}_2)$, where $\boldsymbol{\eta}_{1}=[\Re\{\tilde{\mathbf{u}}_1^T\},\Im\{\tilde{\mathbf{u}}_1^T\},\Re\{\mathbf{t}_1^T\},\Im\{\mathbf{t}_1^T\}]^T$, $\boldsymbol{\eta}_2=[\Re\{\mathbf{u}_2^T\},\Im\{\mathbf{u}_2^T\},\Re\{\mathbf{t}_2^T\},\Im\{\mathbf{t}_2^T\}, \Re\{\mathbf{h}_{\alpha}^T\},\Im\{\mathbf{h}_{\alpha}^T\},\boldsymbol{\Theta}^T,\boldsymbol{\Phi}^T]^T$, $ p_1(\mathbf{y}_{\mathrm{dr}};\boldsymbol{\eta}_1) $ and $ p_2(\mathbf{y}_{\mathrm{da}};\boldsymbol{\eta}_2)$  denote the corresponding probability density functions of $ \mathbf{y}_{\mathrm{dr}} $ and $ \mathbf{y}_{\mathrm{da}} $. Then, based on \eqref{eq:fimdef}, the Fisher information matrix can be further given by
 \fi
 \if\thecol1
 \begin{equation}
 	\begin{split}
 		\boldsymbol{\mathcal{I}}(\boldsymbol{\eta})&=-\mathbb{E}\left\lbrace \frac{\partial}{\partial \boldsymbol{\eta}}\left[\frac{\partial}{\partial \boldsymbol{\eta}}\bigl(\ln p_1(\mathbf{y}_{\mathrm{dr}};\boldsymbol{\eta}_1)+\ln p_2(\mathbf{y}_{\mathrm{da}};\boldsymbol{\eta}_2)\bigr)\right]^T\right\rbrace\\
 		&=-\mathbb{E}\left\lbrace
 		\left(
 		\begin{matrix}
 			\frac{\partial^2 \ln p_1(\mathbf{y}_{\mathrm{dr}};\boldsymbol{\eta}_1)}{\partial \boldsymbol{\eta}_1\partial \boldsymbol{\eta}_1^T}&\frac{\partial^2 \ln p_2(\mathbf{y}_{\mathrm{da}};\boldsymbol{\eta}_2)}{\partial \boldsymbol{\eta}_1\partial \boldsymbol{\eta}_2^T}\\
 			\frac{\partial^2 \ln p_1(\mathbf{y}_{\mathrm{dr}};\boldsymbol{\eta}_1)}{\partial \boldsymbol{\eta}_2\partial \boldsymbol{\eta}_1^T}&\frac{\partial^2 \ln p_2(\mathbf{y}_{\mathrm{da}};\boldsymbol{\eta}_2)}{\partial \boldsymbol{\eta}_2\partial \boldsymbol{\eta}_2^T}
 		\end{matrix}
 		\right)
 		\right\rbrace\\
 		&\overset{(a)}{=}\mathrm{blkdiag}[\boldsymbol{\mathcal{I}}(\boldsymbol{\eta}_1),\boldsymbol{\mathcal{I}}(\boldsymbol{\eta}_2)],
 	\end{split}
 	\label{eq:fimdef2}
 \end{equation}
 \else
 \begin{align}
 	\boldsymbol{\mathcal{I}}(\boldsymbol{\eta})&=-\mathbb{E}\left\lbrace \frac{\partial}{\partial \boldsymbol{\eta}}\left[\frac{\partial}{\partial \boldsymbol{\eta}}\bigl(\ln p_1(\mathbf{y}_{\mathrm{dr}};\boldsymbol{\eta}_1)+\ln p_2(\mathbf{y}_{\mathrm{da}};\boldsymbol{\eta}_2)\bigr)\right]^T\right\rbrace\nonumber\\
 	&=-\mathbb{E}\left\lbrace
 	\left(
 	\begin{matrix}
 		\frac{\partial^2 \ln p_1(\mathbf{y}_{\mathrm{dr}};\boldsymbol{\eta}_1)}{\partial \boldsymbol{\eta}_1\partial \boldsymbol{\eta}_1^T}&\frac{\partial^2 \ln p_2(\mathbf{y}_{\mathrm{da}};\boldsymbol{\eta}_2)}{\partial \boldsymbol{\eta}_1\partial \boldsymbol{\eta}_2^T}\\
 		\frac{\partial^2 \ln p_1(\mathbf{y}_{\mathrm{dr}};\boldsymbol{\eta}_1)}{\partial \boldsymbol{\eta}_2\partial \boldsymbol{\eta}_1^T}&\frac{\partial^2 \ln p_2(\mathbf{y}_{\mathrm{da}};\boldsymbol{\eta}_2)}{\partial \boldsymbol{\eta}_2\partial \boldsymbol{\eta}_2^T}
 	\end{matrix}
 	\right)
 	\right\rbrace\nonumber\\
 	&\overset{(a)}{=}\mathrm{blkdiag}[\boldsymbol{\mathcal{I}}(\boldsymbol{\eta}_1),\boldsymbol{\mathcal{I}}(\boldsymbol{\eta}_2)],
 	\label{eq:fimdef2}
 \end{align}
 \fi
 where $(a)$ holds due to the independence between $ \mathbf{y}_{\mathrm{dr}} $ and $  \boldsymbol{\eta}_2 $ as well as the independence between $ \mathbf{y}_{\mathrm{da}}$ and $ \boldsymbol{\eta}_1$,  $\boldsymbol{\mathcal{I}}(\boldsymbol{\eta}_1)=-\mathbb{E}\left\lbrace \frac{\partial^2 \ln p_1(\mathbf{y}_{\mathrm{dr}};\boldsymbol{\eta}_1)}{\partial \boldsymbol{\eta}_1\partial \boldsymbol{\eta}_1^T} \right\rbrace$ denotes the Fisher information matrix of $\boldsymbol{\eta}_1$, and $\boldsymbol{\mathcal{I}}(\boldsymbol{\eta}_2)=-\mathbb{E}\left\lbrace \frac{\partial^2 \ln p_2(\mathbf{y}_{\mathrm{da}};\boldsymbol{\eta}_2)}{\partial \boldsymbol{\eta}_2\partial \boldsymbol{\eta}_2^T} \right\rbrace$ denotes the Fisher information matrix of $\boldsymbol{\eta}_2$. Thus, Lemma \ref{lem:fimtrn} holds.

 \section{Proof of Lemma \ref{lem:fisherinfomatu1t1}}\label{proof:fisherinfomatu1t1}
 In Section \ref{sec:proforanddecom}, $ \mathbf{u}_1 $ and $ \mathbf{t}_1 $ are estimated from the received signal $ \tilde{\mathbf{y}}_{\mathrm{dr}} $ and $ \mathbf{y}_{\mathrm{dt}} $, respectively. According to Proposition \ref{prop:solutot1u1}, $ \tilde{\mathbf{y}}_{\mathrm{dr}} $ and $ \mathbf{y}_{\mathrm{dt}} $ are denoted as
 \if\thecol1
 \begin{equation}
 	\tilde{\mathbf{y}}_{\mathrm{dr}}=[\mathrm{vec}(\tilde{\mathbf{Y}}_{\mathrm{dr},1})^T,\cdots,\mathrm{vec}(\tilde{\mathbf{Y}}_{\mathrm{dr},P_{\mathrm{dr}}})^T]^T,\text{ and }
 	\mathbf{y}_{\mathrm{dt}}=[\mathbf{y}_{\mathrm{dr},1},\cdots,\mathbf{y}_{\mathrm{dr},(P_{\mathrm{dr}}-1)M_{\mathrm{r}}+1}]^T.
 \end{equation}
 \else
 \begin{equation}
 	\begin{split}
 		&\tilde{\mathbf{y}}_{\mathrm{dr}}=[\mathrm{vec}(\tilde{\mathbf{Y}}_{\mathrm{dr},1})^T,\cdots,\mathrm{vec}(\tilde{\mathbf{Y}}_{\mathrm{dr},P_{\mathrm{dr}}})^T]^T,\\
 		&\mathbf{y}_{\mathrm{dt}}=[\mathbf{y}_{\mathrm{dr},1},\cdots,\mathbf{y}_{\mathrm{dr},(P_{\mathrm{dr}}-1)M_{\mathrm{r}}+1}]^T.
 	\end{split}
 \end{equation}
 \fi
 Thus, similar to Lemma \ref{lem:fimtrn}, we have $ \boldsymbol{\mathcal{I}}(\boldsymbol{\eta}_1)=\mathrm{blkdiag}(\boldsymbol{\mathcal{I}}(\boldsymbol{\eta}_{1,1}),\boldsymbol{\mathcal{I}}(\boldsymbol{\eta}_{1,2})) $ due to the independence between $ \tilde{\mathbf{y}}_{\mathrm{dr}} $ and $ \mathbf{y}_{\mathrm{dt}} $, where  $ \boldsymbol{\eta}_{1,1}=[\Re\{\tilde{\mathbf{u}}_1^T\},\Im\{\tilde{\mathbf{u}}_1^T\}]^T $, and $ \boldsymbol{\eta}_{1,2}=[\Re\{\mathbf{t}_1^T\},\Im\{\mathbf{t}_1^T\}]^T $.

 To derive the fishier information matrices, we first model the received signal $ \bar{\mathbf{Y}}_{\mathrm{dr},p} $. Based on \eqref{eq:recsignaldrpri} and \eqref{eq:recsignaldr}, $\bar{\mathbf{Y}}_{\mathrm{dr},p} $ can be denoted as
 \begin{equation}
 	\bar{\mathbf{Y}}_{\mathrm{dr},p}=\beta_{\mathrm{d}}\mathbf{u}_1\mathbf{t}_1^T \mathbf{X}_{\mathrm{dr}}+\mathbf{u}_1 \mathbf{b}_{\mathrm{dr}}^T\bar{\mathbf{N}}_{\mathrm{dr},p}
 	=\mathbf{u}_1 \mathbf{x}_{\mathrm{tn},p}^T,
 \end{equation}
 where $ \mathbf{x}_{\mathrm{tn},p}=\beta_{\mathrm{d}}\mathbf{X}_{\mathrm{dr}}^T \mathbf{t}_1+\bar{\mathbf{N}}_{\mathrm{dr},p}^T\mathbf{b}_{\mathrm{dr}}$. Then, we have
 \if\thecol1
 \begin{equation}
 	\mathbf{y}_{\mathrm{dr},(p-1)M_{\mathrm{r}}+1}=u_{1,1}\mathbf{x}_{\mathrm{tn},p},\text{ and }
 	\tilde{\mathbf{Y}}_{\mathrm{dr},p}=\tilde{\mathbf{u}}_1 \mathbf{x}_{\mathrm{tn},p}^T,\label{eq:xtnmodel}
 \end{equation}	
 \else
 \begin{align}
 	&\mathbf{y}_{\mathrm{dr},(p-1)M_{\mathrm{r}}+1}=u_{1,1}\mathbf{x}_{\mathrm{tn},p},\\
 	&\tilde{\mathbf{Y}}_{\mathrm{dr},p}=\tilde{\mathbf{u}}_1 \mathbf{x}_{\mathrm{tn},p}^T,\label{eq:xtnmodel}
 \end{align}
 \fi
 As we set the first receive digital RF chain of the UE to be the reference, e.g., $ u_{1,1}=1 $,  $ \mathbf{x}_{\mathrm{tn},p} $ is equal to the first row of $ \mathbf{Y}_{\mathrm{dr},p} $, i.e., $ \mathbf{y}_{\mathrm{dr},(p-1)M_{\mathrm{r}}+1}=\mathbf{x}_{\mathrm{tn},p} $. This result indicates that $\tilde{\mathbf{Y}}_{\mathrm{dr},p}$ is a deterministic signal without noises for computing $\tilde{\mathbf{u}}_1$. To derive the closed-form expression of $ \boldsymbol{\mathcal{I}}(\boldsymbol{\eta}_{1,1}) $, we regard computing $\tilde{\mathbf{u}}_1$ as an asymptotic case of the estimation from the signal in white Gaussian noise with zero mean and zero variance. Thus, By using the vectorization of $\tilde{\mathbf{Y}}_{\mathrm{dr},p}$, \eqref{eq:xtnmodel} can be rewritten as
 \begin{equation}
 	\mathrm{vec}\{\tilde{\mathbf{Y}}_{\mathrm{dr},p}\}=(\mathbf{x}_{\mathrm{tn},p}\otimes \mathbf{I}_{M_\mathrm{r}})\tilde{\mathbf{u}}_1 +\mathbf{n}_{\mathrm{au}},
 \end{equation}
 where each entry of $ \mathbf{n}_{\mathrm{au}} $ is distributed as $ \mathcal{CN}(0,\gamma ) $, and $ \gamma\longrightarrow 0 $. Thus, $\mathrm{vec}\{\tilde{\mathbf{Y}}_{\mathrm{dr},p}\}$ is distributed as $\mathcal{CN}(\boldsymbol{\mu}_{\mathrm{tn},p},\gamma\mathbf{I}_{QM_{\mathrm{r}}})$, where $ \boldsymbol{\mu}_{\mathrm{tn},p}=(\mathbf{x}_{\mathrm{tn},p}\otimes
 \mathbf{I}_{M_\mathrm{r}})\tilde{\mathbf{u}}_1 $ .  Since $ \frac{\partial \boldsymbol{\mu}_{\mathrm{tn},p}}{\partial \boldsymbol{\eta}_{1,1}^T}=[(\mathbf{x}_{\mathrm{tn},p}\otimes \mathbf{I}_{M_{\mathrm{r}}-1}),j(\mathbf{x}_{\mathrm{tn},p}\otimes \mathbf{I}_{M_{\mathrm{r}}-1})] $ and based on \cite[(3.31)]{Kay1993Fundamentalsa}, the Fisher information matrix $\boldsymbol{\mathcal{I}}(\boldsymbol{\eta}_{1,1})$ is given by
 \if\thecol1
 \begin{equation}
 	\boldsymbol{\mathcal{I}}(\boldsymbol{\eta}_{1,1})=\frac{2}{\gamma}\sum_{p=1}^{P_{\mathrm{dr}}}\Re\left\lbrace \left(\frac{\partial \boldsymbol{\mu}_{\mathrm{tn},p}}{\partial \boldsymbol{\eta}_{1,1}^T}\right)^H\frac{\partial \boldsymbol{\mu}_{\mathrm{tn},p}}{\partial \boldsymbol{\eta}_{1,1}^T}\right\rbrace=\frac{2\sum_{p=1}^{P_{\mathrm{dr}}}\|\mathbf{x}_{\mathrm{tn},p}\|^2}{\gamma}\mathbf{I}_{2M_{\mathrm{r}}-2}.
 	\label{eq:infomaeta11}
 \end{equation}
 \else
 \begin{equation}
 	\begin{split}
 		\boldsymbol{\mathcal{I}}(\boldsymbol{\eta}_{1,1})
 		&=\frac{2}{\gamma}\sum_{p=1}^{P_{\mathrm{dr}}}\Re\left\lbrace \left(\frac{\partial \boldsymbol{\mu}_{\mathrm{tn},p}}{\partial \boldsymbol{\eta}_{1,1}^T}\right)^H\frac{\partial \boldsymbol{\mu}_{\mathrm{tn},p}}{\partial \boldsymbol{\eta}_{1,1}^T}\right\rbrace\\
 		&=\frac{2\sum_{p=1}^{P_{\mathrm{dr}}}\|\mathbf{x}_{\mathrm{tn},p}\|^2}{\gamma}\mathbf{I}_{2M_{\mathrm{r}}-2}.
 	\end{split}
 	\label{eq:infomaeta11}
 \end{equation}
 \fi

 Further, to derive the closed-form expression of $ \boldsymbol{\mathcal{I}}(\boldsymbol{\eta}_{1,2}) $, we use the fact that $ \mathbf{y}_{\mathrm{dr},(p-1)M_{\mathrm{r}}+1}=\beta_{\mathrm{d}}\mathbf{X}_{\mathrm{dr}}^T \mathbf{t}_1+\mathbf{b}_{\mathrm{dr}}^T\bar{\mathbf{N}}_{\mathrm{dr},p} $. The received signal $ \mathbf{y}_{\mathrm{dr},(p-1)M_{\mathrm{r}}+1} $ is distributed as $ \mathcal{CN}(\boldsymbol{\mu}_{\mathrm{dt},p},\sigma_{\mathrm{n}}^2 \mathbf{I}_{\mathrm{Q}}) $, where $ \boldsymbol{\mu}_{\mathrm{dt},p}=\beta_{\mathrm{d}}\mathbf{X}_{\mathrm{dr}}^T \mathbf{t}_1 $. Further, since $ \frac{\partial \boldsymbol{\mu}_{\mathrm{dt},p}}{\partial \boldsymbol{\eta}_{1,2}^T}=[\beta_{\mathrm{d}}\mathbf{X}_{\mathrm{dr}}^T,j\beta_{\mathrm{d}}\mathbf{X}_{\mathrm{dr}}^T] $, the closed-form expression of $ \boldsymbol{\mathcal{I}}(\boldsymbol{\eta}_{1,2}) $ can be given by
 \if\thecol1
 \begin{equation}
 	\boldsymbol{\mathcal{I}}(\boldsymbol{\eta}_{1,2})
 	=\frac{2}{\sigma_{\mathrm{n}}^2}\Re\left\lbrace \left(\frac{\partial \boldsymbol{\mu}_{\mathrm{dt},p}}{\partial \boldsymbol{\eta}_{1,2}^T}\right)^H\frac{\partial \boldsymbol{\mu}_{\mathrm{dt},p}}{\partial \boldsymbol{\eta}_{1,2}^T}\right\rbrace
 	=\frac{2P_{\mathrm{dr}}|\beta_{\mathrm{d}}|^2}{\sigma_{\mathrm{n}}^2}\Re(\mathbf{E}_{\mathrm{au}}\otimes \mathbf{X}_{\mathrm{dr}}^*\mathbf{X}_{\mathrm{dr}}^T)
 	\overset{(a)}{=}\frac{2\rho_{\mathrm{c}}Q_{\mathrm{dr}}P_{\mathrm{dr}}|\beta_{\mathrm{d}}|^2}{\sigma_{\mathrm{n}}^2}\mathbf{I}_{2M_{\mathrm{t}}},
 \end{equation}
 \else
 \begin{equation}
 	\begin{split}
 		\boldsymbol{\mathcal{I}}(\boldsymbol{\eta}_{1,2})
 		&=\frac{2}{\sigma_{\mathrm{n}}^2}\Re\left\lbrace \left(\frac{\partial \boldsymbol{\mu}_{\mathrm{dt},p}}{\partial \boldsymbol{\eta}_{1,2}^T}\right)^H\frac{\partial \boldsymbol{\mu}_{\mathrm{dt},p}}{\partial \boldsymbol{\eta}_{1,2}^T}\right\rbrace\\
 		&=\frac{2P_{\mathrm{dr}}|\beta_{\mathrm{d}}|^2}{\sigma_{\mathrm{n}}^2}\Re(\mathbf{E}_{\mathrm{au}}\otimes \mathbf{X}_{\mathrm{dr}}^*\mathbf{X}_{\mathrm{dr}}^T)\\
 		&\overset{(a)}{=}\frac{2\rho_{\mathrm{c}}Q_{\mathrm{dr}}P_{\mathrm{dr}}|\beta_{\mathrm{d}}|^2}{\sigma_{\mathrm{n}}^2}\mathbf{I}_{2M_{\mathrm{t}}},
 	\end{split}
 \end{equation}
 \fi
 where $ \mathbf{E}_{\mathrm{au}}=[1,j;-j,1] $, and the step $ (a) $ holds by assuming that $ \mathbf{X}_{\mathrm{dr}} $ is orthogonal in the time domain and $ \mathbf{X}_{\mathrm{dr}}^*\mathbf{X}_{\mathrm{dr}}^T=\rho_{\mathrm{c}}Q_{\mathrm{dr}}\mathbf{I}_{M_\mathrm{t}} $  since $M_t<Q_{dr} $. Thus, Lemma \ref{lem:fisherinfomatu1t1} holds.

 \section{Proof of Lemma \ref{lem:fisherinfomatu2t2}}\label{proof:fisherinfomatu2t2}
 In Section \ref{sec:solutionCal}, $\mathbf{u}_2$ and $\mathbf{t}_2$ are jointly estimated by using the received signal $\mathbf{y}_{\mathrm{da}}$. According to \eqref{eq:recsignalda}, $\mathbf{y}_{\mathrm{da}}$ can be given by
 \begin{equation}
 	\mathbf{y}_{\mathrm{da}}=\underbrace{\mathrm{vec}(\bar{\mathbf{B}}_{\mathrm{da}}^T\mathbf{U}_2\mathbf{A}_{\mathrm{r}}\mathbf{H}_{\alpha}\mathbf{A}_{\mathrm{t}}^T\mathbf{T}_2\tilde{\mathbf{X}}_{\mathrm{da}})}_{\boldsymbol{\mu}_{\mathrm{da}}}+\mathrm{vec}(\mathbf{N}_{\mathrm{da}}),
 \end{equation}
 which obeys complex Gaussian distribution, i.e., $ \mathbf{y}_{\mathrm{da}}\sim \mathcal{CN}(\boldsymbol{\mu}_{\mathrm{da}},\boldsymbol{\Sigma_{\mathrm{da}}}) $, where
 \begin{equation}
 	\boldsymbol{\Sigma}_{\mathrm{da}}=\mathbb{E}\left\lbrace \mathrm{vec}(\mathbf{N}_{\mathrm{da}})\mathrm{vec}(\mathbf{N}_{\mathrm{da}})^H \right\rbrace
 	=\sigma_{\mathrm{n}}^2\mathbf{I}_{Q_{\mathrm{da}}P_{\mathrm{da}}}.
 \end{equation}

 We first derive the partial derivative of $ \boldsymbol{\mu}_{\mathrm{da}} $ with the respect to $ \boldsymbol{\Phi} $, which is denoted as
 \if\thecol1
 \begin{equation}
 	\begin{split}
 		\frac{\partial \boldsymbol{\mu}_{\mathrm{da}}}{\partial \boldsymbol{\Phi}^T}
 		&=(\tilde{\mathbf{X}}_{\mathrm{da}}^T\mathbf{T}_2\mathbf{A}_{\mathrm{t}}\mathbf{H}_{\alpha}\otimes\bar{\mathbf{B}}_{\mathrm{r}}\mathbf{U}_2)\frac{ \partial \mathrm{vec}(\mathbf{A}_{\mathrm{r}})}{\partial \boldsymbol{\Phi}^T}\\
 		&=(\tilde{\mathbf{X}}_{\mathrm{da}}^T\mathbf{T}_2\mathbf{A}_{\mathrm{t}}\mathbf{H}_{\alpha}\otimes\bar{\mathbf{B}}_{\mathrm{r}}\mathbf{U}_2)\mathrm{blkdiag}(\bar{\mathbf{a}}_{\mathrm{r}}(\phi_1),\cdots,\bar{\mathbf{a}}_{\mathrm{r}}(\phi_K))\\
 		&=\mathbf{\Gamma_{\phi}},
 	\end{split}
 	\label{eq:gammaphi}
 \end{equation}
 \else
 \begin{equation}
 	\begin{split}
 		\frac{\partial \boldsymbol{\mu}_{\mathrm{da}}}{\partial \boldsymbol{\Phi}^T}
 		&=(\tilde{\mathbf{X}}_{\mathrm{da}}^T\mathbf{T}_2\mathbf{A}_{\mathrm{t}}\mathbf{H}_{\alpha}\otimes\bar{\mathbf{B}}_{\mathrm{r}}\mathbf{U}_2)\frac{ \partial \mathrm{vec}(\mathbf{A}_{\mathrm{r}})}{\partial \boldsymbol{\Phi}^T}\\
 		&=(\tilde{\mathbf{X}}_{\mathrm{da}}^T\mathbf{T}_2\mathbf{A}_{\mathrm{t}}\mathbf{H}_{\alpha}\otimes\bar{\mathbf{B}}_{\mathrm{r}}\mathbf{U}_2)\cdot\\
 		&\quad\ \mathrm{blkdiag}(\bar{\mathbf{a}}_{\mathrm{r}}(\phi_1),\cdots,\bar{\mathbf{a}}_{\mathrm{r}}(\phi_K))\\
 		&=\mathbf{\Gamma_{\phi}},
 	\end{split}
 	\label{eq:gammaphi}
 \end{equation}
 \fi
 where $ \bar{\mathbf{a}}_{\mathrm{r}}(\phi_k) =\partial \mathbf{a}_{\mathrm{r}}(\phi_k)/\partial \phi_k= \mathbf{a}_{\mathrm{r}}(\phi_k)\circ[0,-j\frac{2\pi d}{\lambda}\cos\phi_k,\cdots,-j\frac{2\pi d}{\lambda}(N_{\mathrm{r}}-1)\cos\phi_k]^T $. Similarly, the partial derivative of $ \boldsymbol{\mu}_{\mathrm{da}} $ with the respect to $ \boldsymbol{\Theta} $ is denoted as
 \if\thecol1
 \begin{equation}
 	\frac{\partial \boldsymbol{\mu}_{\mathrm{da}}}{\partial \boldsymbol{\Theta}^T}
 	=(\tilde{\mathbf{X}}_{\mathrm{da}}^T\mathbf{T}_2\otimes\bar{\mathbf{B}}_{\mathrm{r}}\mathbf{U}_2\mathbf{A}_{\mathrm{r}}\mathbf{H}_{\alpha})\mathbf{E}_{\mathrm{x},N_{\mathrm{t}}K}\mathrm{blkdiag}(\bar{\mathbf{a}}_{\mathrm{t}}(\theta_1),\cdots,\bar{\mathbf{a}}_{\mathrm{t}}(\theta_K))
 	=\mathbf{\Gamma_{\theta}},
 	\label{eq:gammatheta}
 \end{equation}
 \else
 \begin{equation}
 	\begin{split}
 		\frac{\partial \boldsymbol{\mu}_{\mathrm{da}}}{\partial \boldsymbol{\Theta}^T}
 		&=(\tilde{\mathbf{X}}_{\mathrm{da}}^T\mathbf{T}_2\otimes\bar{\mathbf{B}}_{\mathrm{r}}\mathbf{U}_2\mathbf{A}_{\mathrm{r}}\mathbf{H}_{\alpha})\cdot\\
 		&\quad\ \mathbf{E}_{\mathrm{x},N_{\mathrm{t}}K}\mathrm{blkdiag}(\bar{\mathbf{a}}_{\mathrm{t}}(\theta_1),\cdots,\bar{\mathbf{a}}_{\mathrm{t}}(\theta_K))
 		=\mathbf{\Gamma_{\theta}},
 	\end{split}
 	\label{eq:gammatheta}
 \end{equation}
 \fi
 where $ \bar{\mathbf{a}}_{\mathrm{t}}(\theta_k) = \mathbf{a}_{\mathrm{t}}(\theta_1)\circ[0,-j\frac{2\pi d}{\lambda}\cos\theta_k,\cdots,-j\frac{2\pi d}{\lambda}(N_{\mathrm{t}}-1)\cos\theta_k]^T $, and $ \mathbf{E}_{\mathrm{x},N_{\mathrm{t}},K}=\sum_{k=1}^{K}(\mathbf{e}_k^T\otimes \mathbf{I}_{N_{\mathrm{t}}}\otimes \mathbf{e}_k) $, $ \mathbf{e}_k $ is the $ k $-the column of $ \mathbf{I}_{K} $.

 Based on \eqref{eq:gammah}, \eqref{eq:estu2pro}, \eqref{eq:gammat}, \eqref{eq:gammaphi} and \eqref{eq:gammatheta}, the partial derivative of $ \boldsymbol{\mu}_{\mathrm{da}} $ with the respect to $ \boldsymbol{\eta}_{2} $ can be given by $ \frac{\partial \boldsymbol{\mu}_{\mathrm{da}}}{\partial\boldsymbol{\eta}_2^T}=[\mathbf{\Gamma}_{\mathrm{t}},j\mathbf{\Gamma}_{\mathrm{t}},\mathbf{\Gamma}_{\mathrm{u}},j\mathbf{\Gamma}_{\mathrm{u}},\mathbf{\Gamma}_{\mathrm{h}},j\mathbf{\Gamma}_{\mathrm{h}},\mathbf{\Gamma}_{\mathrm{\theta}},\mathbf{\Gamma}_{\mathrm{\Phi}}]=\boldsymbol{\Upsilon_{\mathrm{\eta}}} $. Finally, based on \cite[(3.31)]{Kay1993Fundamentalsa}, the Fisher information matrix $ \boldsymbol{\mathcal{I}}(\boldsymbol{\eta}_2) $ can be given by
 \begin{equation}
 	\boldsymbol{\mathcal{I}}(\boldsymbol{\eta}_2)=2\Re\left\lbrace \left(\frac{\partial \boldsymbol{\mu}_{\mathrm{da}}}{\partial\boldsymbol{\eta}_2^T}\right)^H\boldsymbol{\Sigma}_{\mathrm{da}}^{-1}\frac{\partial \boldsymbol{\mu}_{\mathrm{da}}}{\partial\boldsymbol{\eta}_2^T}\right\rbrace=\frac{2}{\sigma_{\mathrm{n}}^2}\Re \left\lbrace \boldsymbol{\Upsilon}_{\mathrm{\eta}}^H\boldsymbol{\Upsilon}_{\mathrm{\eta}} \right\rbrace.
 \end{equation}
 Accordingly, Lemma \ref{lem:fisherinfomatu2t2} holds.

 \section{Proof of Proposition \ref{prop:CRLBDL}}\label{proof:CRLBDL}
 The partial derivative of the transformation function $ \mathbf{g}(\boldsymbol{\eta}) $ with respect to $ \boldsymbol{\eta} $ can be denoted as
 \if\thecol1
 \begin{equation}
 	\frac{\partial \mathbf{g}(\boldsymbol{\eta})}{\partial \boldsymbol{\eta}^T}=\mathrm{blkdiag}([\mathbf{I}_{M_{\mathrm{r}}-1},j\mathbf{I}_{M_{\mathrm{r}}-1}],[\mathbf{I}_{M_\mathrm{t}},j\mathbf{I}_{M_{\mathrm{t}}}],[\mathbf{I}_{N_{\mathrm{r}}},j\mathbf{I}_{N_{\mathrm{r}}}],[\mathbf{I}_{N_{\mathrm{t}}},j\mathbf{I}_{N_{\mathrm{t}}}],\mathbf{0}_{4K,4K}).
 \end{equation}
 \else
 \begin{equation}
 	\begin{split}
 		\frac{\partial \mathbf{g}(\boldsymbol{\eta})}{\partial \boldsymbol{\eta}^T}=\mathrm{blkdiag}([\mathbf{I}_{M_{\mathrm{r}}-1},j\mathbf{I}_{M_{\mathrm{r}}-1}],[\mathbf{I}_{M_\mathrm{t}},j\mathbf{I}_{M_{\mathrm{t}}}],\\ [\mathbf{I}_{N_{\mathrm{r}}},j\mathbf{I}_{N_{\mathrm{r}}}],[\mathbf{I}_{N_{\mathrm{t}}},j\mathbf{I}_{N_{\mathrm{t}}}],\mathbf{0}_{4K,4K}).
 	\end{split}
 \end{equation}
 \fi

 Based on the definition of CRLB, we can obtain
 \if\thecol1
 \begin{equation}
 	\begin{split}
 		\frac{\partial \mathbf{g}(\boldsymbol{\eta})}{\partial \boldsymbol{\eta}^T}\boldsymbol{\mathcal{I}}(\boldsymbol{\eta})^{-1}\left(\frac{\partial \mathbf{g}(\boldsymbol{\eta})}{\partial \boldsymbol{\eta}^T}\right)^H
 		=&\mathrm{blkdiag}([\mathbf{I}_{M_{\mathrm{r}}-1},j\mathbf{I}_{M_{\mathrm{r}}-1}]\boldsymbol{\mathcal{I}}(\boldsymbol{\eta}_{1,1})^{-1}[\mathbf{I}_{M_{\mathrm{r}}-1},j\mathbf{I}_{M_{\mathrm{r}}-1}]^H,\\
 		&[\mathbf{I}_{M_\mathrm{t}},j\mathbf{I}_{M_{\mathrm{t}}}]\boldsymbol{\mathcal{I}}(\boldsymbol{\eta}_{1,2})^{-1}[\mathbf{I}_{M_\mathrm{t}},j\mathbf{I}_{M_{\mathrm{t}}}]^H,\boldsymbol{\Pi}\boldsymbol{\mathcal{I}}(\boldsymbol{\eta}_{2})^{-1}\boldsymbol{\Pi}^H),
 	\end{split}
 	\label{eq:CRLBut}
 \end{equation}
 \else
 \begin{equation}
 	\begin{split}
 		&\frac{\partial \mathbf{g}(\boldsymbol{\eta})}{\partial \boldsymbol{\eta}^T}\boldsymbol{\mathcal{I}}(\boldsymbol{\eta})^{-1}\left(\frac{\partial \mathbf{g}(\boldsymbol{\eta})}{\partial \boldsymbol{\eta}^T}\right)^H=\\
 		&\mathrm{blkdiag}([\mathbf{I}_{M_{\mathrm{r}}-1},j\mathbf{I}_{M_{\mathrm{r}}-1}]\boldsymbol{\mathcal{I}}(\boldsymbol{\eta}_{1,1})^{-1}[\mathbf{I}_{M_{\mathrm{r}}-1},j\mathbf{I}_{M_{\mathrm{r}}-1}]^H,\\
 		&[\mathbf{I}_{M_\mathrm{t}},j\mathbf{I}_{M_{\mathrm{t}}}]\boldsymbol{\mathcal{I}}(\boldsymbol{\eta}_{1,2})^{-1}[\mathbf{I}_{M_\mathrm{t}},j\mathbf{I}_{M_{\mathrm{t}}}]^H,\boldsymbol{\Pi}\boldsymbol{\mathcal{I}}(\boldsymbol{\eta}_{2})^{-1}\boldsymbol{\Pi}^H),
 	\end{split}
 	\label{eq:CRLBut}
 \end{equation}
 \fi
 where $ \boldsymbol{\Pi}=[\mathrm{blkdiag}([\mathbf{I}_{N_{\mathrm{r}}},j\mathbf{I}_{N_{\mathrm{r}}}],[\mathbf{I}_{N_{\mathrm{t}}},j\mathbf{I}_{N_{\mathrm{t}}}]),\mathbf{0}_{N_{\mathrm{r}}+N_{\mathrm{t}},4K}] $. By substituting \eqref{eq:FIMt1u1} and \eqref{eq:FIMt2u2} into \eqref{eq:CRLBut}, the CRLB of $ \boldsymbol{\eta}_{\mathrm{ut}} $ can be given by \eqref{eq:CRLBut12}.

 \AtNextBibliography{\small}
 \printbibliography

@Article{Larsson2014Massive,
  author  = {Larsson, Erik G. and Edfors, Ove and Tufvesson, Fredrik and Marzetta, Thomas L.},
  journal = {IEEE Commun. Mag.},
  title   = {Massive {{MIMO}} for next Generation Wireless Systems},
  year    = {2014},
  issn    = {0163-6804, 1558-1896},
  month   = feb,
  number  = {2},
  pages   = {186--195},
  volume  = {52},
  doi     = {10.1109/MCOM.2014.6736761},
}

@Article{Molisch2017Hybrid,
  author  = {Molisch, Andreas F. and Ratnam, Vishnu V. and Han, Shengqian and Li, Zheda and Nguyen, Sinh Le Hong and Li, Linsheng and Haneda, Katsuyuki},
  journal = {IEEE Commun. Mag.},
  month   = sep,
  number  = {9},
  pages   = {134--141},
  title   = {Hybrid {{Beamforming}} for {{Massive MIMO}}: {{A Survey}}},
  volume  = {55},
  year    = {2017},
  issn    = {0163-6804},
  doi     = {10.1109/MCOM.2017.1600400},
}

@Article{Flordelis2018Massive,
  author  = {Flordelis, Jose and Rusek, Fredrik and Tufvesson, Fredrik and Larsson, Erik G. and Edfors, Ove},
  journal = {IEEE Trans. Wireless Commun.},
  title   = {Massive {{MIMO Performance}}\textemdash{{TDD Versus FDD}}: {{What Do Measurements Say}}?},
  year    = {2018},
  issn    = {1536-1276, 1558-2248},
  month   = apr,
  number  = {4},
  pages   = {2247--2261},
  volume  = {17},
  doi     = {10.1109/TWC.2018.2790912},
}

@article{Bezdek2003Convergence,
  title = {Convergence of Alternating Optimization},
  author = {Bezdek, James C. and Hathaway, Richard J.},
  year = {2003},
  month = dec,
  volume = {11},
  pages = {351--368},
  issn = {1061-5369},
  journal = {Neural Parallel Sci. Comput.},
  number = {4}
}

@Article{Shan2018General,
  author  = {Shan, Chuanqiang and Chen, Li and Chen, Xiaohui and Wang, Weidong},
  journal = {IEEE Trans. Veh. Technol.},
  title   = {A {{General Matched Filter Design}} for {{Reciprocity Calibration}} in {{Multiuser Massive MIMO Systems}}},
  year    = {2018},
  issn    = {0018-9545, 1939-9359},
  month   = sep,
  number  = {9},
  pages   = {8939--8943},
  volume  = {67},
  doi     = {10.1109/TVT.2018.2839591},
}

@Article{Nie2021Impact,
  author  = {Nie, Rongjiang and Chen, Li and Zhao, Nan and Chen, Yunfei and Wang, Weidong and Wang, Xianbin},
  journal = {IEEE Trans. Wireless Commun.},
  title   = {Impact and {{Calibration}} of {{Nonlinear Reciprocity Mismatch}} in {{Massive MIMO Systems}}},
  year    = {2021},
  issn    = {1558-2248},
  month   = oct,
  number  = {10},
  pages   = {6418--6435},
  volume  = {20},
  doi     = {10.1109/TWC.2021.3073949},
}

@Article{Zhang2015LargeScale,
  author  = {Zhang, Wence and Ren, Hong and Pan, Cunhua and Chen, Ming and {de Lamare}, Rodrigo C. and Du, Bo and Dai, Jianxin},
  journal = {IEEE Trans. Commun.},
  title   = {Large-{{Scale Antenna Systems With UL}}/{{DL Hardware Mismatch}}: {{Achievable Rates Analysis}} and {{Calibration}}},
  year    = {2015},
  issn    = {0090-6778},
  month   = apr,
  number  = {4},
  pages   = {1216--1229},
  volume  = {63},
  doi     = {10.1109/TCOMM.2015.2395432},
}

@Article{Wei2016Impact,
  author  = {Wei, Hao and Wang, Dongming and Wang, Jiangzhou and You, Xiaohu},
  journal = {Sci. China Inf. Sci.},
  title   = {Impact of {{RF}} Mismatches on the Performance of Massive {{MIMO}} Systems with {{ZF}} Precoding},
  year    = {2016},
  issn    = {1674-733X, 1869-1919},
  month   = feb,
  number  = {2},
  pages   = {1--14},
  volume  = {59},
  doi     = {10.1007/s11432-015-5509-1},
}

@Article{Raeesi2018Performance,
  author  = {Raeesi, Orod and Gokceoglu, Ahmet and Zou, Yaning and Bjornson, Emil and Valkama, Mikko},
  journal = {IEEE Trans. Commun.},
  title   = {Performance {{Analysis}} of {{Multi-User Massive MIMO Downlink Under Channel Non-Reciprocity}} and {{Imperfect CSI}}},
  year    = {2018},
  issn    = {0090-6778, 1558-0857},
  month   = jun,
  number  = {6},
  pages   = {2456--2471},
  volume  = {66},
  doi     = {10.1109/TCOMM.2018.2792017},
}

@InProceedings{Guillaud2005practical,
  author    = {Guillaud, M. and Slock, D.T.M. and Knopp, R.},
  booktitle = {Proc. 8th Int. Symp. Signal Process. Applic. (ISSPA)},
  title     = {A Practical Method for Wireless Channel Reciprocity Exploitation through Relative Calibration},
  year      = {2005},
  address   = {{Sydney, Australia}},
  pages     = {403--406},
  publisher = {{IEEE}},
  volume    = {1},
  doi       = {10.1109/ISSPA.2005.1580281},
  isbn      = {978-0-7803-9243-4},
}

@InProceedings{Kaltenberger2010Relative,
  author    = {Kaltenberger, Florian and Jiang, Haiyong and Guillaud, Maxime and Knopp, Raymond},
  booktitle = {Proc. Futur. Netw. Mob. Summit},
  title     = {Relative Channel Reciprocity Calibration in {{MIMO}}/{{TDD}} Systems},
  year      = {2010},
  month     = jun,
  pages     = {1--10},
}

@InProceedings{Kouassi2012Estimation,
  author    = {Kouassi, Boris and Ghauri, Irfan and Deneire, Luc},
  booktitle = {Proc. 7th Int. ICST Conf. Cogn. Radio Oriented Wirel. Networks Commun.},
  title     = {Estimation of {{Time-Domain Calibration Parameters}} to {{Restore MIMO-TDD Channel Reciprocity}}},
  year      = {2012},
  address   = {{Stockholm, Sweden}},
  publisher = {{IEEE}},
  doi       = {10.4108/icst.crowncom.2012.248472},
  isbn      = {978-1-936968-55-8},
}

@InProceedings{Petermann2010Lowcomplexity,
  author    = {Petermann, Mark and Stefer, Markus and Wubben, Dirk and Schneider, Martin and Kammeyer, Karl-Dirk},
  booktitle = {Proc. 7th Int. Symp. Wireless Commun. Syst.},
  title     = {Low-Complexity Calibration of Mutually Coupled Non-Reciprocal Multi-Antenna {{OFDM}} Transceivers},
  year      = {2010},
  address   = {{York}},
  month     = sep,
  pages     = {285--289},
  publisher = {{IEEE}},
  doi       = {10.1109/ISWCS.2010.5624277},
  isbn      = {978-1-4244-6315-2 978-1-4244-6317-6},
}

@InProceedings{Shepard2012Argos,
  author    = {Shepard, Clayton and Yu, Hang and Anand, Narendra and Li, Erran and Marzetta, Thomas and Yang, Richard and Zhong, Lin},
  booktitle = {Proc. 18th Annu. Int. Conf. Mobile Comput. Networking (Mobicom)},
  title     = {Argos: Practical Many-Antenna Base Stations},
  year      = {2012},
  address   = {{Istanbul, Turkey}},
  pages     = {53},
  publisher = {{ACM Press}},
  doi       = {10.1145/2348543.2348553},
  isbn      = {978-1-4503-1159-5},
}

@Article{Wei2016Mutual,
  author  = {Wei, Hao and Wang, Dongming and Zhu, Huiling and Wang, Jiangzhou and Sun, Shaohui and You, Xiaohu},
  journal = {IEEE Trans. Wireless Commun.},
  title   = {Mutual {{Coupling Calibration}} for {{Multiuser Massive MIMO Systems}}},
  year    = {2016},
  issn    = {1536-1276},
  month   = jan,
  number  = {1},
  pages   = {606--619},
  volume  = {15},
  doi     = {10.1109/TWC.2015.2476467},
}

@Article{Jiang2018Framework,
  author  = {Jiang, Xiwen and Decurninge, Alexis and Gopala, Kalyana and Kaltenberger, Florian and Guillaud, Maxime and Slock, Dirk and Deneire, Luc},
  journal = {IEEE Trans. Wireless Commun.},
  title   = {A {{Framework}} for {{Over-the-Air Reciprocity Calibration}} for {{TDD Massive MIMO Systems}}},
  year    = {2018},
  issn    = {1536-1276, 1558-2248},
  month   = sep,
  number  = {9},
  pages   = {5975--5990},
  volume  = {17},
  doi     = {10.1109/TWC.2018.2853743},
}

@InProceedings{Rogalin2013Hardwareimpairment,
  author    = {Rogalin, R. and Bursalioglu, O. Y. and Papadopoulos, H. C. and Caire, G. and Molisch, A. F.},
  booktitle = {Proc. 2013 Inf. Theory Appl. Work. (ITA)},
  title     = {Hardware-Impairment Compensation for Enabling Distributed Large-Scale {{MIMO}}},
  year      = {2013},
  address   = {{San Diego, CA}},
  month     = feb,
  pages     = {1--10},
  publisher = {{IEEE}},
  doi       = {10.1109/ITA.2013.6502966},
  isbn      = {978-1-4673-4647-4 978-1-4673-4648-1},
}

@Article{Su2014Retrieving,
  author  = {Su, Liyan and Yang, Chenyang and Wang, Gang and Lei, Ming},
  journal = {IEEE Trans. Commun.},
  title   = {Retrieving {{Channel Reciprocity}} for {{Coordinated Multi-Point Transmission}} with {{Joint Processing}}},
  year    = {2014},
  issn    = {0090-6778},
  month   = may,
  number  = {5},
  pages   = {1541--1553},
  volume  = {62},
  doi     = {10.1109/TCOMM.2014.031014.130367},
}

@Article{Nie2020Relaying,
  author  = {Nie, Rongjiang and Chen, Li and Zhao, Nan and Chen, Yunfei and Yu, F. Richard and Wei, Guo},
  journal = {IEEE Trans. Commun.},
  title   = {Relaying {{Systems With Reciprocity Mismatch}}: {{Impact Analysis}} and {{Calibration}}},
  year    = {2020},
  issn    = {0090-6778, 1558-0857},
  month   = jul,
  number  = {7},
  pages   = {4035--4049},
  volume  = {68},
  doi     = {10.1109/TCOMM.2020.2982632},
}

@InProceedings{Jiang2015MIMOTDD,
  author    = {Jiang, Xiwen and {\v C}irki{\'c}, Mirsad and Kaltenberger, Florian and Larsson, Erik G. and Deneire, Luc and Knopp, Raymond},
  booktitle = {IEEE Int. Conf. Commun.},
  title     = {{{MIMO-TDD}} Reciprocity under Hardware Imbalances: {{Experimental}} Results},
  year      = {2015},
  month     = jun,
  pages     = {4949--4953},
  doi       = {10.1109/ICC.2015.7249107},
  issn      = {1938-1883},
}

@Article{Nishimori2001Automatic,
  author  = {Nishimori, K. and Cho, K. and Takatori, Y. and Hori, T.},
  journal = {IEEE Trans. Veh. Technol.},
  title   = {Automatic Calibration Method Using Transmitting Signals of an Adaptive Array for {{TDD}} Systems},
  year    = {2001},
  issn    = {00189545},
  month   = nov,
  number  = {6},
  pages   = {1636--1640},
  volume  = {50},
  doi     = {10.1109/25.966592},
}

@InProceedings{Bourdoux2003Nonreciprocal,
  author    = {Bourdoux, A. and Come, B. and Khaled, N.},
  booktitle = {Proc. IEEE Radio Wirel. Conf. (RAWCON)},
  title     = {Non-Reciprocal Transceivers in {{OFDM}}/{{SDMA}} Systems: {{Impact}} and Mitigation},
  year      = {2003},
  address   = {{Boston, Massachusetts, USA}},
  pages     = {183--186},
  publisher = {{IEEE}},
  doi       = {10.1109/RAWCON.2003.1227923},
  isbn      = {978-0-7803-7829-2},
}

@InProceedings{Benzin2017Internal,
  author    = {Benzin, A. and Caire, G.},
  booktitle = {Proc. Int. ITG Workshop Smart Antennas (WSA)},
  title     = {Internal {{Self-Calibration Methods}} for {{Large Scale Array Transceiver Software-Defined Radios}}},
  year      = {2017},
  month     = mar,
  pages     = {1--8},
}

@Article{Luo2019Massive,
  author  = {Luo, Xiliang and Yang, Fuqian and Zhu, Hanyu},
  journal = {IEEE Trans. Veh. Technol.},
  title   = {Massive {{MIMO Self-Calibration}}: {{Optimal Interconnection}} for {{Full Calibration}}},
  year    = {2019},
  issn    = {0018-9545, 1939-9359},
  month   = nov,
  number  = {11},
  pages   = {10357--10371},
  volume  = {68},
  doi     = {10.1109/TVT.2019.2941544},
}

@Article{Jiang2018Channel,
  author    = {Jiang, Xiwen and Kaltenberger, Florian},
  journal   = {IEEE J. Sel. Topics Signal Process.},
  title     = {Channel {{Reciprocity Calibration}} in {{TDD Hybrid Beamforming Massive MIMO Systems}}},
  year      = {2018},
  issn      = {1932-4553, 1941-0484},
  month     = jun,
  number    = {3},
  pages     = {422--431},
  volume    = {12},
  doi       = {10.1109/JSTSP.2018.2819118},
  language  = {en},
}

@book{Kay1993Fundamentalsa,
	title = {Fundamentals of Statistical Signal Processing},
	author = {Kay, Steven M.},
	year = {1993},
	publisher = {{Prentice-Hall PTR}},
	address = {{Englewood Cliffs, N.J}},
	pages = {45--47},
	isbn = {978-0-13-345711-7 978-0-13-504135-2 978-0-13-280803-3},
	language = {en},
	lccn = {TK5102.5 .K379 1993},
	series = {Prentice {{Hall}} Signal Processing Series}
}

@Article{Qi2019OffGrid,
  author  = {Qi, Biqing and Wang, Wei and Wang, Ben},
  journal = {IEEE Commun. Lett.},
  title   = {Off-{{Grid Compressive Channel Estimation}} for Mm-{{Wave Massive MIMO With Hybrid Precoding}}},
  year    = {2019},
  issn    = {1558-2558},
  month   = jan,
  number  = {1},
  pages   = {108--111},
  volume  = {23},
  doi     = {10.1109/LCOMM.2018.2878557},
}

@article{Horn2020Rank,
	title = {Rank of a {{Hadamard}} Product},
	author = {Horn, Roger A. and Yang, Zai},
	year = {2020},
	month = apr,
	journal = {Linear Algebra and its Applications},
	volume = {591},
	pages = {87--98},
	issn = {0024-3795},
	doi = {10.1016/j.laa.2020.01.005}
}

@Article{Guo2017MillimeterWave,
  author  = {Guo, Z. and Wang, X. and Heng, W.},
  journal = {IEEE Trans. Wireless Commun.},
  title   = {Millimeter-{{Wave Channel Estimation Based}} on 2-{{D Beamspace MUSIC Method}}},
  year    = {2017},
  issn    = {1558-2248},
  month   = aug,
  number  = {8},
  pages   = {5384--5394},
  volume  = {16},
  doi     = {10.1109/TWC.2017.2710049},
}

@Article{Ayach2014Spatially,
  author  = {Ayach, O. E. and Rajagopal, S. and {Abu-Surra}, S. and Pi, Z. and Heath, R. W.},
  journal = {IEEE Trans. Wireless Commun.},
  title   = {Spatially {{Sparse Precoding}} in {{Millimeter Wave MIMO Systems}}},
  year    = {2014},
  issn    = {1558-2248},
  month   = mar,
  number  = {3},
  pages   = {1499--1513},
  volume  = {13},
  doi     = {10.1109/TWC.2014.011714.130846},
  ids     = {Ayach2014Spatiallya,Ayach2014Spatiallyb},
}

@book{Zhang2017Matrix,
	title = {Matrix {{Analysis}} and {{Applications}}},
	author = {Zhang, Xian-Da},
	year = {2017},
	month = oct,
	publisher = {{Cambridge University Press}},
	googlebooks = {YBs0DwAAQBAJ},
	isbn = {978-1-108-41741-9},
	pages = {325--329}
}

@Article{Wei2020Calibration,
  author  = {Wei, X. and Jiang, Y. and Liu, Q. and Wang, X.},
  journal = {IEEE Trans. Signal Process.},
  month   = apr,
  pages   = {2302--2315},
  title   = {Calibration of {{Phase Shifter Network}} for {{Hybrid Beamforming}} in {{mmWave Massive MIMO Systems}}},
  volume  = {68},
  year    = {2020},
  issn    = {1941-0476},
}

@article{Luo2016Multiuser,
	title = {Multiuser {{Massive MIMO Performance With Calibration Errors}}},
	author = {Luo, Xiliang},
	year = {2016},
	month = jul,
	journal = {IEEE Trans. Wirel. Commun.},
	volume = {15},
	number = {7},
	pages = {4521--4534},
	issn = {1558-2248}
}

@article{Zhang2022MMVBased,
	title = {{{MMV-Based Sequential AoA}} and {{AoD Estimation}} for {{Millimeter Wave MIMO Channels}}},
	author = {Zhang, Wei and Dong, Miaomiao and Kim, Taejoon},
	year = {2022},
	month = jun,
	journal = {IEEE Trans. Commun.},
	volume = {70},
	number = {6},
	pages = {4063--4077},
	issn = {1558-0857}
}

@article{Venugopal2017Channel,
	title = {Channel {{Estimation}} for {{Hybrid Architecture-Based Wideband Millimeter Wave Systems}}},
	author = {Venugopal, Kiran and Alkhateeb, Ahmed and Gonz{\'a}lez Prelcic, Nuria and Heath, Robert W.},
	year = {2017},
	month = sep,
	journal = {IEEE J. Sel. Areas Commun.},
	volume = {35},
	number = {9},
	pages = {1996--2009},
	issn = {1558-0008}
}

@article{Sohrabi2016Hybrid,
	title = {Hybrid {{Digital}} and {{Analog Beamforming Design}} for {{Large-Scale Antenna Arrays}}},
	author = {Sohrabi, Foad and Yu, Wei},
	year = {2016},
	month = apr,
	journal = {IEEE J. Sel. Topics Signal Process.},
	volume = {10},
	number = {3},
	pages = {501--513},
	issn = {1941-0484}
}

@article{Yu2016Alternating,
	title = {Alternating {{Minimization Algorithms}} for {{Hybrid Precoding}} in {{Millimeter Wave MIMO Systems}}},
	author = {Yu, Xianghao and Shen, Juei-Chin and Zhang, Jun and Letaief, Khaled B.},
	year = {2016},
	month = apr,
	journal = {IEEE J. Sel. Topics Signal Process.},
	volume = {10},
	number = {3},
	pages = {485--500},
	issn = {1941-0484}
}

@article{Lin2019Hybrida,
	title = {Hybrid {{Beamforming}} for {{Millimeter Wave Systems Using}} the {{MMSE Criterion}}},
	author = {Lin, Tian and Cong, Jiaqi and Zhu, Yu and Zhang, Jun and Ben Letaief, Khaled},
	year = {2019},
	journal = {IEEE Trans. Commun.},
	volume = {67},
	number = {5},
	pages = {3693--3708},
	issn = {1558-0857}
}

@article{Wei2021TxRx,
  title = {Tx-{{Rx Reciprocity Calibration}} for {{Hybrid Massive MIMO Systems}}},
  author = {Wei, Xizixiang and Jiang, Yi and Wang, Xin and Shen, Cong},
  year = {2021},
  journal = {IEEE Wireless  Commun. Lett.},
  pages = {1--1},
  issn = {2162-2345}
}

@inproceedings{Mi2018SelfCalibration,
  title = {Self-{{Calibration}} for {{Massive MIMO}} with {{Channel Reciprocity}} and {{Channel Estimation Errors}}},
  booktitle = {in Proc. IEEE Global Commun. Conf. (GLOBECOM)},
  author = {Mi, De and Zhang, Lei and Dianati, Mehrdad and Muhaidat, Sami and Xiao, Pei and Tafazolli, Rahim},
  year = {2018},
  month = dec,
  pages = {1--7},
  publisher = {{IEEE}},
  address = {{Abu Dhabi, United Arab Emirates}},
  isbn = {978-1-5386-4727-1}
}

@inproceedings{Mi2020Massive,
  title = {Massive {{MIMO}} in {{Mobile Networks}}: {{Self-Calibration}} with {{Channel Estimation Error}}},
  shorttitle = {Massive {{MIMO}} in {{Mobile Networks}}},
  booktitle = {in Proc. ACM MobiArch 15th Workshop Mobility Evolving Internet Archit.},
  author = {Mi, De and Chen, Hongzhi and Gao, Zhen and Zhang, Lei and Xiao, Pei},
  year = {2020},
  month = sep,
  series = {{{MobiArch}}'20},
  pages = {30--35},
  publisher = {{Association for Computing Machinery}},
  address = {{New York, USA}},
  isbn = {978-1-4503-8081-2}
}

@article{Yang2013OffGrid,
  title = {Off-{{Grid Direction}} of {{Arrival Estimation Using Sparse Bayesian Inference}}},
  author = {Yang, Zai and Xie, Lihua and Zhang, Cishen},
  year = {2013},
  journal = {IEEE Trans. Signal Process.},
  volume = {61},
  number = {1},
  pages = {38--43},
  issn = {1941-0476}
}

@article{Mi2017Massive,
  title = {Massive {{MIMO Performance With Imperfect Channel Reciprocity}} and {{Channel Estimation Error}}},
  author = {Mi, De and Dianati, Mehrdad and Zhang, Lei and Muhaidat, Sami and Tafazolli, Rahim},
  year = {2017},
  month = sep,
  journal = {IEEE Trans. Commun.},
  volume = {65},
  number = {9},
  pages = {3734--3749},
  issn = {0090-6778}
}
 \end{refsection}

\end{document}